\documentclass[letterpaper,english,aps,prl,notitlepage,twocolumn,superscriptaddress,longbibliography]{revtex4-1}
\usepackage[T1]{fontenc}
\usepackage[latin9]{inputenc}
\usepackage{color}
\usepackage{array}
\usepackage{booktabs}
\usepackage{mathrsfs}
\usepackage{mathtools}
\usepackage{amsmath}
\usepackage{amssymb}
\usepackage{graphicx}
\usepackage{esint}

\makeatletter

\pdfpageheight\paperheight
\pdfpagewidth\paperwidth

\providecommand{\tabularnewline}{\\}

\makeatother

\usepackage{babel}
\begin{document}
\title{Optimizing Brownian heat engine with shortcut strategy}
\author{Jin-Fu Chen}
\email{chenjinfu@pku.edu.cn}

\affiliation{School of Physics, Peking University, Beijing, 100871, China}
\date{\today}
\begin{abstract}
Shortcuts to isothermality provide a powerful method to speed up quasistatic
thermodynamic processes within finite-time manipulation. We employ
the shortcut strategy to design and optimize Brownian heat engines,
and formulate a geometric description of the energetics with the thermodynamic
length. We obtain a tight and reachable bound of the output power,
which is reached by the optimal protocol to vary the control parameters
with a proper constant velocity of the thermodynamic length. Our results
generalize the previous optimization in the highly underdamped and
the overdamped regimes to the general-damped situation, and are applicable
for arbitrary finite-time cycles.
\end{abstract}
\maketitle
\textit{Introduction.} In the past few decades, the flourishing stochastic
thermodynamics has brought great interest in studying the nonequilibrium
thermodynamics of microscopic systems featured with fluctuations \citep{Sekimoto2010,Jarzynski2011,Seifert2012,Broeck2015,Esposito2009,Campisi2011,Holubec2021}.
Downsized to microscopic systems, the heat engines have been invented
with a single trapped ion or a Brownian particle in experiments \citep{Abah2012,Martinez2015}.
Seeking the optimal control schemes of microscopic systems is crucial
to designing microscopic machines with high accuracy and low irreversibility.
Various methods have been proposed to optimize the control schemes
of heat-engine cycles, for example, the optimal control theory to
find the optimal configuration of ideal-gas \citep{Rubin1979,Rubin1979a}
and two-level heat engines \citep{Cavina2018}, and the thermodynamic
geometry to optimize slow isothermal processes \citep{Salamon1983,Crooks2007,Sivak2012,Gong2016,Scandi2019,Brandner2020,Salazar2020,Chen2021}.
For Brownian heat engines, the optimal control of the cycle is known
in the highly underdamped \citep{Dechant2017,Chen2021b} and the overdamped
regimes \citep{Schmiedl2007a,Schmiedl2007}.

Shortcuts to isothermality were recently proposed for the Brownian
motion model \citep{Martinez2016,Li2017}, and experimentally realized
with the optical tweezers \citep{Albay2019,Albay2020,Albay2020a}.
By implementing an auxiliary Hamiltonian, the system is steered to
evolve along instantaneous equilibrium states of the original Hamiltonian
within finite-time manipulation. Such a strategy is feasible to speed
up heat-engine cycles \citep{Pancotti2020,Nakamura2020,Plata2020}
and the control of biophysical processes \citep{Iram2020,Ilker2021}.
In addition, the temperature can also be treated as a time-dependent
control parameter according to the generalization in Refs. \citep{Jun2021,Jun2022}.
In the shortcut strategy, the thermodynamic cost, as the irreversible
work $W_{\mathrm{irr}}$ to implement the auxiliary Hamiltonian, is
bounded by the thermodynamic length $\mathcal{L}$ as \citep{Li2021a,Jun2022}
\begin{equation}
W_{\mathrm{irr}}\geq\mathcal{L}^{2}/\tau,\label{eq:Wirr>=00003DL^2/tau}
\end{equation}
It is remarkable that this bound holds for arbitrary finite-time shortcut
processes. The equality is saturated by the optimal protocol to vary
the control parameters with a constant velocity of the thermodynamic
length \citep{Li2021a}.

In this Letter, we analyze the thermodynamic cost of shortcut processes
with both work parameter and temperature time-dependent, and formulate
a geometric description of the energetics based on the thermodynamic
length. Previous studies of thermodynamic length associated with the
corresponding optimization are usually limited to the slow-driving
regime \citep{Salamon1983,Crooks2007,Esposito2010,Sivak2012,Tu2012,Wang2012,Gong2016,Ryabov2016,Cavina2017,Ma2018a,Scandi2019,Brandner2020,Salazar2020,Abiuso2020,PhysRevLett.128.230601,Chen2021,Watanabe2022,Yuan2022,Ye2022}.
Our setup does not have this limitation. We employ the shortcut strategy
to design and optimize the Brownian heat engine in the general-damped
situation, and derive the maximum power and the efficiency at the
maximum power (EMP) expressed by the thermodynamic length. We also
obtain the optimal protocol of the cycle to reach the maximum power.

\textit{Setup.} We consider a heat engine with a single Brownian particle
as the working substance. The probability distribution $\rho=\rho(x,p,t)$
for the Brownian particle evolves according to the complete Fokker-Planck
equation (the Kramer equation) \citep{Kramers1940}

\begin{equation}
\frac{\partial\rho}{\partial t}=\mathscr{L}\left[\rho\right]+\mathscr{D}\left[\rho\right],\label{eq:Kramer_eq}
\end{equation}
where $\mathscr{L}\left[\rho\right]=-\partial_{x}\left(\rho\partial_{p}H\right)+\partial_{p}\left(\rho\partial_{x}H\right)$
and $\mathscr{D}\left[\rho\right]=\kappa m\partial_{p}\left(\rho\partial_{p}H+\partial_{p}\rho/\beta\right)$
reflect the deterministic and the dissipative parts of the evolution
with the total Hamiltonian $H$, the mass $m$ of the particle, the
friction coefficient $\kappa$, and the inverse temperature $\beta$
of the environment.

In the shortcut strategy \citep{Li2017}, the total Hamiltonian is
$H=H_{\mathrm{o}}+H_{\mathrm{a}}$. An auxiliary Hamiltonian $H_{\mathrm{a}}(x,p,t)$
is implemented to steer the system evolving along instantaneous equilibrium
states $\rho_{\mathrm{ieq}}\coloneqq\exp\{\beta[F(\lambda,\beta)-H_{\mathrm{o}}(x,p,\lambda)]\}$
of the original Hamiltonian $H_{\mathrm{o}}(x,p,\lambda)=p^{2}/(2m)+U_{\mathrm{o}}(x,\lambda)$
with the potential $U_{\mathrm{o}}(x,\lambda)$ and the work parameter
$\lambda$. The free energy $F(\lambda,\beta)$ is determined by $\beta F(\lambda,\beta)=-\ln\{\iint\exp[-\beta H_{\mathrm{o}}(x,p,\lambda)]dxdp\}$.
The auxiliary Hamiltonian is solved in the form $H_{\mathrm{a}}(x,p,t)=\dot{\lambda}h_{\lambda}(x,p,\lambda)+\dot{\beta}h_{\beta}(x,p,\lambda,\beta)$,
where $h_{\lambda}(x,p,\lambda)$ and $h_{\beta}(x,p,\lambda,\beta)$
are two auxiliary functions related to the original Hamiltonian \citep{Li2017,Jun2021,supplementary_material}.

We employ the shortcut strategy to construct a heat-engine cycle.
The work parameter $\lambda$ and the inverse temperature $\beta$
are time-dependent control parameters, and are cyclical functions
of the operation time. The input power and the heat flux are defined
as $\dot{W}\coloneqq\left\langle \partial_{t}H\right\rangle $ and
$\dot{Q}:=\iint H\partial_{t}\rho_{\mathrm{ieq}}dxdp$ \citep{Sekimoto2010},
where the average is over an instantaneous equilibrium state $\left\langle \cdot\right\rangle \coloneqq\iint(\cdot)\rho_{\mathrm{ieq}}dxdp$.
Positive $\dot{W}$ and $\dot{Q}$ indicate the work is performed
on the system and the heat is absorbed by the system from the environment,
respectively. We divide $\dot{W}$ and $\dot{Q}$ into the quasistatic
and the irreversible parts

\begin{align}
\dot{W} & =\dot{W}_{\mathrm{o}}+\dot{W}_{\mathrm{irr}},\\
\dot{Q} & =\dot{Q}_{\mathrm{o}}+\dot{Q}_{\mathrm{irr}},
\end{align}
according to $H_{\mathrm{o}}$ and $H_{\mathrm{a}}$. The quasistatic
and the irreversible input powers are $\dot{W}_{\mathrm{o}}=\dot{\lambda}\left\langle \partial_{\lambda}H_{\mathrm{o}}\right\rangle $
and $\dot{W}_{\mathrm{irr}}=\left\langle \partial_{t}H_{\mathrm{a}}\right\rangle $.
The quasistatic and the irreversible heat fluxes are $\dot{Q}_{\mathrm{o}}=\iint H_{\mathrm{o}}\partial_{t}\rho_{\mathrm{ieq}}dxdp$
and $\dot{Q}_{\mathrm{irr}}=\iint H_{\mathrm{a}}\partial_{t}\rho_{\mathrm{ieq}}dxdp$.
The first law of thermodynamics is satisfied for each part $\partial_{t}\left\langle H_{\mathrm{o}}\right\rangle =\dot{W}_{\mathrm{o}}+\dot{Q}_{\mathrm{o}}$
and $\partial_{t}\left\langle H_{\mathrm{a}}\right\rangle =\dot{W}_{\mathrm{irr}}+\dot{Q}_{\mathrm{irr}}$.
In isothermal processes ($\beta(t)\equiv\beta$), the quasistatic
work is equal to the free energy change $W_{\mathrm{o}}=F(\lambda(\tau),\beta)-F(\lambda(0),\beta)$
\citep{Li2017}. In general shortcut processes with both $\lambda(t)$
and $\beta(t)$ time-dependent, the quasistatic work $W_{\mathrm{o}}=\int\lambda^{\prime}(s)\left\langle \partial_{\lambda}H_{\mathrm{o}}\right\rangle ds$
relies on the path $(\lambda(s),\beta(s))$ in the control-parameter
space, but is independent of the protocol and the operation time.
The irreversible heat is carried out as \citep{supplementary_material}

\begin{equation}
Q_{\mathrm{irr}}=-\int_{0}^{\tau}\left(\begin{array}{cc}
\dot{\lambda} & \dot{\beta}\end{array}\right)\mathbf{g}\left(\begin{array}{c}
\dot{\lambda}\\
\dot{\beta}
\end{array}\right)dt,\label{eq:Qirr}
\end{equation}
with the metric 
\begin{equation}
\mathbf{g}=\kappa m\left(\begin{array}{cc}
\left\langle \frac{\partial h_{\lambda}}{\partial p}\frac{\partial h_{\lambda}}{\partial p}\right\rangle  & \left\langle \frac{\partial h_{\lambda}}{\partial p}\frac{\partial h_{\beta}}{\partial p}\right\rangle \\
\left\langle \frac{\partial h_{\lambda}}{\partial p}\frac{\partial h_{\beta}}{\partial p}\right\rangle  & \left\langle \frac{\partial h_{\beta}}{\partial p}\frac{\partial h_{\beta}}{\partial p}\right\rangle 
\end{array}\right).
\end{equation}
The irreversible work is

\begin{equation}
W_{\mathrm{irr}}=\left\langle H_{\mathrm{a}}\right\rangle (\tau^{+})-\left\langle H_{\mathrm{a}}\right\rangle (0^{-})-Q_{\mathrm{irr}},
\end{equation}
where $0^{-}$ and $\tau^{+}$ denote the initial time and the final
time of the shortcut process. The auxiliary Hamiltonian $H_{\mathrm{a}}$
is switched on at the beginning and off at the end of the shortcut
process, and thus $\left\langle H_{\mathrm{a}}\right\rangle (\tau^{+})-\left\langle H_{\mathrm{a}}\right\rangle (0^{-})=0$.
This term also vanishes in a heat-engine cycle due to the cyclical
variation of the control parameters.\textbf{ }The metric $\mathbf{g}$
is positive semi-definite \citep{supplementary_material}. Therefore,
the irreversible work is non-negative $W_{\mathrm{irr}}\geq0$. For
a given path $(\lambda(s),\beta(s))$ in the control-parameter space,
the thermodynamic length is defined as

\begin{equation}
\mathcal{L}\coloneqq\int\sqrt{\left(\begin{array}{cc}
\lambda^{\prime}(s) & \beta^{\prime}(s)\end{array}\right)\mathbf{g}\left(\begin{array}{c}
\lambda^{\prime}(s)\\
\beta^{\prime}(s)
\end{array}\right)}ds,
\end{equation}
which provides a tight and reachable bound (\ref{eq:Wirr>=00003DL^2/tau})
of the irreversible work in general shortcut processes. 

By applying the bound (\ref{eq:Wirr>=00003DL^2/tau}) to a heat-engine
cycle, the output power $P\coloneqq-W/\tau$ is bounded by $P\leq-W_{\mathrm{o}}/\tau-\mathcal{L}^{2}/\tau^{2}$.
The equality is saturated with the optimal protocol, and the irreversible
heat flux becomes a negative constant during the whole cycle

\begin{equation}
\dot{Q}_{\mathrm{irr}}\equiv-\frac{\mathcal{L}^{2}}{\tau^{2}}.\label{eq:Qdot_irr}
\end{equation}
By further choosing the operation time $\tau=\tau_{\mathrm{max}}\coloneqq2\mathcal{L}^{2}/(-W_{\mathrm{o}})$,
the maximum power is reached

\begin{equation}
P_{\mathrm{max}}=\frac{(-W_{\mathrm{o}})^{2}}{4\mathcal{L}^{2}}.\label{eq:maximumpower_general}
\end{equation}
Further optimization of the output power of the cycle is converted
to finding out the closed path with a large ratio $-W_{\mathrm{o}}/\mathcal{L}$.

According to positive (negative) quasistatic heat flux $\dot{Q}_{\mathrm{o}}>0$
($\dot{Q}_{\mathrm{o}}<0$), we divide the cycle into the heat absorbed
(released) path with the thermodynamic length $\mathcal{\mathcal{L}}_{+}$
($\mathcal{\mathcal{L}}_{-}$) satisfying $\mathcal{\mathcal{L}}=\mathcal{\mathcal{L}}_{+}+\mathcal{\mathcal{L}}_{-}$.
The overall heat absorbed (released) on each path is $Q_{+}=Q_{\mathrm{o},+}+Q_{\mathrm{irr},+}$
($Q_{-}=Q_{\mathrm{o},-}+Q_{\mathrm{irr},-}$). The efficiency of
the cycle is defined as $\eta\coloneqq-W/Q_{+}$. In the optimal protocol,
the overall heat absorbed\textbf{ }is explicitly $Q_{+}=Q_{\mathrm{o},+}-\mathcal{\mathcal{L}}_{+}\mathcal{\mathcal{L}}/\tau$.
The variation of the temperature is constrained between the lowest
$T_{L}$ and the highest temperatures $T_{H}$. At the maximum power,
the efficiency is expressed by the thermodynamic length as \citep{supplementary_material}

\begin{equation}
\eta_{\mathrm{EMP}}=\frac{\eta_{\mathrm{o}}}{2-\eta_{\mathrm{o}}\mathcal{\mathcal{L}}_{+}/\mathcal{L}},\label{eq:eta_EMP}
\end{equation}
where $\eta_{\mathrm{o}}\coloneqq-W_{\mathrm{o}}/Q_{\mathrm{o},+}$
is the efficiency of the quasistatic cycle and is less than the Carnot
efficiency $\eta_{\mathrm{C}}\coloneqq1-T_{L}/T_{H}$. We also obtain
the maximum power $P_{\eta}$ at given efficiency $\eta$ as \citep{supplementary_material}

\begin{align}
\frac{P_{\eta}}{P_{\mathrm{max}}} & =\frac{4(\eta_{\mathrm{o}}-\eta)(1-\eta_{\mathrm{o}}\mathcal{L}_{+}/\mathcal{L})\eta}{\eta_{\mathrm{o}}^{2}(1-\eta\mathcal{L}_{+}/\mathcal{L})^{2}}.\label{eq:10}
\end{align}
Equations (\ref{eq:maximumpower_general})-(\ref{eq:10}) are the
main result of this work, and the detailed derivations are left in
\citep{supplementary_material}.

\textit{Application to power-law potentials.} We next realize the
Brownian heat engine with the shortcut strategy in a class of power-law
potentials $U_{\mathrm{o}}(x,\lambda)=m\lambda^{n+1}x^{2n}/(2n)$\textcolor{black}{{}
}\footnote{This parameterization simplifies the expressions of the thermodynamic
length, and allows the same optimal protocol to vary $\lambda$ in
the highly underdamped regime. For the harmonic potential ($n=1$),
the work parameter $\lambda$ becomes the frequency}, where $n$ and $\lambda$ characterize the shape and the stiffness
of the potential. The auxiliary Hamiltonian $H_{\mathrm{a}}$ for
these potentials has been derived in Ref. \citep{Jun2021}, with the
auxiliary functions

\begin{align}
h_{\lambda} & =\frac{f_{n}}{4\kappa\lambda m}(p-\kappa mx)^{2}+\frac{f_{n}m}{4\kappa n}\lambda^{n}x^{2n},\\
h_{\beta} & =\frac{np^{2}+\left(p-\kappa mx\right)^{2}+m^{2}f_{n}\lambda^{n+1}x^{2n}}{4\beta\kappa mn},
\end{align}
where $f_{n}=1+1/n$ is the effective degree of freedom \citep{Salazar2020}.

We use a dimensionless work parameter $r=\ln(\lambda/\lambda_{0})$
and the temperature $T=1/\beta$ to represent the control parameters,
and choose a closed path for the cycle. The quasistatic output work
is obtained as $-W_{\mathrm{o}}=-(f_{n}/2)\oint Tdr$, which is proportional
to the area $\mathcal{A}=\left|\oint Tdr\right|$ of the closed path.
The thermodynamic length of the cycle is

\begin{equation}
\mathcal{L}=\frac{f_{n}}{2}\oint\sqrt{\frac{(Tdr-dT)^{2}}{\kappa T}+\frac{\chi}{\kappa T}\left(Tdr-\frac{dT}{n+1}\right)^{2}},\label{eq:thermodynamiclength_a_cycle}
\end{equation}
where $\chi$ is a dimensionless quantity depending on the control
parameters 

\begin{equation}
\chi=\frac{R_{n}\kappa^{2}}{\lambda_{0}^{f_{n}}e^{f_{n}r}}\left(\frac{m}{T}\right)^{1-1/n},
\end{equation}
and $R_{n}=(2n)^{1/n}\Gamma(3/(2n))/\Gamma(1/(2n))$ is a pure number
ranging from $1/3$ (reached by $n\rightarrow\infty$) to $1$ (reached
by $n=1$). $\Gamma(\cdot)$ is the gamma function. The highly underdamped
and the overdamped regimes are reflected by $\chi\ll1$ and $\chi\gg1$,
respectively. The maximum power (\ref{eq:maximumpower_general}) is
explicitly

\begin{equation}
P_{\mathrm{max}}=\frac{f_{n}^{2}\mathcal{A}^{2}}{16\mathcal{L}^{2}},\label{eq:maximumpower_general-1}
\end{equation}
with the operation time $\tau_{\mathrm{max}}=4\mathcal{L}^{2}/(f_{n}\mathcal{A})$.
The area $\mathcal{A}$ and the thermodynamic length $\mathcal{\mathcal{L}}$
of the closed path are obtained geometrically in the $r-T$ diagram.

\begin{figure}
\includegraphics[width=7cm]{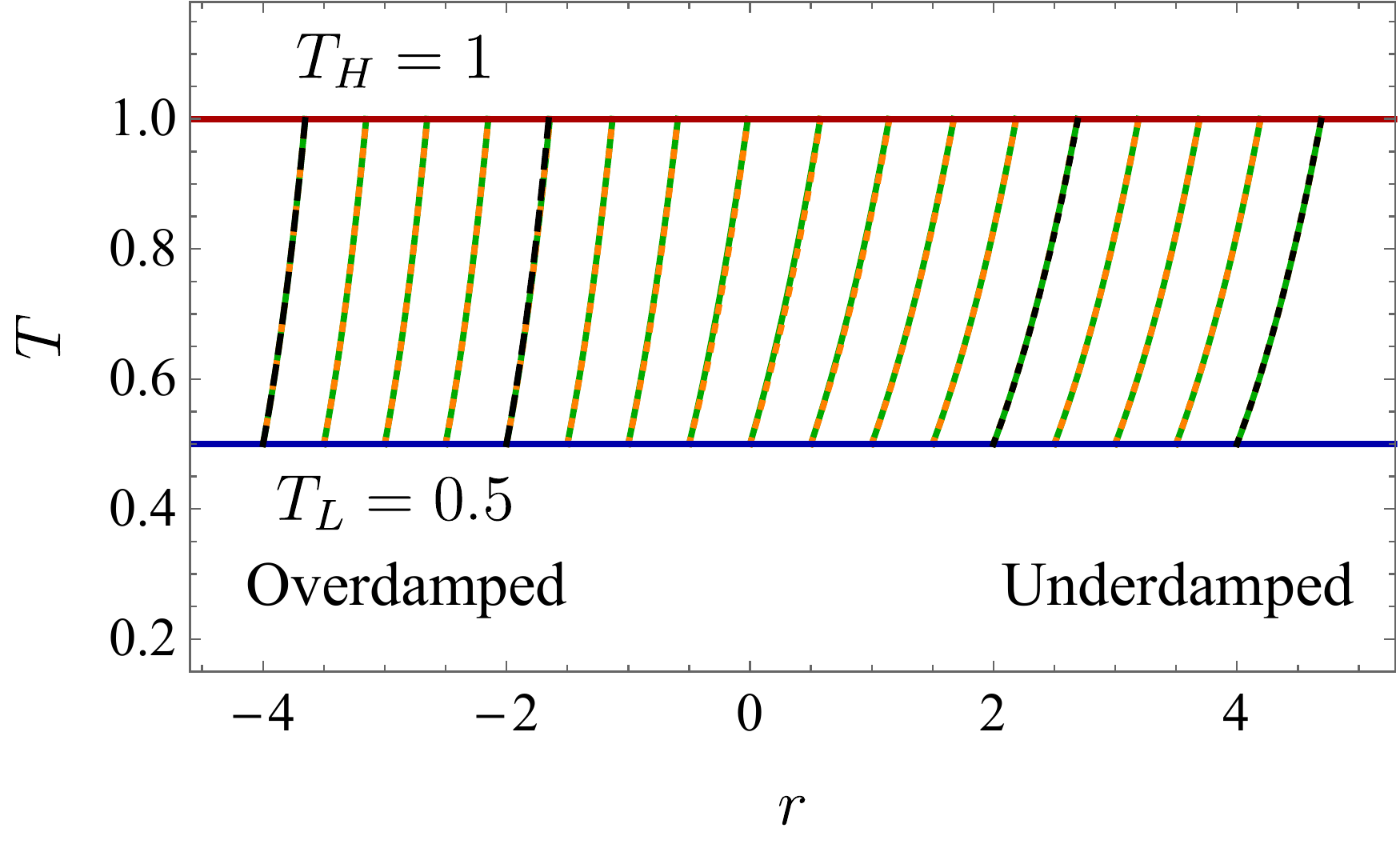}

\caption{The $r-T$ diagram with the harmonic potential $n=1$. We set the
parameters $\lambda_{0}=1$, $m=1$, $\kappa=1$, and the temperatures
$T_{L}=0.5$ and $T_{H}=1$, and use these values in all the later
discussions. The shortest geodesic path (green solid curves) and the
shortest exponential path (orange dashed curves) almost coincide.
We also show the zero-length paths $\alpha=1$ (black dotted curve)
and\textbf{ }$\alpha=n+1$ (black dashed curve) in the highly underdamped
and the overdamped regimes. \label{fig:Optimal_underdamped}}
\end{figure}

To construct a cycle, we pick up four boundary points $(r_{1},T_{L})$,
$(r_{2},T_{L})$, $(r_{1}^{\prime},T_{H})$ and $(r_{2}^{\prime},T_{H})$
with $r_{2}>r_{1}$ and $r_{2}^{\prime}>r_{1}^{\prime}$ in the control-parameter
space, and connect these points to form a closed path consisting of
two isothermal and two connecting paths. In the optimal protocol of
the cycle, the control parameters are varied with the same velocity
of the thermodynamic length in both the isothermal and the connecting
processes. To achieve possibly large output power, we construct the
cycle with a large ratio $\mathcal{A}/\mathcal{L}$. An efficient
choice of the connecting paths is the geodesic paths according to
the metric (\ref{eq:thermodynamiclength_a_cycle}). The explicit expressions
of the geodesic equations are left in \citep{supplementary_material}.
Initiated from one point on the low-temperature line $T=T_{L}$, we
solve the shortest geodesic path to reach the high-temperature line
$T=T_{H}$ with the shooting method \citep{Berger2007} (green solid
curves in Fig. \ref{fig:Optimal_underdamped}). We choose the two
connecting paths as the shortest geodesic paths. Thus, $r_{i}^{\prime}$
is determined by $r_{i}$. The area $\mathcal{A}_{\mathrm{geo}}(r_{1},r_{2})$
and the thermodynamic length $\mathcal{L}_{\mathrm{geo}}(r_{1},r_{2})$
of the cycle are functions of $r_{1}$ and $r_{2}$, and can be numerically
calculated.

\begin{table}
\begin{tabular}{>{\centering}m{0.1\linewidth}>{\centering}m{0.76\linewidth}}
\toprule 
$\alpha$ & $\mathcal{L}_{\alpha}(r_{0})$\tabularnewline
\midrule 
$1$ & $\sqrt{R_{n}\frac{\kappa m(T_{L}/m)^{1/n}}{\lambda_{0}^{f_{n}}e^{f_{n}r_{0}}}}(1-\sqrt{\frac{T_{L}}{T_{H}}})$\tabularnewline
$n+1$ & $\sqrt{\frac{T_{H}}{\kappa}}(1-\sqrt{\frac{T_{L}}{T_{H}}})$\tabularnewline
\bottomrule
\end{tabular}

\caption{Expressions of the thermodynamic lengths of the exponential paths
with $\alpha=1$ and $\alpha=n+1$. \label{tab:Thermodynamic-length-of}}
\end{table}

We can alternatively choose the connecting paths as the exponential
path $T(r)=T_{L}\exp[\alpha(r-r_{0})]$, $r_{0}\leq r\leq r_{0}^{\prime}\coloneqq r_{0}+\ln(T_{H}/T_{L})/\alpha$.
The thermodynamic length $\mathcal{L}_{\alpha}(r_{0})$ of the exponential
path is obtained analytically in \citep{supplementary_material},
and given in Table \ref{tab:Thermodynamic-length-of} for $\alpha=1$
and $n+1$. For given $r_{0}$, we can also optimize $\alpha$ to
minimize the thermodynamic length $\mathcal{L}_{\alpha}(r_{0})$ of
the exponential path, and obtain the shortest exponential path (orange
dotted curve in Fig. \ref{fig:Optimal_underdamped}), which almost
coincides with the shortest geodesic path (but not exactly). We remark
that the cycles with the exponential paths contain several cycles
in classical thermodynamics, e.g., the Carnot ($\alpha_{1,2}=1$)
and the Stirling cycles ($\alpha_{1,2}=\infty$). The quasistatic
output work $-W_{\mathrm{o}}=(f_{n}/2)\mathcal{A}_{\mathrm{exp}}(r_{1},\alpha_{1},r_{2},\alpha_{2})$,
the quasistatic efficiency $\eta_{\mathrm{o}}$, and the thermodynamic
length $\mathcal{L}_{\mathrm{exp}}(r_{1},\alpha_{1},r_{2},\alpha_{2})$
of these cycles are derived analytically in \citep{supplementary_material}.
We compare the shortest geodesic paths and the shortest exponential
paths for the harmonic potential $n=1$ in Fig. \ref{fig:Optimal_underdamped}.

\begin{figure}
\includegraphics[width=8.3cm]{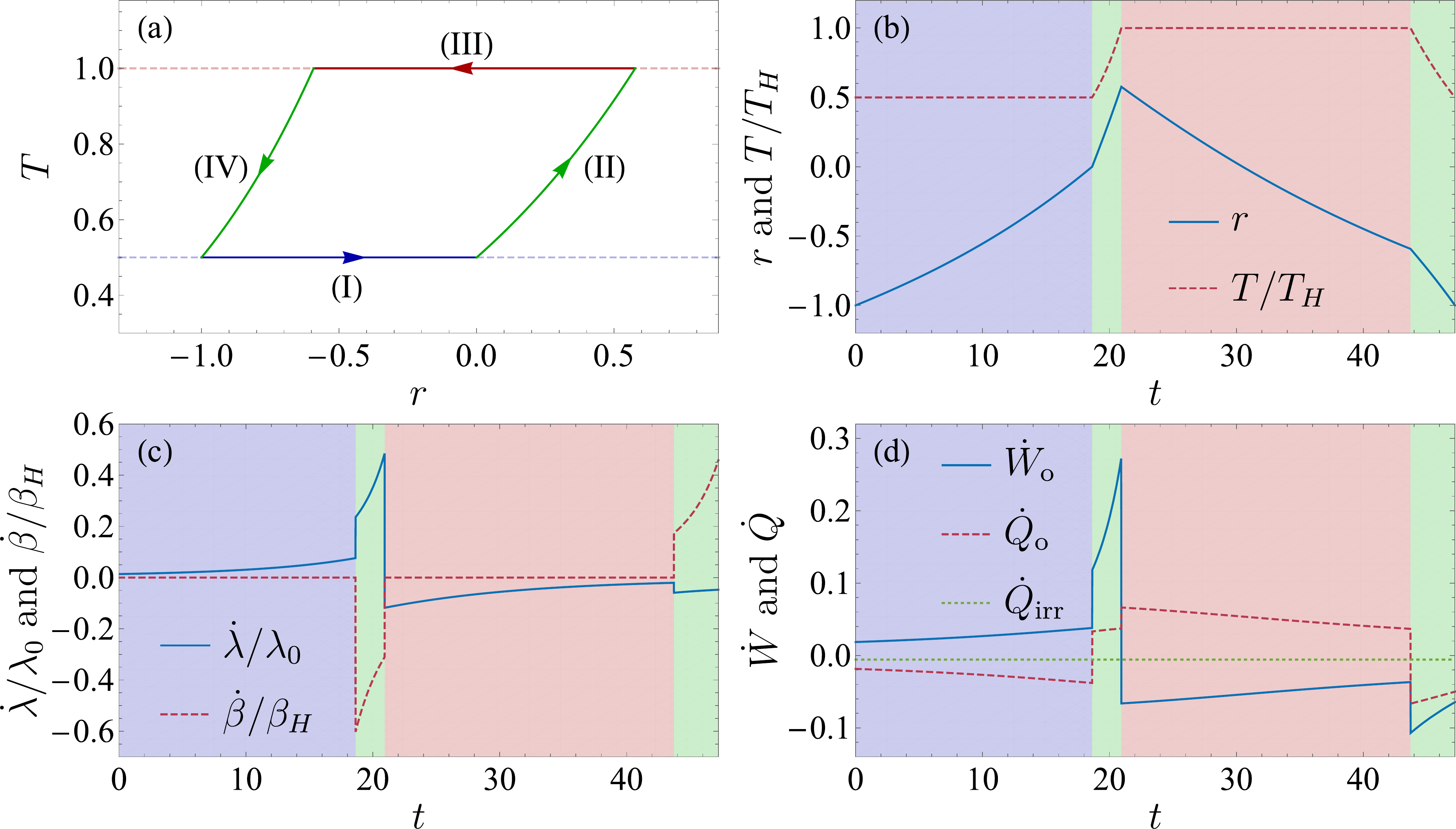}

\caption{The control scheme to achieve the maximum power. (a) The $r-T$ diagram
with $r_{1}=-1$ and $r_{2}=0$. The connecting paths (II and IV)
are the shortest geodesic paths. The whole cycle is divided into four
processes (I)--(IV). (b) The variation of the control parameters
$r(t)$ and $T(t)$ in a cycle. (c) The implementation of the auxiliary
Hamiltonian represented by $\dot{\lambda}$ and $\dot{\beta}$. (d)
The input power $\dot{W}$ and the heat flux $\dot{Q}$. \label{fig:cyclic-control-of}}
\end{figure}

Figure \ref{fig:cyclic-control-of} shows the control scheme of the
heat engine with the shortcut strategy to achieve the maximum power
with the boundary values $r_{1}=-1$ and $r_{2}=0$ of the work parameter.
The connecting paths are chosen as the shortest geodesic paths (II
and IV). The area and the thermodynamic length of the cycle are $\mathcal{A}_{\mathrm{geo}}(-1,0)=0.546$
and $\mathcal{L}_{\mathrm{geo}}(-1,0)=3.59$. The maximum power $P_{\mathrm{max}}=0.00578$
is reached with the operation time $\tau_{\mathrm{max}}=47.2$. The
cycle contains four processes, (I) isothermal compression, (II) connecting
compression, (III) isothermal expansion, and (IV) connecting expansion.\textbf{
}The control scheme to vary the control parameters $r(t)$ and $T(t)$
is shown in Fig. \ref{fig:cyclic-control-of}(b). Figure \ref{fig:cyclic-control-of}(c)
shows the implementation of the auxiliary Hamiltonian characterized
by $\dot{\lambda}$ and $\dot{\beta}$. Figure \ref{fig:cyclic-control-of}(d)
plots the input power and the heat flux in a cycle. The quasistatic
heat is absorbed $\dot{Q}_{\mathrm{o}}>0$ in processes (II) and (III),
and released $\dot{Q}_{\mathrm{o}}<0$ in processes (I) and (IV).
The irreversible heat flux (\ref{eq:Qdot_irr}) is \textbf{$\dot{Q}_{\mathrm{irr}}=-0.00578$}
during the whole cycle.

\textit{Discussions.} In the highly underdamped regime $\chi\ll1$,
the thermodynamic length {[}Eq. (\ref{eq:thermodynamiclength_a_cycle}){]}
is simplified into $\mathcal{L}^{\mathrm{under}}=(f_{n}/2)\oint\sqrt{(T-dT/dr)^{2}/(\kappa T)}dr$,
which becomes zero on the exponential paths with $\alpha=1$. The
connecting processes on these paths are free of the irreversible work
and does not cost any time (compared to the isothermal processes).
Thus, we choose two of these exponential paths $T(r)=T_{L}\exp(r-r_{i}),i=1,2$
to construct the optimal cycle in the highly underdamped regime. The
area of the closed path is

\begin{align}
\mathcal{A} & =(T_{H}-T_{L})(r_{2}-r_{1}).\label{eq:26}
\end{align}
The thermodynamic length of the cycle is \citep{supplementary_material}

\begin{equation}
\mathcal{L}^{\mathrm{under}}=\frac{f_{n}}{2}\left(\frac{\sqrt{T_{H}}+\sqrt{T_{L}}}{\sqrt{\kappa}}\right)(r_{2}-r_{1}).
\end{equation}
The maximum power (\ref{eq:maximumpower_general-1}) is

\begin{equation}
P_{\mathrm{max}}^{\mathrm{under}}=\frac{\kappa}{4}(\sqrt{T_{H}}-\sqrt{T_{L}})^{2}.\label{eq:maximumpowerunderdamped}
\end{equation}
The quasistatic cycle is the Carnot cycle with the efficiency $\eta_{\mathrm{o}}=\eta_{\mathrm{C}}$.
The EMP (\ref{eq:eta_EMP}) becomes the Curzon-Ahlborn efficiency
\citep{Curzon1975} 
\begin{equation}
\eta_{\mathrm{EMP}}^{\mathrm{under}}=\eta_{\mathrm{CA}}\coloneqq1-\sqrt{T_{L}/T_{H}},\label{eq:emp_under}
\end{equation}
since the highly underdamped regime satisfies the preconditions of
the Curzon-Ahlborn heat engines \citep{Chen2021b}. Both the maximum
power (\ref{eq:maximumpowerunderdamped}) and the EMP (\ref{eq:emp_under})
agree with the results in Refs. \citep{Dechant2017,Chen2021b}, and
are independent of the shape of the potential $n$ and the choices
of $r_{1}$ and $r_{2}$. With the shortcut strategy, the optimal
protocol of the cycle and the maximum power at given efficiency differ
from those in Refs. \citep{Dechant2017,Chen2021b,Chen1989} (see \citep{supplementary_material}
for the discussion).

In the overdamped regime $\chi\gg1$, the thermodynamic length {[}Eq.
(\ref{eq:thermodynamiclength_a_cycle}){]} is simplified into 
\begin{align}
\mathcal{L}^{\mathrm{over}} & =\frac{f_{n}}{2}\oint\sqrt{\frac{\chi}{\kappa T}\left(Tdr-\frac{1}{n+1}\frac{dT}{dr}\right)^{2}}dr,
\end{align}
which becomes zero on the exponential paths with $\alpha=n+1$. The
optimal cycle in the overdamped regime are constructed with two of
these exponential paths $T(r)=T_{L}\exp[(n+1)(r-r_{i})],i=1,2$. The
area of the closed path is also given by Eq. (\ref{eq:26}), while
the thermodynamic length of the cycle is

\begin{align}
\mathcal{L}^{\mathrm{over}} & =2\sqrt{R_{n}\frac{\kappa m}{\lambda_{0}}\left(\frac{T_{L}}{\lambda_{0}m}\right)^{1/n}}\left(e^{-f_{n}r_{1}/2}-e^{-f_{n}r_{2}/2}\right).
\end{align}
The maximum power (\ref{eq:maximumpower_general-1}) is

\begin{align}
P_{\mathrm{max}}^{\mathrm{over}} & =\frac{f_{n}^{2}}{64R_{n}}\frac{\lambda_{0}^{f_{n}}m^{1/n}}{\kappa mT_{L}^{1/n}}\frac{(T_{H}-T_{L})^{2}(r_{2}-r_{1})^{2}}{\left(e^{-f_{n}r_{1}/2}-e^{-f_{n}r_{2}/2}\right)^{2}}.\label{eq:maximumpower_overdamped}
\end{align}
The efficiency of the quasistatic cycle is \citep{supplementary_material}
\begin{equation}
\eta_{\mathrm{o}}^{\mathrm{over}}=\frac{\eta_{\mathrm{C}}}{1+\eta_{\mathrm{C}}/[f_{n}(r_{2}-r_{1})]\}},
\end{equation}
which deviates from the Carnot efficiency due to extra heat exchange
from the kinetic energy change. The EMP (\ref{eq:eta_EMP}) is

\begin{equation}
\eta_{\mathrm{EMP}}^{\mathrm{over}}=\frac{\eta_{\mathrm{C}}}{2+\{4/[f_{n}(r_{2}-r_{1})]-1\}\eta_{\mathrm{C}}/2}.\label{eq:emp_overdamped}
\end{equation}
Both the maximum power (\ref{eq:maximumpower_overdamped}) and the
EMP (\ref{eq:emp_overdamped}) agree with the results in Ref. \citep{Schmiedl2007}.
The optimal protocol with the shortcut strategy in the overdamped
regime also converges to the control scheme of Ref. \citep{Schmiedl2007}
since the auxiliary Hamiltonian becomes a potential of the position
in the overdamped regime (see \citep{supplementary_material} for
the discussion). 

\begin{figure}
\includegraphics[width=7cm]{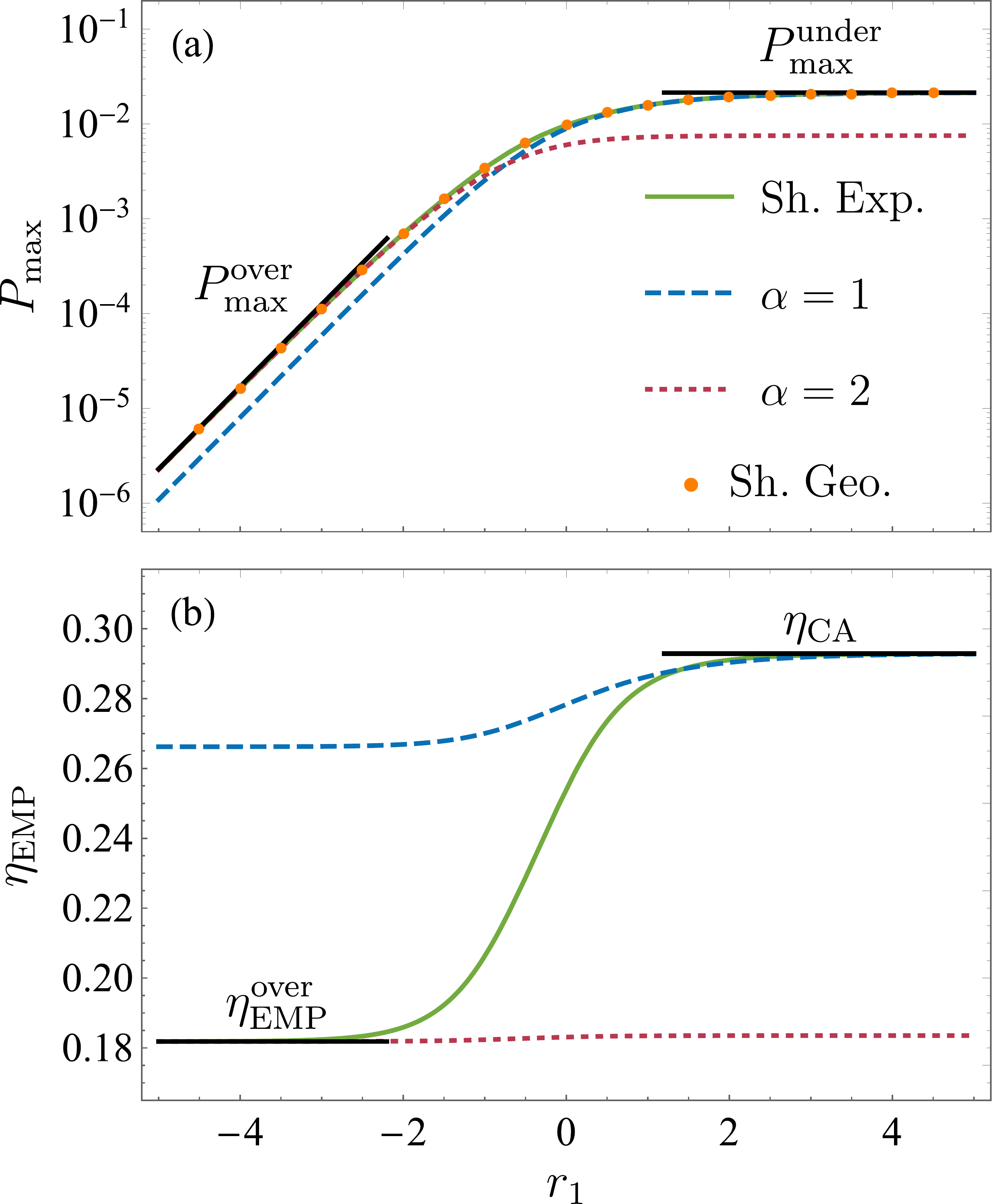}

\caption{The maximum power and the EMP as functions of $r_{1}$. We set $r_{2}=r_{1}+0.5$,
and choose the connecting paths as the shortest exponential paths
(green solid curve) and the exponential paths with fixed $\alpha_{1,2}=1$
(blue dashed curve) and $\alpha_{1,2}=2$ (red dotted curve). (a)
The maximum power $P_{\mathrm{max}}$. The orange dots are the results
for the shortest geodesic paths. (b) The EMP $\eta_{\mathrm{EMP}}$
of the cycle. \label{fig:The-maximum-power}}
\end{figure}

Figure \ref{fig:The-maximum-power} shows the maximum power $P_{\mathrm{max}}$
and the EMP $\eta_{\mathrm{EMP}}$ of different cycles in the general-damped
situation. We choose the connecting paths as the shortest exponential
paths or the exponential paths with fixed $\alpha_{1,2}=1$ and $\alpha_{1,2}=2$.
In Fig. \ref{fig:The-maximum-power}(a), almost the same maximum power
is reached for the shortest exponential paths and the shortest geodesic
paths (orange dots). The black solid lines represent the bounds (\ref{eq:maximumpowerunderdamped})
and (\ref{eq:maximumpower_overdamped}) of the output power obtained
in the highly underdamped and the overdamped regimes. In Fig. \ref{fig:The-maximum-power}(b),
the EMP $\eta_{\mathrm{EMP}}$ of the cycles with the shortest exponential
paths approaches the Curzon-Ahlborn efficiency (\ref{eq:emp_under})
in the highly underdamped regime and Eq. (\ref{eq:emp_overdamped})
in the overdamped regime.

\textit{Conclusion.} Based on the shortcut strategy \citep{Martinez2016,Li2017,Jun2021},
we formulate a geometric description of the energetics in shortcut
processes with both work parameter and temperature time-dependent.
The thermodynamic length not only provides a tight and reachable bound
of the irreversible work, but also guides designing the optimal protocol
of either a single process or a heat-engine cycle. The optimal protocol
is to vary the control parameters with a constant velocity of the
thermodynamic length.

We employ the shortcut strategy to design and optimize the Brownian
heat engine, and obtain the optimal protocol of the cycle. The maximum
power, the EMP, and the maximum power at given efficiency are all
related to the thermodynamic length. Such optimization can be further
extended to specific cycles, e.g., Carnot, Otto, and Stirling cycles,
etc. We illustrate that the EMP approaches the Curzon-Ahlborn efficiency
in the highly underdamped regime \citep{Dechant2017,Chen2021b}, and
converges to the result \citep{Schmiedl2007} in the overdamped regime.
We conclude this Letter that the shortcut strategy is a powerful method
for optimizing the performance of heat engines, and may also be beneficial
for the experimental realization of Brownian heat engines \citep{Martinez2015}.

\textit{Note added: }We notice a related recent work \citep{Zhao2022},
where shortcuts to adiabaticity and shortcuts to isothermality are
introduced simultaneously to realize finite-time Carnot cycles.
\begin{acknowledgments}
J.F.C thanks Hui Dong and H.T. Quan for carefully reading the manuscript.
This work is supported by the National Natural Science Foundation
of China (NSFC) under Grants No. 11775001, No. 11825501, and No. 12147157.
\end{acknowledgments}

\bibliographystyle{apsrev4-1_addtitle}
\bibliography{ref}

\begin{thebibliography}{55}%
\makeatletter
\providecommand \@ifxundefined [1]{%
 \@ifx{#1\undefined}
}%
\providecommand \@ifnum [1]{%
 \ifnum #1\expandafter \@firstoftwo
 \else \expandafter \@secondoftwo
 \fi
}%
\providecommand \@ifx [1]{%
 \ifx #1\expandafter \@firstoftwo
 \else \expandafter \@secondoftwo
 \fi
}%
\providecommand \natexlab [1]{#1}%
\providecommand \enquote  [1]{``#1''}%
\providecommand \bibnamefont  [1]{#1}%
\providecommand \bibfnamefont [1]{#1}%
\providecommand \citenamefont [1]{#1}%
\providecommand \href@noop [0]{\@secondoftwo}%
\providecommand \href [0]{\begingroup \@sanitize@url \@href}%
\providecommand \@href[1]{\@@startlink{#1}\@@href}%
\providecommand \@@href[1]{\endgroup#1\@@endlink}%
\providecommand \@sanitize@url [0]{\catcode `\\12\catcode `\$12\catcode
  `\&12\catcode `\#12\catcode `\^12\catcode `\_12\catcode `\%12\relax}%
\providecommand \@@startlink[1]{}%
\providecommand \@@endlink[0]{}%
\providecommand \url  [0]{\begingroup\@sanitize@url \@url }%
\providecommand \@url [1]{\endgroup\@href {#1}{\urlprefix }}%
\providecommand \urlprefix  [0]{URL }%
\providecommand \Eprint [0]{\href }%
\providecommand \doibase [0]{http://dx.doi.org/}%
\providecommand \selectlanguage [0]{\@gobble}%
\providecommand \bibinfo  [0]{\@secondoftwo}%
\providecommand \bibfield  [0]{\@secondoftwo}%
\providecommand \translation [1]{[#1]}%
\providecommand \BibitemOpen [0]{}%
\providecommand \bibitemStop [0]{}%
\providecommand \bibitemNoStop [0]{.\EOS\space}%
\providecommand \EOS [0]{\spacefactor3000\relax}%
\providecommand \BibitemShut  [1]{\csname bibitem#1\endcsname}%
\let\auto@bib@innerbib\@empty
\bibitem [{\citenamefont {Sekimoto}(2010)}]{Sekimoto2010}%
  \BibitemOpen
  \bibfield  {author} {\bibinfo {author} {\bibfnamefont {K.}~\bibnamefont
  {Sekimoto}},\ }\href
  {https://www.ebook.de/de/product/25407654/ken_sekimoto_stochastic_energetics.html}
  {\emph {\bibinfo {title} {Stochastic Energetics}}}\ (\bibinfo  {publisher}
  {Springer-Verlag GmbH},\ \bibinfo {year} {2010})\BibitemShut {NoStop}%
\bibitem [{\citenamefont {Jarzynski}(2011)}]{Jarzynski2011}%
  \BibitemOpen
  \bibfield  {author} {\bibinfo {author} {\bibfnamefont {C.}~\bibnamefont
  {Jarzynski}},\ }\bibinfo {title} {Equalities and Inequalities:
  Irreversibility and the Second Law of Thermodynamics at the Nanoscale},\
  \href {\doibase 10.1146/annurev-conmatphys-062910-140506} {\bibfield
  {journal} {\bibinfo  {journal} {Annu. Rev. Condens. Matter Phys.}\ }\textbf
  {\bibinfo {volume} {2}},\ \bibinfo {pages} {329} (\bibinfo {year}
  {2011})}\BibitemShut {NoStop}%
\bibitem [{\citenamefont {Seifert}(2012)}]{Seifert2012}%
  \BibitemOpen
  \bibfield  {author} {\bibinfo {author} {\bibfnamefont {U.}~\bibnamefont
  {Seifert}},\ }\bibinfo {title} {Stochastic thermodynamics, fluctuation
  theorems and molecular machines},\ \href {\doibase
  10.1088/0034-4885/75/12/126001} {\bibfield  {journal} {\bibinfo  {journal}
  {Rep. Prog. Phys.}\ }\textbf {\bibinfo {volume} {75}},\ \bibinfo {pages}
  {126001} (\bibinfo {year} {2012})}\BibitemShut {NoStop}%
\bibitem [{\citenamefont {den Broeck}\ and\ \citenamefont
  {Esposito}(2015)}]{Broeck2015}%
  \BibitemOpen
  \bibfield  {author} {\bibinfo {author} {\bibfnamefont {C.~V.}\ \bibnamefont
  {den Broeck}}\ and\ \bibinfo {author} {\bibfnamefont {M.}~\bibnamefont
  {Esposito}},\ }\bibinfo {title} {Ensemble and trajectory thermodynamics: A
  brief introduction},\ \href {\doibase 10.1016/j.physa.2014.04.035} {\bibfield
   {journal} {\bibinfo  {journal} {Phys. A (Amsterdam, Neth.)}\ }\textbf
  {\bibinfo {volume} {418}},\ \bibinfo {pages} {6} (\bibinfo {year}
  {2015})}\BibitemShut {NoStop}%
\bibitem [{\citenamefont {Esposito}\ \emph {et~al.}(2009)\citenamefont
  {Esposito}, \citenamefont {Harbola},\ and\ \citenamefont
  {Mukamel}}]{Esposito2009}%
  \BibitemOpen
  \bibfield  {author} {\bibinfo {author} {\bibfnamefont {M.}~\bibnamefont
  {Esposito}}, \bibinfo {author} {\bibfnamefont {U.}~\bibnamefont {Harbola}}, \
  and\ \bibinfo {author} {\bibfnamefont {S.}~\bibnamefont {Mukamel}},\
  }\bibinfo {title} {Nonequilibrium fluctuations, fluctuation theorems, and
  counting statistics in quantum systems},\ \href {\doibase
  10.1103/revmodphys.81.1665} {\bibfield  {journal} {\bibinfo  {journal} {Rev.
  Mod. Phys.}\ }\textbf {\bibinfo {volume} {81}},\ \bibinfo {pages} {1665}
  (\bibinfo {year} {2009})}\BibitemShut {NoStop}%
\bibitem [{\citenamefont {Campisi}\ \emph {et~al.}(2011)\citenamefont
  {Campisi}, \citenamefont {H\"{a}nggi},\ and\ \citenamefont
  {Talkner}}]{Campisi2011}%
  \BibitemOpen
  \bibfield  {author} {\bibinfo {author} {\bibfnamefont {M.}~\bibnamefont
  {Campisi}}, \bibinfo {author} {\bibfnamefont {P.}~\bibnamefont {H\"{a}nggi}},
  \ and\ \bibinfo {author} {\bibfnamefont {P.}~\bibnamefont {Talkner}},\
  }\bibinfo {title} {Colloquium: Quantum fluctuation relations: Foundations and
  applications},\ \href {\doibase 10.1103/revmodphys.83.771} {\bibfield
  {journal} {\bibinfo  {journal} {Rev. Mod. Phys.}\ }\textbf {\bibinfo {volume}
  {83}},\ \bibinfo {pages} {771} (\bibinfo {year} {2011})}\BibitemShut
  {NoStop}%
\bibitem [{\citenamefont {Holubec}\ and\ \citenamefont
  {Ryabov}(2021)}]{Holubec2021}%
  \BibitemOpen
  \bibfield  {author} {\bibinfo {author} {\bibfnamefont {V.}~\bibnamefont
  {Holubec}}\ and\ \bibinfo {author} {\bibfnamefont {A.}~\bibnamefont
  {Ryabov}},\ }\bibinfo {title} {Fluctuations in heat engines},\ \href
  {\doibase 10.1088/1751-8121/ac3aac} {\bibfield  {journal} {\bibinfo
  {journal} {J. Phys. A: Math. Theor.}\ }\textbf {\bibinfo {volume} {55}},\
  \bibinfo {pages} {013001} (\bibinfo {year} {2021})}\BibitemShut {NoStop}%
\bibitem [{\citenamefont {Abah}\ \emph {et~al.}(2012)\citenamefont {Abah},
  \citenamefont {Ro{\ss}nagel}, \citenamefont {Jacob}, \citenamefont {Deffner},
  \citenamefont {Schmidt-Kaler}, \citenamefont {Singer},\ and\ \citenamefont
  {Lutz}}]{Abah2012}%
  \BibitemOpen
  \bibfield  {author} {\bibinfo {author} {\bibfnamefont {O.}~\bibnamefont
  {Abah}}, \bibinfo {author} {\bibfnamefont {J.}~\bibnamefont {Ro{\ss}nagel}},
  \bibinfo {author} {\bibfnamefont {G.}~\bibnamefont {Jacob}}, \bibinfo
  {author} {\bibfnamefont {S.}~\bibnamefont {Deffner}}, \bibinfo {author}
  {\bibfnamefont {F.}~\bibnamefont {Schmidt-Kaler}}, \bibinfo {author}
  {\bibfnamefont {K.}~\bibnamefont {Singer}}, \ and\ \bibinfo {author}
  {\bibfnamefont {E.}~\bibnamefont {Lutz}},\ }\bibinfo {title} {Single-Ion Heat
  Engine at Maximum Power},\ \href {\doibase 10.1103/physrevlett.109.203006}
  {\bibfield  {journal} {\bibinfo  {journal} {Phys. Rev. Lett.}\ }\textbf
  {\bibinfo {volume} {109}},\ \bibinfo {pages} {203006} (\bibinfo {year}
  {2012})}\BibitemShut {NoStop}%
\bibitem [{\citenamefont {Mart{\'{\i}}nez}\ \emph {et~al.}(2015)\citenamefont
  {Mart{\'{\i}}nez}, \citenamefont {Rold{\'{a}}n}, \citenamefont {Dinis},
  \citenamefont {Petrov}, \citenamefont {Parrondo},\ and\ \citenamefont
  {Rica}}]{Martinez2015}%
  \BibitemOpen
  \bibfield  {author} {\bibinfo {author} {\bibfnamefont {I.~A.}\ \bibnamefont
  {Mart{\'{\i}}nez}}, \bibinfo {author} {\bibfnamefont {{\'{E}}.}~\bibnamefont
  {Rold{\'{a}}n}}, \bibinfo {author} {\bibfnamefont {L.}~\bibnamefont {Dinis}},
  \bibinfo {author} {\bibfnamefont {D.}~\bibnamefont {Petrov}}, \bibinfo
  {author} {\bibfnamefont {J.~M.~R.}\ \bibnamefont {Parrondo}}, \ and\ \bibinfo
  {author} {\bibfnamefont {R.~A.}\ \bibnamefont {Rica}},\ }\bibinfo {title}
  {Brownian Carnot engine},\ \href {\doibase 10.1038/nphys3518} {\bibfield
  {journal} {\bibinfo  {journal} {Nat. Phys.}\ }\textbf {\bibinfo {volume}
  {12}},\ \bibinfo {pages} {67} (\bibinfo {year} {2015})}\BibitemShut {NoStop}%
\bibitem [{\citenamefont {Rubin}(1979{\natexlab{a}})}]{Rubin1979}%
  \BibitemOpen
  \bibfield  {author} {\bibinfo {author} {\bibfnamefont {M.~H.}\ \bibnamefont
  {Rubin}},\ }\bibinfo {title} {Optimal configuration of a class of
  irreversible heat engines. {I}},\ \href {\doibase 10.1103/physreva.19.1272}
  {\bibfield  {journal} {\bibinfo  {journal} {Phys. Rev. A}\ }\textbf {\bibinfo
  {volume} {19}},\ \bibinfo {pages} {1272} (\bibinfo {year}
  {1979}{\natexlab{a}})}\BibitemShut {NoStop}%
\bibitem [{\citenamefont {Rubin}(1979{\natexlab{b}})}]{Rubin1979a}%
  \BibitemOpen
  \bibfield  {author} {\bibinfo {author} {\bibfnamefont {M.~H.}\ \bibnamefont
  {Rubin}},\ }\bibinfo {title} {Optimal configuration of a class of
  irreversible heat engines. {II}},\ \href {\doibase 10.1103/physreva.19.1277}
  {\bibfield  {journal} {\bibinfo  {journal} {Phys. Rev. A}\ }\textbf {\bibinfo
  {volume} {19}},\ \bibinfo {pages} {1277} (\bibinfo {year}
  {1979}{\natexlab{b}})}\BibitemShut {NoStop}%
\bibitem [{\citenamefont {Cavina}\ \emph {et~al.}(2018)\citenamefont {Cavina},
  \citenamefont {Mari}, \citenamefont {Carlini},\ and\ \citenamefont
  {Giovannetti}}]{Cavina2018}%
  \BibitemOpen
  \bibfield  {author} {\bibinfo {author} {\bibfnamefont {V.}~\bibnamefont
  {Cavina}}, \bibinfo {author} {\bibfnamefont {A.}~\bibnamefont {Mari}},
  \bibinfo {author} {\bibfnamefont {A.}~\bibnamefont {Carlini}}, \ and\
  \bibinfo {author} {\bibfnamefont {V.}~\bibnamefont {Giovannetti}},\ }\bibinfo
  {title} {Optimal thermodynamic control in open quantum systems},\ \href
  {\doibase 10.1103/physreva.98.012139} {\bibfield  {journal} {\bibinfo
  {journal} {Phys. Rev. A}\ }\textbf {\bibinfo {volume} {98}},\ \bibinfo
  {pages} {012139} (\bibinfo {year} {2018})}\BibitemShut {NoStop}%
\bibitem [{\citenamefont {Salamon}\ and\ \citenamefont
  {Berry}(1983)}]{Salamon1983}%
  \BibitemOpen
  \bibfield  {author} {\bibinfo {author} {\bibfnamefont {P.}~\bibnamefont
  {Salamon}}\ and\ \bibinfo {author} {\bibfnamefont {R.~S.}\ \bibnamefont
  {Berry}},\ }\bibinfo {title} {Thermodynamic Length and Dissipated
  Availability},\ \href {\doibase 10.1103/physrevlett.51.1127} {\bibfield
  {journal} {\bibinfo  {journal} {Phys. Rev. Lett.}\ }\textbf {\bibinfo
  {volume} {51}},\ \bibinfo {pages} {1127} (\bibinfo {year}
  {1983})}\BibitemShut {NoStop}%
\bibitem [{\citenamefont {Crooks}(2007)}]{Crooks2007}%
  \BibitemOpen
  \bibfield  {author} {\bibinfo {author} {\bibfnamefont {G.~E.}\ \bibnamefont
  {Crooks}},\ }\bibinfo {title} {Measuring Thermodynamic Length},\ \href
  {\doibase 10.1103/physrevlett.99.100602} {\bibfield  {journal} {\bibinfo
  {journal} {Phys. Rev. Lett.}\ }\textbf {\bibinfo {volume} {99}},\ \bibinfo
  {pages} {100602} (\bibinfo {year} {2007})}\BibitemShut {NoStop}%
\bibitem [{\citenamefont {Sivak}\ and\ \citenamefont
  {Crooks}(2012)}]{Sivak2012}%
  \BibitemOpen
  \bibfield  {author} {\bibinfo {author} {\bibfnamefont {D.~A.}\ \bibnamefont
  {Sivak}}\ and\ \bibinfo {author} {\bibfnamefont {G.~E.}\ \bibnamefont
  {Crooks}},\ }\bibinfo {title} {Thermodynamic Metrics and Optimal Paths},\
  \href {\doibase 10.1103/physrevlett.108.190602} {\bibfield  {journal}
  {\bibinfo  {journal} {Phys. Rev. Lett.}\ }\textbf {\bibinfo {volume} {108}},\
  \bibinfo {pages} {190602} (\bibinfo {year} {2012})}\BibitemShut {NoStop}%
\bibitem [{\citenamefont {Gong}\ \emph {et~al.}(2016)\citenamefont {Gong},
  \citenamefont {Lan},\ and\ \citenamefont {Quan}}]{Gong2016}%
  \BibitemOpen
  \bibfield  {author} {\bibinfo {author} {\bibfnamefont {Z.}~\bibnamefont
  {Gong}}, \bibinfo {author} {\bibfnamefont {Y.}~\bibnamefont {Lan}}, \ and\
  \bibinfo {author} {\bibfnamefont {H.~T.}\ \bibnamefont {Quan}},\ }\bibinfo
  {title} {Stochastic Thermodynamics of a Particle in a Box},\ \href {\doibase
  10.1103/physrevlett.117.180603} {\bibfield  {journal} {\bibinfo  {journal}
  {Phys. Rev. Lett.}\ }\textbf {\bibinfo {volume} {117}},\ \bibinfo {pages}
  {180603} (\bibinfo {year} {2016})}\BibitemShut {NoStop}%
\bibitem [{\citenamefont {Scandi}\ and\ \citenamefont
  {Perarnau-Llobet}(2019)}]{Scandi2019}%
  \BibitemOpen
  \bibfield  {author} {\bibinfo {author} {\bibfnamefont {M.}~\bibnamefont
  {Scandi}}\ and\ \bibinfo {author} {\bibfnamefont {M.}~\bibnamefont
  {Perarnau-Llobet}},\ }\bibinfo {title} {Thermodynamic length in open quantum
  systems},\ \href {\doibase 10.22331/q-2019-10-24-197} {\bibfield  {journal}
  {\bibinfo  {journal} {Quantum}\ }\textbf {\bibinfo {volume} {3}},\ \bibinfo
  {pages} {197} (\bibinfo {year} {2019})}\BibitemShut {NoStop}%
\bibitem [{\citenamefont {Brandner}\ and\ \citenamefont
  {Saito}(2020)}]{Brandner2020}%
  \BibitemOpen
  \bibfield  {author} {\bibinfo {author} {\bibfnamefont {K.}~\bibnamefont
  {Brandner}}\ and\ \bibinfo {author} {\bibfnamefont {K.}~\bibnamefont
  {Saito}},\ }\bibinfo {title} {Thermodynamic Geometry of Microscopic Heat
  Engines},\ \href {\doibase 10.1103/physrevlett.124.040602} {\bibfield
  {journal} {\bibinfo  {journal} {Phys. Rev. Lett.}\ }\textbf {\bibinfo
  {volume} {124}},\ \bibinfo {pages} {040602} (\bibinfo {year}
  {2020})}\BibitemShut {NoStop}%
\bibitem [{\citenamefont {Salazar}(2020)}]{Salazar2020}%
  \BibitemOpen
  \bibfield  {author} {\bibinfo {author} {\bibfnamefont {D.~S.~P.}\
  \bibnamefont {Salazar}},\ }\bibinfo {title} {Work distribution in thermal
  processes},\ \href {\doibase 10.1103/physreve.101.030101} {\bibfield
  {journal} {\bibinfo  {journal} {Phys. Rev. E}\ }\textbf {\bibinfo {volume}
  {101}},\ \bibinfo {pages} {030101(R)} (\bibinfo {year} {2020})}\BibitemShut
  {NoStop}%
\bibitem [{\citenamefont {Chen}\ \emph
  {et~al.}(2021{\natexlab{a}})\citenamefont {Chen}, \citenamefont {Sun},\ and\
  \citenamefont {Dong}}]{Chen2021}%
  \BibitemOpen
  \bibfield  {author} {\bibinfo {author} {\bibfnamefont {J.-F.}\ \bibnamefont
  {Chen}}, \bibinfo {author} {\bibfnamefont {C.~P.}\ \bibnamefont {Sun}}, \
  and\ \bibinfo {author} {\bibfnamefont {H.}~\bibnamefont {Dong}},\ }\bibinfo
  {title} {Extrapolating the thermodynamic length with finite-time
  measurements},\ \href {\doibase 10.1103/PhysRevE.104.034117} {\bibfield
  {journal} {\bibinfo  {journal} {Phys. Rev. E}\ }\textbf {\bibinfo {volume}
  {104}},\ \bibinfo {pages} {034117} (\bibinfo {year}
  {2021}{\natexlab{a}})}\BibitemShut {NoStop}%
\bibitem [{\citenamefont {Dechant}\ \emph {et~al.}(2017)\citenamefont
  {Dechant}, \citenamefont {Kiesel},\ and\ \citenamefont {Lutz}}]{Dechant2017}%
  \BibitemOpen
  \bibfield  {author} {\bibinfo {author} {\bibfnamefont {A.}~\bibnamefont
  {Dechant}}, \bibinfo {author} {\bibfnamefont {N.}~\bibnamefont {Kiesel}}, \
  and\ \bibinfo {author} {\bibfnamefont {E.}~\bibnamefont {Lutz}},\ }\bibinfo
  {title} {Underdamped stochastic heat engine at maximum efficiency},\ \href
  {\doibase 10.1209/0295-5075/119/50003} {\bibfield  {journal} {\bibinfo
  {journal} {Europhys. Lett.}\ }\textbf {\bibinfo {volume} {119}},\ \bibinfo
  {pages} {50003} (\bibinfo {year} {2017})}\BibitemShut {NoStop}%
\bibitem [{\citenamefont {Chen}\ \emph
  {et~al.}(2021{\natexlab{b}})\citenamefont {Chen}, \citenamefont {Chen},
  \citenamefont {Fei},\ and\ \citenamefont {Quan}}]{Chen2021b}%
  \BibitemOpen
  \bibfield  {author} {\bibinfo {author} {\bibfnamefont {Y.~H.}\ \bibnamefont
  {Chen}}, \bibinfo {author} {\bibfnamefont {J.-F.}\ \bibnamefont {Chen}},
  \bibinfo {author} {\bibfnamefont {Z.}~\bibnamefont {Fei}}, \ and\ \bibinfo
  {author} {\bibfnamefont {H.~T.}\ \bibnamefont {Quan}},\ }\bibinfo {title} {A
  microscopic theory of Curzon-Ahlborn heat engine},\ \href@noop {} {\
  (\bibinfo {year} {2021}{\natexlab{b}})},\ \Eprint
  {http://arxiv.org/abs/2108.04128} {arXiv:2108.04128 [cond-mat.stat-mech]}
  \BibitemShut {NoStop}%
\bibitem [{\citenamefont {Schmiedl}\ and\ \citenamefont
  {Seifert}(2007{\natexlab{a}})}]{Schmiedl2007a}%
  \BibitemOpen
  \bibfield  {author} {\bibinfo {author} {\bibfnamefont {T.}~\bibnamefont
  {Schmiedl}}\ and\ \bibinfo {author} {\bibfnamefont {U.}~\bibnamefont
  {Seifert}},\ }\bibinfo {title} {Optimal Finite-Time Processes In Stochastic
  Thermodynamics},\ \href {\doibase 10.1103/physrevlett.98.108301} {\bibfield
  {journal} {\bibinfo  {journal} {Phys. Rev. Lett.}\ }\textbf {\bibinfo
  {volume} {98}},\ \bibinfo {pages} {108301} (\bibinfo {year}
  {2007}{\natexlab{a}})}\BibitemShut {NoStop}%
\bibitem [{\citenamefont {Schmiedl}\ and\ \citenamefont
  {Seifert}(2007{\natexlab{b}})}]{Schmiedl2007}%
  \BibitemOpen
  \bibfield  {author} {\bibinfo {author} {\bibfnamefont {T.}~\bibnamefont
  {Schmiedl}}\ and\ \bibinfo {author} {\bibfnamefont {U.}~\bibnamefont
  {Seifert}},\ }\bibinfo {title} {Efficiency at maximum power: An analytically
  solvable model for stochastic heat engines},\ \href {\doibase
  10.1209/0295-5075/81/20003} {\bibfield  {journal} {\bibinfo  {journal}
  {Europhys. Lett.}\ }\textbf {\bibinfo {volume} {81}},\ \bibinfo {pages}
  {20003} (\bibinfo {year} {2007}{\natexlab{b}})}\BibitemShut {NoStop}%
\bibitem [{\citenamefont {Mart{\'{\i}}nez}\ \emph {et~al.}(2016)\citenamefont
  {Mart{\'{\i}}nez}, \citenamefont {Petrosyan}, \citenamefont
  {Gu{\'{e}}ry-Odelin}, \citenamefont {Trizac},\ and\ \citenamefont
  {Ciliberto}}]{Martinez2016}%
  \BibitemOpen
  \bibfield  {author} {\bibinfo {author} {\bibfnamefont {I.~A.}\ \bibnamefont
  {Mart{\'{\i}}nez}}, \bibinfo {author} {\bibfnamefont {A.}~\bibnamefont
  {Petrosyan}}, \bibinfo {author} {\bibfnamefont {D.}~\bibnamefont
  {Gu{\'{e}}ry-Odelin}}, \bibinfo {author} {\bibfnamefont {E.}~\bibnamefont
  {Trizac}}, \ and\ \bibinfo {author} {\bibfnamefont {S.}~\bibnamefont
  {Ciliberto}},\ }\bibinfo {title} {Engineered swift equilibration of a
  Brownian~particle},\ \href {\doibase 10.1038/nphys3758} {\bibfield  {journal}
  {\bibinfo  {journal} {Nat. Phys.}\ }\textbf {\bibinfo {volume} {12}},\
  \bibinfo {pages} {843} (\bibinfo {year} {2016})}\BibitemShut {NoStop}%
\bibitem [{\citenamefont {Li}\ \emph {et~al.}(2017)\citenamefont {Li},
  \citenamefont {Quan},\ and\ \citenamefont {Tu}}]{Li2017}%
  \BibitemOpen
  \bibfield  {author} {\bibinfo {author} {\bibfnamefont {G.}~\bibnamefont
  {Li}}, \bibinfo {author} {\bibfnamefont {H.~T.}\ \bibnamefont {Quan}}, \ and\
  \bibinfo {author} {\bibfnamefont {Z.~C.}\ \bibnamefont {Tu}},\ }\bibinfo
  {title} {Shortcuts to isothermality and nonequilibrium work relations},\
  \href {\doibase 10.1103/physreve.96.012144} {\bibfield  {journal} {\bibinfo
  {journal} {Phys. Rev. E}\ }\textbf {\bibinfo {volume} {96}},\ \bibinfo
  {pages} {012144} (\bibinfo {year} {2017})}\BibitemShut {NoStop}%
\bibitem [{\citenamefont {Albay}\ \emph {et~al.}(2019)\citenamefont {Albay},
  \citenamefont {Wulaningrum}, \citenamefont {Kwon}, \citenamefont {Lai},\ and\
  \citenamefont {Jun}}]{Albay2019}%
  \BibitemOpen
  \bibfield  {author} {\bibinfo {author} {\bibfnamefont {J.~A.~C.}\
  \bibnamefont {Albay}}, \bibinfo {author} {\bibfnamefont {S.~R.}\ \bibnamefont
  {Wulaningrum}}, \bibinfo {author} {\bibfnamefont {C.}~\bibnamefont {Kwon}},
  \bibinfo {author} {\bibfnamefont {P.-Y.}\ \bibnamefont {Lai}}, \ and\
  \bibinfo {author} {\bibfnamefont {Y.}~\bibnamefont {Jun}},\ }\bibinfo {title}
  {Thermodynamic cost of a shortcuts-to-isothermal transport of a Brownian
  particle},\ \href {\doibase 10.1103/physrevresearch.1.033122} {\bibfield
  {journal} {\bibinfo  {journal} {Phys. Rev. Research}\ }\textbf {\bibinfo
  {volume} {1}},\ \bibinfo {pages} {033122} (\bibinfo {year}
  {2019})}\BibitemShut {NoStop}%
\bibitem [{\citenamefont {Albay}\ \emph
  {et~al.}(2020{\natexlab{a}})\citenamefont {Albay}, \citenamefont {Lai},\ and\
  \citenamefont {Jun}}]{Albay2020}%
  \BibitemOpen
  \bibfield  {author} {\bibinfo {author} {\bibfnamefont {J.~A.~C.}\
  \bibnamefont {Albay}}, \bibinfo {author} {\bibfnamefont {P.-Y.}\ \bibnamefont
  {Lai}}, \ and\ \bibinfo {author} {\bibfnamefont {Y.}~\bibnamefont {Jun}},\
  }\bibinfo {title} {Realization of finite-rate isothermal compression and
  expansion using optical feedback trap},\ \href {\doibase 10.1063/1.5143602}
  {\bibfield  {journal} {\bibinfo  {journal} {Appl. Phys. Lett.}\ }\textbf
  {\bibinfo {volume} {116}},\ \bibinfo {pages} {103706} (\bibinfo {year}
  {2020}{\natexlab{a}})}\BibitemShut {NoStop}%
\bibitem [{\citenamefont {Albay}\ \emph
  {et~al.}(2020{\natexlab{b}})\citenamefont {Albay}, \citenamefont {Kwon},
  \citenamefont {Lai},\ and\ \citenamefont {Jun}}]{Albay2020a}%
  \BibitemOpen
  \bibfield  {author} {\bibinfo {author} {\bibfnamefont {J.~A.~C.}\
  \bibnamefont {Albay}}, \bibinfo {author} {\bibfnamefont {C.}~\bibnamefont
  {Kwon}}, \bibinfo {author} {\bibfnamefont {P.-Y.}\ \bibnamefont {Lai}}, \
  and\ \bibinfo {author} {\bibfnamefont {Y.}~\bibnamefont {Jun}},\ }\bibinfo
  {title} {Work relation in instantaneous-equilibrium transition of forward and
  reverse processes},\ \href {\doibase 10.1088/1367-2630/abce78} {\bibfield
  {journal} {\bibinfo  {journal} {New J. Phys.}\ }\textbf {\bibinfo {volume}
  {22}},\ \bibinfo {pages} {123049} (\bibinfo {year}
  {2020}{\natexlab{b}})}\BibitemShut {NoStop}%
\bibitem [{\citenamefont {Pancotti}\ \emph {et~al.}(2020)\citenamefont
  {Pancotti}, \citenamefont {Scandi}, \citenamefont {Mitchison},\ and\
  \citenamefont {Perarnau-Llobet}}]{Pancotti2020}%
  \BibitemOpen
  \bibfield  {author} {\bibinfo {author} {\bibfnamefont {N.}~\bibnamefont
  {Pancotti}}, \bibinfo {author} {\bibfnamefont {M.}~\bibnamefont {Scandi}},
  \bibinfo {author} {\bibfnamefont {M.~T.}\ \bibnamefont {Mitchison}}, \ and\
  \bibinfo {author} {\bibfnamefont {M.}~\bibnamefont {Perarnau-Llobet}},\
  }\bibinfo {title} {Speed-Ups to Isothermality: Enhanced Quantum Thermal
  Machines through Control of the System-Bath Coupling},\ \href {\doibase
  10.1103/physrevx.10.031015} {\bibfield  {journal} {\bibinfo  {journal} {Phys.
  Rev. X}\ }\textbf {\bibinfo {volume} {10}},\ \bibinfo {pages} {031015}
  (\bibinfo {year} {2020})}\BibitemShut {NoStop}%
\bibitem [{\citenamefont {Nakamura}\ \emph {et~al.}(2020)\citenamefont
  {Nakamura}, \citenamefont {Matrasulov},\ and\ \citenamefont
  {Izumida}}]{Nakamura2020}%
  \BibitemOpen
  \bibfield  {author} {\bibinfo {author} {\bibfnamefont {K.}~\bibnamefont
  {Nakamura}}, \bibinfo {author} {\bibfnamefont {J.}~\bibnamefont
  {Matrasulov}}, \ and\ \bibinfo {author} {\bibfnamefont {Y.}~\bibnamefont
  {Izumida}},\ }\bibinfo {title} {Fast-forward approach to stochastic heat
  engine},\ \href {\doibase 10.1103/physreve.102.012129} {\bibfield  {journal}
  {\bibinfo  {journal} {Phys. Rev. E}\ }\textbf {\bibinfo {volume} {102}},\
  \bibinfo {pages} {012129} (\bibinfo {year} {2020})}\BibitemShut {NoStop}%
\bibitem [{\citenamefont {Plata}\ \emph {et~al.}(2020)\citenamefont {Plata},
  \citenamefont {Gu{\'{e}}ry-Odelin}, \citenamefont {Trizac},\ and\
  \citenamefont {Prados}}]{Plata2020}%
  \BibitemOpen
  \bibfield  {author} {\bibinfo {author} {\bibfnamefont {C.~A.}\ \bibnamefont
  {Plata}}, \bibinfo {author} {\bibfnamefont {D.}~\bibnamefont
  {Gu{\'{e}}ry-Odelin}}, \bibinfo {author} {\bibfnamefont {E.}~\bibnamefont
  {Trizac}}, \ and\ \bibinfo {author} {\bibfnamefont {A.}~\bibnamefont
  {Prados}},\ }\bibinfo {title} {Building an irreversible Carnot-like heat
  engine with an overdamped harmonic oscillator},\ \href {\doibase
  10.1088/1742-5468/abb0e1} {\bibfield  {journal} {\bibinfo  {journal} {J.
  Stat. Mech.: Theory Exp.}\ }\textbf {\bibinfo {volume} {2020}},\ \bibinfo
  {pages} {093207} (\bibinfo {year} {2020})}\BibitemShut {NoStop}%
\bibitem [{\citenamefont {Iram}\ \emph {et~al.}(2020)\citenamefont {Iram},
  \citenamefont {Dolson}, \citenamefont {Chiel}, \citenamefont {Pelesko},
  \citenamefont {Krishnan}, \citenamefont {\"{O}zen\c{c} G\"{u}ng\"{o}r},
  \citenamefont {Kuznets-Speck}, \citenamefont {Deffner}, \citenamefont
  {Ilker}, \citenamefont {Scott},\ and\ \citenamefont {Hinczewski}}]{Iram2020}%
  \BibitemOpen
  \bibfield  {author} {\bibinfo {author} {\bibfnamefont {S.}~\bibnamefont
  {Iram}}, \bibinfo {author} {\bibfnamefont {E.}~\bibnamefont {Dolson}},
  \bibinfo {author} {\bibfnamefont {J.}~\bibnamefont {Chiel}}, \bibinfo
  {author} {\bibfnamefont {J.}~\bibnamefont {Pelesko}}, \bibinfo {author}
  {\bibfnamefont {N.}~\bibnamefont {Krishnan}}, \bibinfo {author} {\bibnamefont
  {\"{O}zen\c{c} G\"{u}ng\"{o}r}}, \bibinfo {author} {\bibfnamefont
  {B.}~\bibnamefont {Kuznets-Speck}}, \bibinfo {author} {\bibfnamefont
  {S.}~\bibnamefont {Deffner}}, \bibinfo {author} {\bibfnamefont
  {E.}~\bibnamefont {Ilker}}, \bibinfo {author} {\bibfnamefont {J.~G.}\
  \bibnamefont {Scott}}, \ and\ \bibinfo {author} {\bibfnamefont
  {M.}~\bibnamefont {Hinczewski}},\ }\bibinfo {title} {Controlling the speed
  and trajectory of evolution with counterdiabatic driving},\ \href {\doibase
  10.1038/s41567-020-0989-3} {\bibfield  {journal} {\bibinfo  {journal} {Nat.
  Phys.}\ }\textbf {\bibinfo {volume} {17}},\ \bibinfo {pages} {135} (\bibinfo
  {year} {2020})}\BibitemShut {NoStop}%
\bibitem [{\citenamefont {Ilker}\ \emph {et~al.}(2021)\citenamefont {Ilker},
  \citenamefont {\"{O}zen\c{c} G\"{u}ng\"{o}r}, \citenamefont {Kuznets-Speck},
  \citenamefont {Chiel}, \citenamefont {Deffner},\ and\ \citenamefont
  {Hinczewski}}]{Ilker2021}%
  \BibitemOpen
  \bibfield  {author} {\bibinfo {author} {\bibfnamefont {E.}~\bibnamefont
  {Ilker}}, \bibinfo {author} {\bibnamefont {\"{O}zen\c{c} G\"{u}ng\"{o}r}},
  \bibinfo {author} {\bibfnamefont {B.}~\bibnamefont {Kuznets-Speck}}, \bibinfo
  {author} {\bibfnamefont {J.}~\bibnamefont {Chiel}}, \bibinfo {author}
  {\bibfnamefont {S.}~\bibnamefont {Deffner}}, \ and\ \bibinfo {author}
  {\bibfnamefont {M.}~\bibnamefont {Hinczewski}},\ }\bibinfo {title} {Shortcuts
  in stochastic systems and control of biophysical processes},\ \href@noop {}
  {\  (\bibinfo {year} {2021})},\ \Eprint {http://arxiv.org/abs/2106.07130}
  {arXiv:2106.07130 [cond-mat.stat-mech]} \BibitemShut {NoStop}%
\bibitem [{\citenamefont {Jun}\ and\ \citenamefont {Lai}(2021)}]{Jun2021}%
  \BibitemOpen
  \bibfield  {author} {\bibinfo {author} {\bibfnamefont {Y.}~\bibnamefont
  {Jun}}\ and\ \bibinfo {author} {\bibfnamefont {P.-Y.}\ \bibnamefont {Lai}},\
  }\bibinfo {title} {Instantaneous equilibrium transition for Brownian systems
  under time-dependent temperature and potential variations: Reversibility,
  heat and work relations, and fast isentropic process},\ \href {\doibase
  10.1103/physrevresearch.3.033130} {\bibfield  {journal} {\bibinfo  {journal}
  {Phys. Rev. Research}\ }\textbf {\bibinfo {volume} {3}},\ \bibinfo {pages}
  {033130} (\bibinfo {year} {2021})}\BibitemShut {NoStop}%
\bibitem [{\citenamefont {Jun}\ and\ \citenamefont {Lai}(2022)}]{Jun2022}%
  \BibitemOpen
  \bibfield  {author} {\bibinfo {author} {\bibfnamefont {Y.}~\bibnamefont
  {Jun}}\ and\ \bibinfo {author} {\bibfnamefont {P.-Y.}\ \bibnamefont {Lai}},\
  }\bibinfo {title} {Minimal dissipation protocols of an instantaneous
  equilibrium Brownian particle under time-dependent temperature and potential
  variations},\ \href {\doibase 10.1103/physrevresearch.4.023157} {\bibfield
  {journal} {\bibinfo  {journal} {Phys. Rev. Research}\ }\textbf {\bibinfo
  {volume} {4}},\ \bibinfo {pages} {023157} (\bibinfo {year}
  {2022})}\BibitemShut {NoStop}%
\bibitem [{\citenamefont {Li}\ \emph {et~al.}(2022)\citenamefont {Li},
  \citenamefont {Chen}, \citenamefont {Sun},\ and\ \citenamefont
  {Dong}}]{Li2021a}%
  \BibitemOpen
  \bibfield  {author} {\bibinfo {author} {\bibfnamefont {G.}~\bibnamefont
  {Li}}, \bibinfo {author} {\bibfnamefont {J.-F.}\ \bibnamefont {Chen}},
  \bibinfo {author} {\bibfnamefont {C.~P.}\ \bibnamefont {Sun}}, \ and\
  \bibinfo {author} {\bibfnamefont {H.}~\bibnamefont {Dong}},\ }\bibinfo
  {title} {Geodesic Path for the Minimal Energy Cost in Shortcuts to
  Isothermality},\ \href {\doibase 10.1103/PhysRevLett.128.230603} {\bibfield
  {journal} {\bibinfo  {journal} {Phys. Rev. Lett.}\ }\textbf {\bibinfo
  {volume} {128}},\ \bibinfo {pages} {230603} (\bibinfo {year}
  {2022})}\BibitemShut {NoStop}%
\bibitem [{\citenamefont {Esposito}\ \emph {et~al.}(2010)\citenamefont
  {Esposito}, \citenamefont {Kawai}, \citenamefont {Lindenberg},\ and\
  \citenamefont {den Broeck}}]{Esposito2010}%
  \BibitemOpen
  \bibfield  {author} {\bibinfo {author} {\bibfnamefont {M.}~\bibnamefont
  {Esposito}}, \bibinfo {author} {\bibfnamefont {R.}~\bibnamefont {Kawai}},
  \bibinfo {author} {\bibfnamefont {K.}~\bibnamefont {Lindenberg}}, \ and\
  \bibinfo {author} {\bibfnamefont {C.~V.}\ \bibnamefont {den Broeck}},\
  }\bibinfo {title} {Efficiency at Maximum Power of Low-Dissipation Carnot
  Engines},\ \href {\doibase 10.1103/physrevlett.105.150603} {\bibfield
  {journal} {\bibinfo  {journal} {Phys. Rev. Lett.}\ }\textbf {\bibinfo
  {volume} {105}},\ \bibinfo {pages} {150603} (\bibinfo {year}
  {2010})}\BibitemShut {NoStop}%
\bibitem [{\citenamefont {Tu}(2012)}]{Tu2012}%
  \BibitemOpen
  \bibfield  {author} {\bibinfo {author} {\bibfnamefont {Z.-C.}\ \bibnamefont
  {Tu}},\ }\bibinfo {title} {Recent advance on the efficiency at maximum power
  of heat engines},\ \href {\doibase 10.1088/1674-1056/21/2/020513} {\bibfield
  {journal} {\bibinfo  {journal} {Chin. Phys. B}\ }\textbf {\bibinfo {volume}
  {21}},\ \bibinfo {pages} {020513} (\bibinfo {year} {2012})}\BibitemShut
  {NoStop}%
\bibitem [{\citenamefont {Wang}\ and\ \citenamefont {Tu}(2012)}]{Wang2012}%
  \BibitemOpen
  \bibfield  {author} {\bibinfo {author} {\bibfnamefont {Y.}~\bibnamefont
  {Wang}}\ and\ \bibinfo {author} {\bibfnamefont {Z.~C.}\ \bibnamefont {Tu}},\
  }\bibinfo {title} {Efficiency at maximum power output of linear irreversible
  Carnot-like heat engines},\ \href {\doibase 10.1103/physreve.85.011127}
  {\bibfield  {journal} {\bibinfo  {journal} {Phys. Rev. E}\ }\textbf {\bibinfo
  {volume} {85}},\ \bibinfo {pages} {011127} (\bibinfo {year}
  {2012})}\BibitemShut {NoStop}%
\bibitem [{\citenamefont {Ryabov}\ and\ \citenamefont
  {Holubec}(2016)}]{Ryabov2016}%
  \BibitemOpen
  \bibfield  {author} {\bibinfo {author} {\bibfnamefont {A.}~\bibnamefont
  {Ryabov}}\ and\ \bibinfo {author} {\bibfnamefont {V.}~\bibnamefont
  {Holubec}},\ }\bibinfo {title} {Maximum efficiency of steady-state heat
  engines at arbitrary power},\ \href {\doibase 10.1103/physreve.93.050101}
  {\bibfield  {journal} {\bibinfo  {journal} {Phys. Rev. E}\ }\textbf {\bibinfo
  {volume} {93}},\ \bibinfo {pages} {050101(R)} (\bibinfo {year}
  {2016})}\BibitemShut {NoStop}%
\bibitem [{\citenamefont {Cavina}\ \emph {et~al.}(2017)\citenamefont {Cavina},
  \citenamefont {Mari},\ and\ \citenamefont {Giovannetti}}]{Cavina2017}%
  \BibitemOpen
  \bibfield  {author} {\bibinfo {author} {\bibfnamefont {V.}~\bibnamefont
  {Cavina}}, \bibinfo {author} {\bibfnamefont {A.}~\bibnamefont {Mari}}, \ and\
  \bibinfo {author} {\bibfnamefont {V.}~\bibnamefont {Giovannetti}},\ }\bibinfo
  {title} {Slow Dynamics and Thermodynamics of Open Quantum Systems},\ \href
  {\doibase 10.1103/physrevlett.119.050601} {\bibfield  {journal} {\bibinfo
  {journal} {Phys. Rev. Lett.}\ }\textbf {\bibinfo {volume} {119}},\ \bibinfo
  {pages} {050601} (\bibinfo {year} {2017})}\BibitemShut {NoStop}%
\bibitem [{\citenamefont {Ma}\ \emph {et~al.}(2018)\citenamefont {Ma},
  \citenamefont {Xu}, \citenamefont {Dong},\ and\ \citenamefont
  {Sun}}]{Ma2018a}%
  \BibitemOpen
  \bibfield  {author} {\bibinfo {author} {\bibfnamefont {Y.-H.}\ \bibnamefont
  {Ma}}, \bibinfo {author} {\bibfnamefont {D.}~\bibnamefont {Xu}}, \bibinfo
  {author} {\bibfnamefont {H.}~\bibnamefont {Dong}}, \ and\ \bibinfo {author}
  {\bibfnamefont {C.-P.}\ \bibnamefont {Sun}},\ }\bibinfo {title} {Universal
  constraint for efficiency and power of a low-dissipation heat engine},\ \href
  {\doibase 10.1103/physreve.98.042112} {\bibfield  {journal} {\bibinfo
  {journal} {Phys. Rev. E}\ }\textbf {\bibinfo {volume} {98}},\ \bibinfo
  {pages} {042112} (\bibinfo {year} {2018})}\BibitemShut {NoStop}%
\bibitem [{\citenamefont {Abiuso}\ and\ \citenamefont
  {Perarnau-Llobet}(2020)}]{Abiuso2020}%
  \BibitemOpen
  \bibfield  {author} {\bibinfo {author} {\bibfnamefont {P.}~\bibnamefont
  {Abiuso}}\ and\ \bibinfo {author} {\bibfnamefont {M.}~\bibnamefont
  {Perarnau-Llobet}},\ }\bibinfo {title} {Optimal Cycles for Low-Dissipation
  Heat Engines},\ \href {\doibase 10.1103/physrevlett.124.110606} {\bibfield
  {journal} {\bibinfo  {journal} {Phys. Rev. Lett.}\ }\textbf {\bibinfo
  {volume} {124}},\ \bibinfo {pages} {110606} (\bibinfo {year}
  {2020})}\BibitemShut {NoStop}%
\bibitem [{\citenamefont {Frim}\ and\ \citenamefont
  {DeWeese}(2022)}]{PhysRevLett.128.230601}%
  \BibitemOpen
  \bibfield  {author} {\bibinfo {author} {\bibfnamefont {A.~G.}\ \bibnamefont
  {Frim}}\ and\ \bibinfo {author} {\bibfnamefont {M.~R.}\ \bibnamefont
  {DeWeese}},\ }\bibinfo {title} {Geometric Bound on the Efficiency of
  Irreversible Thermodynamic Cycles},\ \href {\doibase
  10.1103/PhysRevLett.128.230601} {\bibfield  {journal} {\bibinfo  {journal}
  {Phys. Rev. Lett.}\ }\textbf {\bibinfo {volume} {128}},\ \bibinfo {pages}
  {230601} (\bibinfo {year} {2022})}\BibitemShut {NoStop}%
\bibitem [{\citenamefont {Watanabe}\ and\ \citenamefont
  {Minami}(2022)}]{Watanabe2022}%
  \BibitemOpen
  \bibfield  {author} {\bibinfo {author} {\bibfnamefont {G.}~\bibnamefont
  {Watanabe}}\ and\ \bibinfo {author} {\bibfnamefont {Y.}~\bibnamefont
  {Minami}},\ }\bibinfo {title} {Finite-time thermodynamics of fluctuations in
  microscopic heat engines},\ \href {\doibase
  10.1103/physrevresearch.4.l012008} {\bibfield  {journal} {\bibinfo  {journal}
  {Phys. Rev. Research}\ }\textbf {\bibinfo {volume} {4}},\ \bibinfo {pages}
  {L012008} (\bibinfo {year} {2022})}\BibitemShut {NoStop}%
\bibitem [{\citenamefont {Yuan}\ \emph {et~al.}(2022)\citenamefont {Yuan},
  \citenamefont {Ma},\ and\ \citenamefont {Sun}}]{Yuan2022}%
  \BibitemOpen
  \bibfield  {author} {\bibinfo {author} {\bibfnamefont {H.}~\bibnamefont
  {Yuan}}, \bibinfo {author} {\bibfnamefont {Y.-H.}\ \bibnamefont {Ma}}, \ and\
  \bibinfo {author} {\bibfnamefont {C.~P.}\ \bibnamefont {Sun}},\ }\bibinfo
  {title} {Optimizing thermodynamic cycles with two finite-sized reservoirs},\
  \href {\doibase 10.1103/physreve.105.l022101} {\bibfield  {journal} {\bibinfo
   {journal} {Phys. Rev. E}\ }\textbf {\bibinfo {volume} {105}},\ \bibinfo
  {pages} {L022101} (\bibinfo {year} {2022})}\BibitemShut {NoStop}%
\bibitem [{\citenamefont {Ye}\ \emph {et~al.}(2022)\citenamefont {Ye},
  \citenamefont {Cerisola}, \citenamefont {Abiuso}, \citenamefont {Anders},
  \citenamefont {Perarnau-Llobet},\ and\ \citenamefont {Holubec}}]{Ye2022}%
  \BibitemOpen
  \bibfield  {author} {\bibinfo {author} {\bibfnamefont {Z.}~\bibnamefont
  {Ye}}, \bibinfo {author} {\bibfnamefont {F.}~\bibnamefont {Cerisola}},
  \bibinfo {author} {\bibfnamefont {P.}~\bibnamefont {Abiuso}}, \bibinfo
  {author} {\bibfnamefont {J.}~\bibnamefont {Anders}}, \bibinfo {author}
  {\bibfnamefont {M.}~\bibnamefont {Perarnau-Llobet}}, \ and\ \bibinfo {author}
  {\bibfnamefont {V.}~\bibnamefont {Holubec}},\ }\bibinfo {title} {Optimal
  finite-time heat engines under constrained control},\ \href@noop {} {\
  (\bibinfo {year} {2022})},\ \Eprint {http://arxiv.org/abs/2202.12953}
  {arXiv:2202.12953 [cond-mat.stat-mech]} \BibitemShut {NoStop}%
\bibitem [{\citenamefont {Kramers}(1940)}]{Kramers1940}%
  \BibitemOpen
  \bibfield  {author} {\bibinfo {author} {\bibfnamefont {H.~A.}\ \bibnamefont
  {Kramers}},\ }\bibinfo {title} {Brownian motion in a field of force and the
  diffusion model of chemical reactions},\ \href {\doibase
  10.1016/s0031-8914(40)90098-2} {\bibfield  {journal} {\bibinfo  {journal}
  {Physica}\ }\textbf {\bibinfo {volume} {7}},\ \bibinfo {pages} {284}
  (\bibinfo {year} {1940})}\BibitemShut {NoStop}%
\bibitem [{sup()}]{supplementary_material}%
  \BibitemOpen
  \href@noop {} {\enquote {\bibinfo {title} {See supplementary material.}}\
  }\BibitemShut {NoStop}%
\bibitem [{Note1()}]{Note1}%
  \BibitemOpen
  \bibinfo {note} {This parameterization simplifies the expressions of the
  thermodynamic length, and allows the same optimal protocol to vary $\lambda $
  in the highly underdamped regime. For the harmonic potential ($n=1$), the
  work parameter $\lambda $ becomes the frequency}\BibitemShut {NoStop}%
\bibitem [{\citenamefont {Berger}(2007)}]{Berger2007}%
  \BibitemOpen
  \bibfield  {author} {\bibinfo {author} {\bibfnamefont {M.}~\bibnamefont
  {Berger}},\ }\href
  {https://www.ebook.de/de/product/2047791/marcel_berger_a_panoramic_view_of_riemannian_geometry.html}
  {\emph {\bibinfo {title} {A Panoramic View of Riemannian Geometry}}}\
  (\bibinfo  {publisher} {Springer Berlin Heidelberg},\ \bibinfo {year}
  {2007})\BibitemShut {NoStop}%
\bibitem [{\citenamefont {Curzon}\ and\ \citenamefont
  {Ahlborn}(1975)}]{Curzon1975}%
  \BibitemOpen
  \bibfield  {author} {\bibinfo {author} {\bibfnamefont {F.~L.}\ \bibnamefont
  {Curzon}}\ and\ \bibinfo {author} {\bibfnamefont {B.}~\bibnamefont
  {Ahlborn}},\ }\bibinfo {title} {Efficiency of a Carnot engine at maximum
  power output},\ \href {\doibase 10.1119/1.10023} {\bibfield  {journal}
  {\bibinfo  {journal} {Am. J. Phys.}\ }\textbf {\bibinfo {volume} {43}},\
  \bibinfo {pages} {22} (\bibinfo {year} {1975})}\BibitemShut {NoStop}%
\bibitem [{\citenamefont {Chen}\ and\ \citenamefont {Yan}(1989)}]{Chen1989}%
  \BibitemOpen
  \bibfield  {author} {\bibinfo {author} {\bibfnamefont {L.}~\bibnamefont
  {Chen}}\ and\ \bibinfo {author} {\bibfnamefont {Z.}~\bibnamefont {Yan}},\
  }\bibinfo {title} {The effect of heat-transfer law on performance of a
  two-heat-source endoreversible cycle},\ \href {\doibase 10.1063/1.455832}
  {\bibfield  {journal} {\bibinfo  {journal} {J. Chem. Phys.}\ }\textbf
  {\bibinfo {volume} {90}},\ \bibinfo {pages} {3740} (\bibinfo {year}
  {1989})}\BibitemShut {NoStop}%
\bibitem [{\citenamefont {Zhao}\ \emph {et~al.}(2022)\citenamefont {Zhao},
  \citenamefont {Gong},\ and\ \citenamefont {Tu}}]{Zhao2022}%
  \BibitemOpen
  \bibfield  {author} {\bibinfo {author} {\bibfnamefont {X.-H.}\ \bibnamefont
  {Zhao}}, \bibinfo {author} {\bibfnamefont {Z.-N.}\ \bibnamefont {Gong}}, \
  and\ \bibinfo {author} {\bibfnamefont {Z.~C.}\ \bibnamefont {Tu}},\ }\bibinfo
  {title} {Microscopic low-dissipation heat engine via shortcuts to
  adiabaticity and shortcuts to isothermality},\ \href@noop {} {\  (\bibinfo
  {year} {2022})},\ \Eprint {http://arxiv.org/abs/2206.02337} {arXiv:2206.02337
  [cond-mat.stat-mech]} \BibitemShut {NoStop}%
\end{thebibliography}%


\begin{thebibliography}{8}%
\makeatletter
\providecommand \@ifxundefined [1]{%
 \@ifx{#1\undefined}
}%
\providecommand \@ifnum [1]{%
 \ifnum #1\expandafter \@firstoftwo
 \else \expandafter \@secondoftwo
 \fi
}%
\providecommand \@ifx [1]{%
 \ifx #1\expandafter \@firstoftwo
 \else \expandafter \@secondoftwo
 \fi
}%
\providecommand \natexlab [1]{#1}%
\providecommand \enquote  [1]{``#1''}%
\providecommand \bibnamefont  [1]{#1}%
\providecommand \bibfnamefont [1]{#1}%
\providecommand \citenamefont [1]{#1}%
\providecommand \href@noop [0]{\@secondoftwo}%
\providecommand \href [0]{\begingroup \@sanitize@url \@href}%
\providecommand \@href[1]{\@@startlink{#1}\@@href}%
\providecommand \@@href[1]{\endgroup#1\@@endlink}%
\providecommand \@sanitize@url [0]{\catcode `\\12\catcode `\$12\catcode
  `\&12\catcode `\#12\catcode `\^12\catcode `\_12\catcode `\%12\relax}%
\providecommand \@@startlink[1]{}%
\providecommand \@@endlink[0]{}%
\providecommand \url  [0]{\begingroup\@sanitize@url \@url }%
\providecommand \@url [1]{\endgroup\@href {#1}{\urlprefix }}%
\providecommand \urlprefix  [0]{URL }%
\providecommand \Eprint [0]{\href }%
\providecommand \doibase [0]{http://dx.doi.org/}%
\providecommand \selectlanguage [0]{\@gobble}%
\providecommand \bibinfo  [0]{\@secondoftwo}%
\providecommand \bibfield  [0]{\@secondoftwo}%
\providecommand \translation [1]{[#1]}%
\providecommand \BibitemOpen [0]{}%
\providecommand \bibitemStop [0]{}%
\providecommand \bibitemNoStop [0]{.\EOS\space}%
\providecommand \EOS [0]{\spacefactor3000\relax}%
\providecommand \BibitemShut  [1]{\csname bibitem#1\endcsname}%
\let\auto@bib@innerbib\@empty
\bibitem [{\citenamefont {Li}\ \emph {et~al.}(2017)\citenamefont {Li},
  \citenamefont {Quan},\ and\ \citenamefont {Tu}}]{Li2017}%
  \BibitemOpen
  \bibfield  {author} {\bibinfo {author} {\bibfnamefont {G.}~\bibnamefont
  {Li}}, \bibinfo {author} {\bibfnamefont {H.~T.}\ \bibnamefont {Quan}}, \ and\
  \bibinfo {author} {\bibfnamefont {Z.~C.}\ \bibnamefont {Tu}},\ }\bibinfo
  {title} {Shortcuts to isothermality and nonequilibrium work relations},\
  \href {\doibase 10.1103/physreve.96.012144} {\bibfield  {journal} {\bibinfo
  {journal} {Phys. Rev. E}\ }\textbf {\bibinfo {volume} {96}},\ \bibinfo
  {pages} {012144} (\bibinfo {year} {2017})}\BibitemShut {NoStop}%
\bibitem [{\citenamefont {Jun}\ and\ \citenamefont {Lai}(2021)}]{Jun2021}%
  \BibitemOpen
  \bibfield  {author} {\bibinfo {author} {\bibfnamefont {Y.}~\bibnamefont
  {Jun}}\ and\ \bibinfo {author} {\bibfnamefont {P.-Y.}\ \bibnamefont {Lai}},\
  }\bibinfo {title} {Instantaneous equilibrium transition for Brownian systems
  under time-dependent temperature and potential variations: Reversibility,
  heat and work relations, and fast isentropic process},\ \href {\doibase
  10.1103/physrevresearch.3.033130} {\bibfield  {journal} {\bibinfo  {journal}
  {Phys. Rev. Research}\ }\textbf {\bibinfo {volume} {3}},\ \bibinfo {pages}
  {033130} (\bibinfo {year} {2021})}\BibitemShut {NoStop}%
\bibitem [{\citenamefont {Berger}(2007)}]{Berger2007}%
  \BibitemOpen
  \bibfield  {author} {\bibinfo {author} {\bibfnamefont {M.}~\bibnamefont
  {Berger}},\ }\href
  {https://www.ebook.de/de/product/2047791/marcel_berger_a_panoramic_view_of_riemannian_geometry.html}
  {\emph {\bibinfo {title} {A Panoramic View of Riemannian Geometry}}}\
  (\bibinfo  {publisher} {Springer Berlin Heidelberg},\ \bibinfo {year}
  {2007})\BibitemShut {NoStop}%
\bibitem [{\citenamefont {Curzon}\ and\ \citenamefont
  {Ahlborn}(1975)}]{Curzon1975}%
  \BibitemOpen
  \bibfield  {author} {\bibinfo {author} {\bibfnamefont {F.~L.}\ \bibnamefont
  {Curzon}}\ and\ \bibinfo {author} {\bibfnamefont {B.}~\bibnamefont
  {Ahlborn}},\ }\bibinfo {title} {Efficiency of a Carnot engine at maximum
  power output},\ \href {\doibase 10.1119/1.10023} {\bibfield  {journal}
  {\bibinfo  {journal} {Am. J. Phys.}\ }\textbf {\bibinfo {volume} {43}},\
  \bibinfo {pages} {22} (\bibinfo {year} {1975})}\BibitemShut {NoStop}%
\bibitem [{\citenamefont {Schmiedl}\ and\ \citenamefont
  {Seifert}(2007)}]{Schmiedl2007}%
  \BibitemOpen
  \bibfield  {author} {\bibinfo {author} {\bibfnamefont {T.}~\bibnamefont
  {Schmiedl}}\ and\ \bibinfo {author} {\bibfnamefont {U.}~\bibnamefont
  {Seifert}},\ }\bibinfo {title} {Efficiency at maximum power: An analytically
  solvable model for stochastic heat engines},\ \href {\doibase
  10.1209/0295-5075/81/20003} {\bibfield  {journal} {\bibinfo  {journal}
  {Europhys. Lett.}\ }\textbf {\bibinfo {volume} {81}},\ \bibinfo {pages}
  {20003} (\bibinfo {year} {2007})}\BibitemShut {NoStop}%
\bibitem [{\citenamefont {Chen}\ and\ \citenamefont {Yan}(1989)}]{Chen1989}%
  \BibitemOpen
  \bibfield  {author} {\bibinfo {author} {\bibfnamefont {L.}~\bibnamefont
  {Chen}}\ and\ \bibinfo {author} {\bibfnamefont {Z.}~\bibnamefont {Yan}},\
  }\bibinfo {title} {The effect of heat-transfer law on performance of a
  two-heat-source endoreversible cycle},\ \href {\doibase 10.1063/1.455832}
  {\bibfield  {journal} {\bibinfo  {journal} {J. Chem. Phys.}\ }\textbf
  {\bibinfo {volume} {90}},\ \bibinfo {pages} {3740} (\bibinfo {year}
  {1989})}\BibitemShut {NoStop}%
\bibitem [{\citenamefont {Dechant}\ \emph {et~al.}(2017)\citenamefont
  {Dechant}, \citenamefont {Kiesel},\ and\ \citenamefont {Lutz}}]{Dechant2017}%
  \BibitemOpen
  \bibfield  {author} {\bibinfo {author} {\bibfnamefont {A.}~\bibnamefont
  {Dechant}}, \bibinfo {author} {\bibfnamefont {N.}~\bibnamefont {Kiesel}}, \
  and\ \bibinfo {author} {\bibfnamefont {E.}~\bibnamefont {Lutz}},\ }\bibinfo
  {title} {Underdamped stochastic heat engine at maximum efficiency},\ \href
  {\doibase 10.1209/0295-5075/119/50003} {\bibfield  {journal} {\bibinfo
  {journal} {Europhys. Lett.}\ }\textbf {\bibinfo {volume} {119}},\ \bibinfo
  {pages} {50003} (\bibinfo {year} {2017})}\BibitemShut {NoStop}%
\bibitem [{\citenamefont {Chen}\ \emph {et~al.}(2021)\citenamefont {Chen},
  \citenamefont {Chen}, \citenamefont {Fei},\ and\ \citenamefont
  {Quan}}]{Chen2021b}%
  \BibitemOpen
  \bibfield  {author} {\bibinfo {author} {\bibfnamefont {Y.~H.}\ \bibnamefont
  {Chen}}, \bibinfo {author} {\bibfnamefont {J.-F.}\ \bibnamefont {Chen}},
  \bibinfo {author} {\bibfnamefont {Z.}~\bibnamefont {Fei}}, \ and\ \bibinfo
  {author} {\bibfnamefont {H.~T.}\ \bibnamefont {Quan}},\ }\bibinfo {title} {A
  microscopic theory of Curzon-Ahlborn heat engine},\ \href@noop {} {\
  (\bibinfo {year} {2021})},\ \Eprint {http://arxiv.org/abs/2108.04128v2}
  {arXiv:2108.04128v2 [cond-mat.stat-mech]} \BibitemShut {NoStop}%
\end{thebibliography}%

\end{document}


\title{Supplementary material: Optimizing Brownian heat engine with shortcut
strategy}
\author{Jin-Fu Chen}
\email{chenjinfu@pku.edu.cn}

\affiliation{School of Physics, Peking University, Beijing, 100871, China}

\maketitle
We show derivations to the results of the main text, and discuss the
optimal protocols of the cycles in the highly underdamped regime and
the overdamped regime. In Section \ref{sec:work and heat}, we show
the implementation of the auxiliary Hamiltonian in the shortcut strategy,
and illustrate work and heat in shortcut processes. Section \ref{sec:Explicit-results-for}
gives the explicit results for the power-law potentials. In Section
\ref{sec:Output-work,-efficiency}, we derive the analytical results
of the output work, the quasistatic efficiency, and the thermodynamic
length for the heat-engine cycles with exponential connecting paths.
In Section \ref{sec:Maximum-power-and}, we derive the maximum power
and the efficiency at the maximum power expressed by the thermodynamic
length. In Section \ref{sec:Maximum-power-at}, we derive the maximum
power at given efficiency. In Sections \ref{sec:Underdamped-limit}
and \ref{sec:Overdamped-regime}, we show the optimal control schemes
of the cycles in the highly underdamped regime and the overdamped
regimes. They are compared with the ones obtained in previous works.

\tableofcontents{}

\section{Work and heat in shortcut processes \label{sec:work and heat}}

We rewrite the Kramer equation {[}Eq. (1) in the main text{]} as

\begin{equation}
\frac{\partial\rho}{\partial t}=\mathscr{L}_{\mathrm{o}}\left[\rho\right]+\mathscr{D}_{\mathrm{o}}\left[\rho\right]+\mathscr{L}_{\mathrm{a}}\left[\rho\right]+\mathscr{D}_{\mathrm{a}}\left[\rho\right].\label{eq:Kramer_eq_in_shortcut_regime}
\end{equation}
The operators $\mathscr{L}_{\mathrm{o}}$ and $\mathscr{D}_{\mathrm{o}}$
reflect the deterministic and the dissipative parts of the evolution
subject to the original Hamiltonian $H_{\mathrm{o}}(x,p,\lambda)=p^{2}/(2m)+U_{\mathrm{o}}(x,\lambda)$,
and are explicitly

\begin{align}
\mathscr{L}_{\mathrm{o}}\left[\rho\right] & =-\frac{\partial}{\partial x}\left(\frac{\partial H_{\mathrm{o}}}{\partial p}\rho\right)+\frac{\partial}{\partial p}\left(\frac{\partial H_{\mathrm{o}}}{\partial x}\rho\right),\\
\mathscr{D}_{\mathrm{o}}\left[\rho\right] & =\kappa m\frac{\partial}{\partial p}\left(\frac{\partial H_{\mathrm{o}}}{\partial p}\rho+\frac{1}{\beta}\frac{\partial\rho}{\partial p}\right).
\end{align}
The operators $\mathscr{L}_{\mathrm{a}}$ and $\mathscr{D}_{\mathrm{a}}$
reflect the deterministic and the dissipative parts of the evolution
subject to the auxiliary Hamiltonian $H_{\mathrm{a}}$, and are explicitly 

\begin{align}
\mathscr{L}_{\mathrm{a}}\left[\rho\right] & =-\frac{\partial}{\partial x}\left(\frac{\partial H_{\mathrm{a}}}{\partial p}\rho\right)+\frac{\partial}{\partial p}\left(\frac{\partial H_{\mathrm{a}}}{\partial x}\rho\right),\\
\mathscr{D}_{\mathrm{a}}\left[\rho\right] & =\kappa m\frac{\partial}{\partial p}\left(\frac{\partial H_{\mathrm{a}}}{\partial p}\rho\right).
\end{align}
The instantaneous equilibrium state $\rho_{\mathrm{ieq}}\coloneqq\exp\{\beta[F(\lambda,\beta)-H_{\mathrm{o}}(x,p,\lambda)]\}$
nullifies the operators $\mathscr{L}_{\mathrm{o}}\left[\rho_{\mathrm{ieq}}\right]=0$
and $\mathscr{D}_{\mathrm{o}}\left[\rho_{\mathrm{ieq}}\right]=0$,
and its time derivative is 

\begin{equation}
\frac{\partial\rho_{\mathrm{ieq}}}{\partial t}=\beta\left[\frac{\dot{\beta}}{\beta}\left(F-H_{\mathrm{o}}\right)+\dot{\beta}\frac{\partial F}{\partial\beta}+\dot{\lambda}\left(\frac{\partial F}{\partial\lambda}-\frac{\partial U_{\mathrm{o}}}{\partial\lambda}\right)\right]\rho_{\mathrm{ieq}}.\label{eq:timederivativeofrho_ieq}
\end{equation}

Equation (\ref{eq:Kramer_eq_in_shortcut_regime}) gives the principle
to construct the auxiliary Hamiltonian \citep{Li2017}

\begin{equation}
\frac{\partial\rho_{\mathrm{ieq}}}{\partial t}=\mathscr{L}_{\mathrm{a}}\left[\rho_{\mathrm{ieq}}\right]+\mathscr{D}_{\mathrm{a}}\left[\rho_{\mathrm{ieq}}\right],\label{eq:equationforauxillary}
\end{equation}
which is explicitly

\begin{equation}
\frac{\partial\rho_{\mathrm{ieq}}}{\partial t}=-\frac{\partial H_{\mathrm{a}}}{\partial p}\frac{\partial\rho_{\mathrm{ieq}}}{\partial x}+\frac{\partial H_{\mathrm{a}}}{\partial x}\frac{\partial\rho_{\mathrm{ieq}}}{\partial p}+\kappa m\frac{\partial}{\partial p}\left(\frac{\partial H_{\mathrm{a}}}{\partial p}\rho_{\mathrm{ieq}}\right).\label{eq:21}
\end{equation}
As done in Ref. \citep{Jun2021}, both the inverse temperature $\beta(t)$
and the work parameter $\lambda(t)$ are treated as the time-dependent
control parameters. By setting the auxiliary Hamiltonian as 

\begin{equation}
H_{\mathrm{a}}=\dot{\lambda}h_{\lambda}(x,p,\lambda)+\dot{\beta}h_{\beta}(x,p,\lambda,\beta),
\end{equation}
Eqs. (\ref{eq:timederivativeofrho_ieq}) and (\ref{eq:21}) lead to
the partial differential equations for the auxiliary functions $h_{\lambda}(x,p,\lambda)$
and $h_{\beta}(x,p,\lambda,\beta)$ as 

\begin{align}
\frac{\kappa m}{\beta}\frac{\partial^{2}h_{\lambda}}{\partial p^{2}}-\kappa p\frac{\partial h_{\lambda}}{\partial p}+\frac{\partial U_{\mathrm{o}}}{\partial x}\frac{\partial h_{\lambda}}{\partial p}-\frac{p}{m}\frac{\partial h_{\lambda}}{\partial x} & =\frac{\partial F}{\partial\lambda}-\frac{\partial U_{\mathrm{o}}}{\partial\lambda},\\
\frac{\kappa m}{\beta}\frac{\partial^{2}h_{\beta}}{\partial p^{2}}-\kappa p\frac{\partial h_{\beta}}{\partial p}+\frac{\partial U_{\mathrm{o}}}{\partial x}\frac{\partial h_{\beta}}{\partial p}-\frac{p}{m}\frac{\partial h_{\beta}}{\partial x} & =\frac{1}{\beta}(F-H_{\mathrm{o}})+\frac{\partial F}{\partial\beta}.
\end{align}

The quasistatic work, the quasistatic heat, the irreversible work
and the irreversible heat are defined as

\begin{align}
\dot{W}_{\mathrm{o}} & =\dot{\lambda}\iint\frac{\partial U_{\mathrm{o}}}{\partial\lambda}\rho_{\mathrm{ieq}}dxdp,\\
\dot{Q}_{\mathrm{o}} & =\iint H_{\mathrm{o}}\frac{\partial\rho_{\mathrm{ieq}}}{\partial t}dxdp,\\
\dot{W}_{\mathrm{irr}} & =\iint\frac{\partial H_{\mathrm{a}}}{\partial t}\rho_{\mathrm{ieq}}dxdp,\\
\dot{Q}_{\mathrm{irr}} & =\iint H_{\mathrm{a}}\frac{\partial\rho_{\mathrm{ieq}}}{\partial t}dxdp.\label{eq:irreversible_heat}
\end{align}
Plugging Eq. (\ref{eq:21}) into Eq. (\ref{eq:irreversible_heat}),
we obtain the irreversible heat as

\begin{align}
\frac{dQ_{\mathrm{irr}}}{dt} & =\iint\left\{ \begin{array}{c}
\left(\dot{\lambda}h_{\lambda}+\dot{\beta}h_{\beta}\right)\dot{\lambda}\left[\frac{\partial h_{\lambda}}{\partial x}\frac{\partial\rho_{\mathrm{ieq}}}{\partial p}-\frac{\partial h_{\lambda}}{\partial p}\frac{\partial\rho_{\mathrm{ieq}}}{\partial x}+\kappa m\frac{\partial}{\partial p}\left(\frac{\partial h_{\lambda}}{\partial p}\rho_{\mathrm{ieq}}\right)\right]\\
+\left(\dot{\lambda}h_{\lambda}+\dot{\beta}h_{\beta}\right)\dot{\beta}\left[\frac{\partial h_{\beta}}{\partial x}\frac{\partial\rho_{\mathrm{ieq}}}{\partial p}-\frac{\partial h_{\beta}}{\partial p}\frac{\partial\rho_{\mathrm{ieq}}}{\partial x}+\kappa m\frac{\partial}{\partial p}\left(\frac{\partial h_{\beta}}{\partial p}\rho_{\mathrm{ieq}}\right)\right]
\end{array}\right\} dxdp.
\end{align}
We show the calculation for each term 

\begin{align}
 & \iint\left(\dot{\lambda}h_{\lambda}+\dot{\beta}h_{\beta}\right)\dot{\lambda}\left[\frac{\partial h_{\lambda}}{\partial x}\frac{\partial\rho_{\mathrm{ieq}}}{\partial p}-\frac{\partial h_{\lambda}}{\partial p}\frac{\partial\rho_{\mathrm{ieq}}}{\partial x}\right]dxdp\label{eq:first_term}\\
= & \iint-\dot{\lambda}\frac{\partial}{\partial p}\left[\left(\dot{\lambda}h_{\lambda}+\dot{\beta}h_{\beta}\right)\frac{\partial h_{\lambda}}{\partial x}\right]\rho_{\mathrm{ieq}}+\dot{\lambda}\frac{\partial}{\partial x}\left[\left(\dot{\lambda}h_{\lambda}+\dot{\beta}h_{\beta}\right)\frac{\partial h_{\lambda}}{\partial p}\right]\rho_{\mathrm{ieq}}dxdp\\
= & \dot{\lambda}\dot{\beta}\iint\left(\frac{\partial h_{\beta}}{\partial x}\frac{\partial h_{\lambda}}{\partial p}-\frac{\partial h_{\beta}}{\partial p}\frac{\partial h_{\lambda}}{\partial x}\right)\rho_{\mathrm{ieq}}dxdp\\
= & \dot{\lambda}\dot{\beta}\left[\left\langle \frac{\partial h_{\lambda}}{\partial p}\frac{\partial h_{\beta}}{\partial x}\right\rangle -\left\langle \frac{\partial h_{\lambda}}{\partial x}\frac{\partial h_{\beta}}{\partial p}\right\rangle \right],
\end{align}
where we do the partial integral in the second line, and the boundary
term vanishes for the trapped potential. In the last line, the average
$\left\langle \cdot\right\rangle $ is over the instantaneous equilibrium
state $\rho_{\mathrm{ieq}}$. We next calculate 

\begin{align}
 & \iint\left(\dot{\lambda}h_{\lambda}+\dot{\beta}h_{\beta}\right)\dot{\lambda}\kappa m\frac{\partial}{\partial p}\left(\frac{\partial h_{\lambda}}{\partial p}\rho_{\mathrm{ieq}}\right)dxdp\\
= & \iint-\left(\dot{\lambda}\frac{\partial h_{\lambda}}{\partial p}+\dot{\beta}\frac{\partial h_{\beta}}{\partial p}\right)\dot{\lambda}\kappa m\frac{\partial h_{\lambda}}{\partial p}\rho_{\mathrm{ieq}}dxdp\\
= & -\dot{\lambda}\kappa m\left(\dot{\lambda}\left\langle \frac{\partial h_{\lambda}}{\partial p}\frac{\partial h_{\lambda}}{\partial p}\right\rangle +\dot{\beta}\left\langle \frac{\partial h_{\lambda}}{\partial p}\frac{\partial h_{\beta}}{\partial p}\right\rangle \right).
\end{align}
It is similar to obtain 

\begin{equation}
\iint\left(\dot{\lambda}h_{\lambda}+\dot{\beta}h_{\beta}\right)\dot{\beta}\left[\frac{\partial h_{\beta}}{\partial x}\frac{\partial\rho_{\mathrm{ieq}}}{\partial p}-\frac{\partial h_{\beta}}{\partial p}\frac{\partial\rho_{\mathrm{ieq}}}{\partial x}\right]dxdp=\dot{\lambda}\dot{\beta}\left(\left\langle \frac{\partial h_{\lambda}}{\partial x}\frac{\partial h_{\beta}}{\partial p}\right\rangle -\left\langle \frac{\partial h_{\lambda}}{\partial p}\frac{\partial h_{\beta}}{\partial x}\right\rangle \right),
\end{equation}
which cancels out the first term {[}Eq. (\ref{eq:first_term}){]}.
The last term is 

\begin{equation}
\iint\left(\dot{\lambda}h_{\lambda}+\dot{\beta}h_{\beta}\right)\kappa m\frac{\partial}{\partial p}\left(\frac{\partial h_{\beta}}{\partial p}\rho_{\mathrm{ieq}}\right)dxdp=-\kappa m\left(\dot{\lambda}\left\langle \frac{\partial h_{\lambda}}{\partial p}\frac{\partial h_{\beta}}{\partial p}\right\rangle +\dot{\beta}\left\langle \frac{\partial h_{\beta}}{\partial p}\frac{\partial h_{\beta}}{\partial p}\right\rangle \right).
\end{equation}
Collecting all the terms, we obtain the explicit expression of the
irreversible heat as

\begin{equation}
\frac{dQ_{\mathrm{irr}}}{dt}=-\left(\begin{array}{cc}
\dot{\lambda} & \dot{\beta}\end{array}\right)\mathbf{g}\left(\begin{array}{c}
\dot{\lambda}\\
\dot{\beta}
\end{array}\right),
\end{equation}
with the metric 

\begin{equation}
\mathbf{g}=\kappa m\left(\begin{array}{cc}
\left\langle \frac{\partial h_{\lambda}}{\partial p}\frac{\partial h_{\lambda}}{\partial p}\right\rangle  & \left\langle \frac{\partial h_{\lambda}}{\partial p}\frac{\partial h_{\beta}}{\partial p}\right\rangle \\
\left\langle \frac{\partial h_{\lambda}}{\partial p}\frac{\partial h_{\beta}}{\partial p}\right\rangle  & \left\langle \frac{\partial h_{\beta}}{\partial p}\frac{\partial h_{\beta}}{\partial p}\right\rangle 
\end{array}\right).\label{eq:metric}
\end{equation}
The metric $\mathbf{g}$ is positive semi-definite since

\begin{align}
\left(\begin{array}{cc}
\dot{\lambda} & \dot{\beta}\end{array}\right)\mathbf{g}\left(\begin{array}{c}
\dot{\lambda}\\
\dot{\beta}
\end{array}\right) & =\kappa m\left\langle \left(\dot{\lambda}\frac{\partial h_{\lambda}}{\partial p}+\dot{\beta}\frac{\partial h_{\beta}}{\partial p}\right)^{2}\right\rangle \\
 & =\kappa m\iint\left(\dot{\lambda}\frac{\partial h_{\lambda}}{\partial p}+\dot{\beta}\frac{\partial h_{\beta}}{\partial p}\right)^{2}\rho_{\mathrm{ieq}}dxdp\geq0.
\end{align}
The irreversible work is

\begin{equation}
\frac{dW_{\mathrm{irr}}}{dt}=\frac{d\left\langle H_{\mathrm{a}}\right\rangle }{dt}+\left(\begin{array}{cc}
\dot{\lambda} & \dot{\beta}\end{array}\right)\mathbf{g}\left(\begin{array}{c}
\dot{\lambda}\\
\dot{\beta}
\end{array}\right).\label{eq:dW_irr/dt}
\end{equation}
Notice that the total differential term $d\left\langle H_{\mathrm{a}}\right\rangle /dt$
does not contribute to the irreversible work. The auxiliary Hamiltonian
is switched on at the beginning and off at the end of the process,
or changes cyclically in a heat-engine cycle. By using the Cauchy-Schwarz
inequality, the integral of the second term is bounded by

\begin{equation}
\int_{0}^{\tau}\left(\begin{array}{cc}
\dot{\lambda} & \dot{\beta}\end{array}\right)\mathbf{g}\left(\begin{array}{c}
\dot{\lambda}\\
\dot{\beta}
\end{array}\right)dt\cdot\tau\geq\left(\int_{0}^{\tau}\sqrt{\left(\begin{array}{cc}
\dot{\lambda} & \dot{\beta}\end{array}\right)\mathbf{g}\left(\begin{array}{c}
\dot{\lambda}\\
\dot{\beta}
\end{array}\right)}dt\right)^{2}=\mathcal{L}^{2},\label{eq:inequality}
\end{equation}
where the thermodynamic length is defined as 

\begin{equation}
\mathcal{L}\coloneqq\int_{0}^{1}\sqrt{\left(\begin{array}{cc}
\lambda^{\prime}(s) & \beta^{\prime}(s)\end{array}\right)\mathbf{g}\left(\begin{array}{c}
\lambda^{\prime}(s)\\
\beta^{\prime}(s)
\end{array}\right)}ds.\label{eq:length}
\end{equation}
The path in the control-parameter space is parameterized by $\left(\lambda(s),\beta(s)\right)$
with $s\in[0,1]$. The integral in Eq. (\ref{eq:length}) is independent
of the protocol and the operation time. The bound (1) of the irreversible
work in the main text follows immediately from the inequality (\ref{eq:inequality}).

\section{Explicit results for the power-law potentials\label{sec:Explicit-results-for}}

For the power-law potential $U_{\mathrm{o}}(x,\lambda)=m\lambda^{n+1}x^{2n}/(2n)$,
the partition function is
\begin{equation}
Z_{\lambda}(\beta)\coloneqq\iint e^{-\beta H_{\mathrm{o}}}dxdp=2^{\frac{1}{2n}+\frac{3}{2}}\sqrt{\frac{m\pi}{\beta}}\lambda^{-\frac{n+1}{2n}}\Gamma\left(1+\frac{1}{2n}\right)\left(\frac{\beta m}{n}\right)^{-\frac{1}{2n}},
\end{equation}
with the gamma function $\Gamma(x)=\int_{0}^{+\infty}t^{x-1}e^{-t}dt$.
The moments of the momentum and position are obtained as 

\begin{align}
\left\langle p^{2}\right\rangle  & =\frac{m}{\beta},\\
\left\langle x^{2}\right\rangle  & =\frac{R_{n}}{\lambda}\frac{1}{\left(\beta\lambda m\right)^{1/n}},\label{eq:x^2_average}\\
\left\langle x^{2n}\right\rangle  & =\frac{1}{\beta m\lambda^{n+1}},
\end{align}
where $R_{n}$ is a pure number
\begin{equation}
R_{n}=\left(2n\right)^{\frac{1}{n}}\frac{\Gamma\left(\frac{3}{2n}\right)}{\Gamma\left(\frac{1}{2n}\right)}.
\end{equation}

The auxiliary functions $h_{\lambda}(x,p,\lambda)$ and $h_{\beta}(x,p,\lambda,\beta)$
are given in the main text, and the average values of them are 

\begin{align}
\left\langle h_{\lambda}\right\rangle  & =\frac{(n+1)^{2}}{4\kappa n^{2}\lambda\beta}+\frac{n+1}{4\lambda^{2}n}\frac{R_{n}\kappa m}{\left(\beta\lambda m\right)^{1/n}},\\
\left\langle h_{\beta}\right\rangle  & =\frac{(n+1)^{2}}{4n^{2}}\frac{1}{\kappa\beta^{2}}+\frac{\kappa m}{4n\lambda\beta}\frac{R_{n}}{\left(\beta\lambda m\right)^{1/n}}.
\end{align}
The average values of the original Hamiltonian and the auxiliary Hamiltonian
are 

\begin{align}
\left\langle H_{\mathrm{o}}\right\rangle  & =\frac{n+1}{2n\beta},\\
\left\langle H_{\mathrm{a}}\right\rangle  & =\frac{(n+1)^{2}}{4\kappa n^{2}\beta}\left(\frac{\dot{\lambda}}{\lambda}+\frac{\dot{\beta}}{\beta}\right)+\frac{\kappa m}{4n\lambda}\frac{R_{n}}{\left(\beta\lambda m\right)^{1/n}}\left[(n+1)\frac{\dot{\lambda}}{\lambda}+\frac{\dot{\beta}}{\beta}\right].
\end{align}
The explicit results of the quasistatic work, the quasistatic heat,
the irreversible work and the irreversible heat are 

\begin{align}
\frac{dW_{\mathrm{o}}}{dt} & =\frac{n+1}{2n\beta}\frac{\dot{\lambda}}{\lambda},\\
\frac{dQ_{\mathrm{o}}}{dt} & =-\frac{n+1}{2n\beta}\left(\frac{\dot{\lambda}}{\lambda}+\frac{\dot{\beta}}{\beta}\right),\label{eq:Qo_dot}\\
\frac{dW_{\mathrm{irr}}}{dt} & =\frac{d\left\langle H_{\mathrm{a}}\right\rangle }{dt}+\frac{(n+1)^{2}}{4n^{2}}\left[\frac{1}{\kappa\beta}\left(\frac{\dot{\lambda}}{\lambda}+\frac{\dot{\beta}}{\beta}\right)^{2}+\frac{\kappa m}{\lambda}\frac{R_{n}}{(\lambda m\beta)^{1/n}}\left(\frac{\dot{\lambda}}{\lambda}+\frac{\dot{\beta}}{(n+1)\beta}\right)^{2}\right],\\
\frac{dQ_{\mathrm{\mathrm{irr}}}}{dt} & =-\frac{(n+1)^{2}}{4n^{2}}\left[\frac{1}{\kappa\beta}\left(\frac{\dot{\lambda}}{\lambda}+\frac{\dot{\beta}}{\beta}\right)^{2}+\frac{\kappa m}{\lambda}\frac{R_{n}}{(\lambda m\beta)^{1/n}}\left(\frac{\dot{\lambda}}{\lambda}+\frac{\dot{\beta}}{(n+1)\beta}\right)^{2}\right].
\end{align}

For a heat-engine cycle, the irreversible work of a cycle is 

\begin{equation}
W_{\mathrm{irr}}=\frac{(n+1)^{2}}{4n^{2}}\int_{0}^{\tau}\left[\frac{1}{\kappa\beta}\left(\frac{\dot{\lambda}}{\lambda}+\frac{\dot{\beta}}{\beta}\right)^{2}+\frac{\kappa m}{\lambda}\frac{R_{n}}{(\lambda m\beta)^{1/n}}\left(\frac{\dot{\lambda}}{\lambda}+\frac{\dot{\beta}}{(n+1)\beta}\right)^{2}\right]dt,\label{eq:Wirr}
\end{equation}
and the thermodynamic length is 

\begin{align}
\mathcal{L} & =\frac{n+1}{2n}\oint\left\{ \frac{1}{\kappa\beta}\left[\frac{\lambda^{\prime}(s)}{\lambda}+\frac{\beta^{\prime}(s)}{\beta}\right]^{2}+\frac{\kappa m}{\lambda}\frac{R_{n}}{(\lambda m\beta)^{1/n}}\left[\frac{\lambda^{\prime}(s)}{\lambda}+\frac{\beta^{\prime}(s)}{(n+1)\beta}\right]^{2}\right\} ^{\frac{1}{2}}ds,\label{eq:thermodynamic_length}
\end{align}
where $\lambda(s)$ and $\beta(s)$ give the path of the cycle in
the control parameter space, and the primes denote their derivatives
in respect to arc-length parameter $s$. Let us define a dimensionless
quantity 
\begin{align}
\chi & \coloneqq R_{n}\frac{\kappa^{2}(m\beta)^{1-1/n}}{\lambda^{1+1/n}}.
\end{align}
The thermodynamic length can be rewritten as 

\begin{equation}
\mathcal{L}=\frac{n+1}{2n}\oint\frac{1}{\sqrt{\kappa\beta}}\left\{ \left[\frac{\lambda^{\prime}(s)}{\lambda}+\frac{\beta^{\prime}(s)}{\beta}\right]^{2}+\chi\left[\frac{\lambda^{\prime}(s)}{\lambda}+\frac{\beta^{\prime}(s)}{(n+1)\beta}\right]^{2}\right\} ^{\frac{1}{2}}ds,\label{eq:lengthrewrittenwithchi}
\end{equation}
The highly underdamped regime is reached when $\chi\ll1$, and the
first term dominates in Eq. (\ref{eq:lengthrewrittenwithchi}). The
overdamped regime is reached when $\chi\gg1$, and the second term
dominates in Eq. (\ref{eq:lengthrewrittenwithchi}).

Equation (15) of the main text is obtained by representing $\lambda=\lambda_{0}e^{r},\beta=1/T$
with the new parameters $r$ and $T$. The elements of the metric
(\ref{eq:metric}) expressed by $r$ and $T$ are

\begin{align}
g_{rr} & =\frac{(n+1)^{2}}{4n^{2}}\left[\frac{T}{\kappa}+\frac{\kappa mR_{n}e^{-r}}{\lambda_{0}}\left(\frac{Te^{-r}}{\lambda_{0}m}\right)^{1/n}\right],\\
g_{rT} & =-\frac{(n+1)^{2}}{4n^{2}T}\left[\frac{T}{\kappa}+\frac{\kappa mR_{n}e^{-r}}{\left(n+1\right)\lambda_{0}}\left(\frac{Te^{-r}}{\lambda_{0}m}\right)^{1/n}\right],\\
g_{TT} & =\frac{(n+1)^{2}}{4n^{2}T^{2}}\left[\frac{T}{\kappa}+\frac{\kappa mR_{n}e^{-r}}{\left(n+1\right)^{2}\lambda_{0}}\left(\frac{Te^{-r}}{\lambda_{0}m}\right)^{1/n}\right].
\end{align}
The Christoffel symbol is determined by the metric as 

\begin{equation}
\Gamma_{\;kl}^{i}=\frac{1}{2}g^{in}\left(\frac{\partial g_{nk}}{\partial x^{l}}+\frac{\partial g_{nl}}{\partial x^{k}}-\frac{\partial g_{kl}}{\partial x^{n}}\right),
\end{equation}
and the geodesic equation is 

\begin{equation}
\ddot{x}^{i}+\Gamma_{\;kl}^{i}\dot{x}^{k}\dot{x}^{l}=0.
\end{equation}
Here the coordinates $x^{i}$ are the control parameters $r$ and
$T$. The geodesic equation is explicitly obtained as 

\begin{align}
\ddot{r} & =\frac{(n+1)^{2}}{2n^{2}\kappa^{2}m^{2}R_{n}}\left(\frac{\lambda_{0}me^{r}}{T}\right)^{1+\frac{1}{n}}\left(\dot{T}-T\dot{r}\right)^{2}+\frac{(n+1)(n+2)}{2n^{2}}\dot{r}^{2}-\frac{(n+2)}{n^{2}}\dot{r}\frac{\dot{T}}{T}-\frac{(n-2)}{2n^{2}}\frac{\dot{T}^{2}}{T^{2}},\\
\frac{\ddot{T}}{T} & =\frac{(n+1)^{2}}{2n^{2}\kappa^{2}m^{2}R_{n}}\left(\frac{\lambda_{0}me^{r}}{T}\right)^{1+\frac{1}{n}}\left(\dot{T}-T\dot{r}\right)^{2}+\frac{(n+1)^{2}}{n^{2}}\dot{r}^{2}-\frac{2(n+1)}{n^{2}}\dot{r}\frac{\dot{T}}{T}+\frac{\left(n^{2}+2\right)}{2n^{2}}\frac{\dot{T}^{2}}{T^{2}}.
\end{align}
By solving the geodesic equation with the boundary conditions $(r_{\mathrm{i}},T_{\mathrm{i}})$
and $(r_{\mathrm{f}},T_{\mathrm{f}})$, we obtain the optimal protocol
to minimize the irreversible work by varying the control parameters
with a constant velocity of the thermodynamic length on the geodesic
path. It is difficult to solve the boundary-value problem of the above
nonlinear ordinary differential equations, but we can convert it into
an initial-value problem. Such a method is known as the shooting method
\citep{Berger2007}. By solving the geodesic equation with different
directions of the initial velocity, we find the shortest geodesic
path connecting the point $(r_{0},T_{L})$ with the high-temperature
isothermal line $T=T_{H}$.

\section{Output work, efficiency and thermodynamic length for cycles with
exponential connecting paths\label{sec:Output-work,-efficiency}}

For a realistic heat engine, the control parameters $r$ and $T$
are varied under constraints. We consider the temperature $T$ can
be arbitrarily tuned within the range $T_{L}\leq T\leq T_{H}$. A
typical heat-engine cycle with the shortcut strategy is constructed
by two isothermal processes (processes I and III) and two connecting
processes (processes II and IV). 
\begin{itemize}
\item Process I: the temperature is $T=T_{L}$ with the work parameter varied
from $r_{1}$ to $r_{2}$ ($r_{2}>r_{1}$).
\item Process II: the control parameters are varied from $(r_{2},T_{L})$
to $(r_{2}^{\prime},T_{H})$.
\item Process III: the temperature is $T=T_{H}$ with the work parameter
varied from $r_{2}^{\prime}$ to $r_{1}^{\prime}$ ($r_{2}^{\prime}>r_{1}^{\prime}$).
\item Process IV: the control parameters are varied from $(r_{1}^{\prime},T_{H})$
to $(r_{1},T_{L})$.
\end{itemize}
We choose the connecting processes with the exponential paths $T(r)=T_{L}\exp[\alpha_{1}(r-r_{1})],\:r_{1}\leq r\leq r_{1}^{\prime}\coloneqq r_{1}+\ln(T_{H}/T_{L})/\alpha_{1}$
(process IV) and $T(r)=T_{L}\exp[\alpha_{2}(r-r_{2})],\:r_{2}\leq r\leq r_{2}^{\prime}\coloneqq r_{2}+\ln(T_{H}/T_{L})/\alpha_{2}$
(process II), and calculate the analytical results of the work and
heat. The quasistatic work of the whole cycle is 
\begin{equation}
W_{\mathrm{o}}=-\frac{n+1}{2n}\mathcal{A}_{\mathrm{exp}}(r_{1},\alpha_{1},r_{2},\alpha_{2})<0,
\end{equation}
which is proportional to the area of the closed path

\begin{equation}
\mathcal{A}_{\mathrm{exp}}(r_{1},\alpha_{1},r_{2},\alpha_{2})=\left(r_{2}-r_{1}\right)\left(T_{H}-T_{L}\right)+\left(\frac{1}{\alpha_{1}}-\frac{1}{\alpha_{2}}\right)\left[T_{H}-T_{L}-T_{H}\ln\left(T_{H}/T_{L}\right)\right].
\end{equation}
In a quasistatic cycle, the heat is absorbed in process III, and can
also be absorbed in processes II and IV according to the values of
$\alpha_{2}$ and $\alpha_{1}$. The quasistatic heat of the whole
cycle is

\begin{equation}
Q_{\mathrm{o},+}=\frac{n+1}{2n}\left\{ \left(r_{2}-r_{1}\right)T_{H}-\left(\frac{1}{\alpha_{1}}-\frac{1}{\alpha_{2}}\right)T_{H}\ln\left(\frac{T_{H}}{T_{L}}\right)+\mathrm{max}[\frac{\alpha_{2}-1}{\alpha_{2}}(T_{H}-T_{L}),0]+\mathrm{max}[-\frac{\alpha_{1}-1}{\alpha_{1}}(T_{H}-T_{L}),0]\right\} .\label{eq:Qo+}
\end{equation}
The quasistatic efficiency is defined as 
\begin{equation}
\eta_{\mathrm{o}}\coloneqq-\frac{W_{\mathrm{o}}}{Q_{\mathrm{o},+}},\label{eq:quasistaticeffciency}
\end{equation}
which is explicitly

\begin{equation}
\eta_{\mathrm{o}}=\frac{\left(r_{2}-r_{1}\right)\left(T_{H}-T_{L}\right)+\left(\frac{1}{\alpha_{1}}-\frac{1}{\alpha_{2}}\right)\left[T_{H}-T_{L}-T_{H}\ln\left(T_{H}/T_{L}\right)\right]}{\left(r_{2}-r_{1}\right)\left(T_{H}\right)-\left(\frac{1}{\alpha_{1}}-\frac{1}{\alpha_{2}}\right)T_{H}\ln\left(T_{H}/T_{L}\right)+\mathrm{max}[\frac{\alpha_{2}-1}{\alpha_{2}}(T_{H}-T_{L}),0]+\mathrm{max}[-\frac{\alpha_{1}-1}{\alpha_{1}}(T_{H}-T_{L}),0]}.
\end{equation}
By setting $\alpha_{1}=\alpha_{2}=1$, the quasistatic cycle is the
Carnot cycle, and the quasistatic efficiency becomes the Carnot efficiency
$\eta_{\mathrm{o}}=\eta_{\mathrm{C}}\coloneqq1-T_{L}/T_{H}$.

We next calculate the thermodynamic length. The thermodynamic lengths
of the isothermal processes are 

\begin{align}
\mathcal{L}_{\mathrm{I}} & =\frac{n+1}{2n}\int_{r_{1}}^{r_{2}}\sqrt{\frac{T_{L}}{\kappa}+R_{n}\frac{\kappa m}{\lambda_{0}}e^{-r}\left(\frac{T_{L}e^{-r}}{\lambda_{0}m}\right)^{1/n}}dr\\
 & =\sqrt{\frac{T_{L}}{\kappa}}\left.\left[\sinh^{-1}\left(\sqrt{\frac{\lambda_{0}T_{L}e^{r}}{\kappa^{2}mR_{n}}\left(\frac{\lambda_{0}me^{r}}{T_{L}}\right)^{1/n}}\right)-\sqrt{1+\frac{\kappa^{2}mR_{n}}{\lambda_{0}T_{L}e^{r}}\left(\frac{T_{L}}{\lambda_{0}me^{r}}\right)^{1/n}}\right]\right|_{r=r_{1}}^{r_{2}}\label{eq:lengthI}
\end{align}
and 

\begin{align}
\mathcal{L}_{\mathrm{III}} & =\frac{n+1}{2n}\int_{r_{1}^{\prime}}^{r_{2}^{\prime}}\sqrt{\frac{T_{H}}{\kappa}+R_{n}\frac{\kappa m}{\lambda_{0}}e^{-r}\left(\frac{T_{H}e^{-r}}{\lambda_{0}m}\right)^{1/n}}dr\\
 & =\sqrt{\frac{T_{H}}{\kappa}}\left.\left[\sinh^{-1}\left(\sqrt{\frac{\lambda_{0}T_{H}e^{r}}{\kappa^{2}mR_{n}}\left(\frac{\lambda_{0}me^{r}}{T_{H}}\right)^{1/n}}\right)-\sqrt{1+\frac{\kappa^{2}mR_{n}}{\lambda_{0}T_{H}e^{r}}\left(\frac{T_{H}}{\lambda_{0}me^{r}}\right)^{1/n}}\right]\right|_{r=r_{1}^{\prime}}^{r_{2}^{\prime}}.\label{eq:lengthIII}
\end{align}
The thermodynamic length of process II is 

\begin{align}
\mathcal{L}_{\mathrm{II}} & =\mathcal{L}_{\alpha_{2}}(r_{2})\\
 & =\frac{n+1}{2n}\int_{r_{2}}^{r_{2}^{\prime}}\sqrt{(1-\alpha_{2})^{2}\frac{T_{L}}{\kappa}e^{\alpha_{2}(r-r_{2})}+(1-\frac{\alpha_{2}}{n+1})^{2}R_{n}\frac{\kappa m}{\lambda_{0}}e^{-r}\left(\frac{T_{L}}{\lambda_{0}m}e^{-r+\alpha_{2}r-\alpha_{2}r_{2}}\right)^{1/n}}dr\\
 & =\frac{n+1}{n}\sqrt{(\frac{1}{\alpha_{2}}-\frac{1}{n+1})^{2}R_{n}\frac{\kappa m}{\lambda_{0}}e^{-r}\left(\frac{T_{L}}{\lambda_{0}m}e^{-r+\alpha_{2}r-\alpha_{2}r_{2}}\right)^{1/n}\left(1+\Theta_{\mathrm{II}}e^{\left(1+\alpha_{2}+\frac{1-\alpha_{2}}{n}\right)r}\right)}\times\nonumber \\
 & \left.\left[1+\frac{1}{2}\Gamma\left(\frac{-1-\frac{1-\alpha_{2}}{n}}{2\left(1+\alpha_{2}+\frac{1-\alpha_{2}}{n}\right)}\right)\,_{2}\tilde{F}_{1}\left(1,\frac{\alpha_{2}}{2\left(1+\alpha_{2}+\frac{1-\alpha_{2}}{n}\right)};\frac{1+2\alpha_{2}+\frac{1-\alpha_{2}}{n}}{2\left(1+\alpha_{2}+\frac{1-\alpha_{2}}{n}\right)};-\Theta_{\mathrm{II}}e^{\left(1+\alpha_{2}+\frac{1-\alpha_{2}}{n}\right)r}\right)\right]\right|_{r=r_{2}}^{r_{2}^{\prime}},
\end{align}
where the constant is 

\begin{align}
\Theta_{\mathrm{II}} & =\frac{(1-\alpha_{2})^{2}}{(1-\frac{\alpha_{2}}{n+1})^{2}}\frac{\lambda_{0}T_{L}e^{-\alpha_{2}r_{2}}}{\kappa^{2}mR_{n}}\left(\frac{\lambda_{0}m}{T_{L}}e^{\alpha_{2}r_{2}}\right)^{1/n},
\end{align}
and the regularized hyper-geometric function is 

\begin{align}
\,_{2}\tilde{F}_{1}(a,b;c;z) & =\frac{\,_{2}F_{1}(a,b;c;z)}{\Gamma(c)}=\frac{1}{\Gamma(c)}\sum_{k=0}^{\infty}\frac{(a)_{k}(b)_{k}}{k!(c)_{k}}z^{k},
\end{align}
with the Pochhammer symbol $(a)_{k}=\Gamma(a+k)/\Gamma(a)$. The thermodynamic
length of process IV is

\begin{align}
\mathcal{L}_{\mathrm{IV}} & =\mathcal{L}_{\alpha_{1}}(r_{1})\\
 & =\frac{n+1}{2n}\int_{r_{1}}^{r_{1}^{\prime}}\sqrt{(1-\alpha_{1})^{2}\frac{T_{L}}{\kappa}e^{\alpha_{1}(r-r_{1})}+(1-\frac{\alpha_{1}}{n+1})^{2}R_{n}\frac{\kappa m}{\lambda_{0}}e^{-r}\left(\frac{T_{L}}{\lambda_{0}m}e^{-r+\alpha_{1}r-\alpha_{1}r_{1}}\right)^{1/n}}dr\\
 & =\frac{n+1}{n}\sqrt{(\frac{1}{\alpha_{1}}-\frac{1}{n+1})^{2}R_{n}\frac{\kappa m}{\lambda_{0}}e^{-r}\left(\frac{T_{L}}{\lambda_{0}m}e^{-r+\alpha_{1}r-\alpha_{1}r_{1}}\right)^{1/n}\left(1+\Theta_{\mathrm{IV}}e^{\left(1+\alpha_{1}+\frac{1-\alpha_{1}}{n}\right)r}\right)}\times\nonumber \\
 & \left.\left[1+\frac{1}{2}\Gamma\left(\frac{-1-\frac{1-\alpha_{1}}{n}}{2\left(1+\alpha_{1}+\frac{1-\alpha_{1}}{n}\right)}\right)\,_{2}\tilde{F}_{1}\left(1,\frac{\alpha_{1}}{2\left(1+\alpha_{1}+\frac{1-\alpha_{1}}{n}\right)};\frac{1+2\alpha_{1}+\frac{1-\alpha_{1}}{n}}{2\left(1+\alpha_{1}+\frac{1-\alpha_{1}}{n}\right)};-\Theta_{\mathrm{IV}}e^{\left(1+\alpha_{1}+\frac{1-\alpha_{1}}{n}\right)r}\right)\right]\right|_{r=r_{1}}^{r_{1}^{\prime}},
\end{align}
where the constant is

\begin{align}
\Theta_{\mathrm{IV}} & =\frac{(1-\alpha_{1})^{2}}{(1-\frac{\alpha_{1}}{n+1})^{2}}\frac{\lambda_{0}T_{L}e^{-\alpha_{1}r_{1}}}{\kappa^{2}mR_{n}}\left(\frac{\lambda_{0}m}{T_{L}}e^{\alpha_{1}r_{1}}\right)^{1/n}.
\end{align}
The thermodynamic length of the whole cycle is 

\begin{equation}
\mathcal{L}_{\mathrm{exp}}(r_{1},\alpha_{1},r_{2},\alpha_{2})=\mathcal{L}_{\mathrm{I}}+\mathcal{L}_{\mathrm{II}}+\mathcal{L}_{\mathrm{III}}+\mathcal{L}_{\mathrm{IV}}.
\end{equation}
In the optimal protocol of the cycle, the control parameters are varied
with a constant velocity of the thermodynamic length. The operation
time consumed in each process is thus proportional to the thermodynamic
length of the corresponding path, i.e.,

\begin{equation}
\tau_{l}=\frac{\mathcal{L}_{l}}{\mathcal{L}}\tau,\label{eq:operation_time_in_each process}
\end{equation}
with $l=\mathrm{I},\:\mathrm{II},\:\mathrm{III},$and $\mathrm{IV}$.
Here, $\mathcal{L}$ is the thermodynamic length of the closed path,
and $\tau$ is the operation time of the whole cycle.

According to positive and negative sign of the quasistatic heat flux
$\dot{Q}_{\mathrm{o}}>0$ and $\dot{Q}_{\mathrm{o}}<0$, we divide
the whole cycle into the heat absorbed and the heat released paths
with the thermodynamic lengths $\mathcal{L}_{+}$ and $\mathcal{L}_{-}$,
respectively. Here we focus on the optimal protocols on the cycles
with exponential connecting paths, and the quasistatic heat flux $\dot{Q}_{\mathrm{o}}$
does not change sign in each process. In the optimal protocol, the
irreversible heat flux $\dot{Q}_{\mathrm{irr}}=-\mathcal{L}^{2}/\tau^{2}$
is a negative constant during the whole cycle. To ensure positive
output work, the heat must be absorbed in process $\mathrm{III}$
and released in process $\mathrm{I}$. The sign of the quasistatic
heat flux \textbf{$\dot{Q}_{\mathrm{o}}$} in the two connecting processes
$\mathrm{II}$ and $\mathrm{IV}$ is determined by $\alpha$ according
to Eq. (\ref{eq:Qo+}). Then, the thermodynamic length is obtained
as 

\begin{align}
\mathcal{L}_{+} & =\begin{cases}
\mathcal{L}_{\mathrm{III}} & \alpha_{1}\geq1,\alpha_{2}\leq1\\
\mathcal{L}_{\mathrm{III}}+\mathcal{L}_{\mathrm{II}} & \alpha_{1}\geq1,\alpha_{2}>1\\
\mathcal{L}_{\mathrm{III}}+\mathcal{L}_{\mathrm{IV}} & \alpha_{1}<1,\alpha_{2}\leq1\\
\mathcal{L}_{\mathrm{III}}+\mathcal{L}_{\mathrm{II}}+\mathcal{L}_{\mathrm{IV}} & \alpha_{1}<1,\alpha_{2}>1
\end{cases}.
\end{align}
We can check the sign by plotting $\dot{Q}_{\mathrm{o}}$ in a whole
cycle {[}Fig. 2(d) in the main text{]}. The analytical results obtained
in this section are used to plot Fig. 3 of the main text.

We remark that the thermodynamic lengths of the connecting paths are
always nonzero in the general damped situation. Thus, the connecting
processes consume finite operation time. In the highly underdamped
regime $\chi\ll1$ and the overdamped regime $\chi\gg1$, we can find
specific exponential paths with zero thermodynamic length. The closed
path can be constructed using these zero-length paths. In Sections
\ref{sec:Underdamped-limit} and \ref{sec:Overdamped-regime}, we
show explicit results for the highly underdamped regime and the overdamped
regime respectively.

\section{Maximum power and efficiency at maximum power \label{sec:Maximum-power-and}}

Since the irreversible work is bounded by $W_{\mathrm{irr}}\geq\mathcal{L}^{2}/\tau$,
the output work in a cycle is bounded by

\begin{equation}
-W\leq-W_{\mathrm{o}}-\frac{\mathcal{L}^{2}}{\tau}.
\end{equation}
The output power $P\coloneqq-W/\tau$ is bounded by 

\begin{equation}
P\leq-\frac{W_{\mathrm{o}}}{\tau}-\frac{\mathcal{L}^{2}}{\tau^{2}},\label{eq:boundofpower}
\end{equation}
where the equality is saturated by the optimal protocol of the cycle.
By choosing the operation time 
\begin{equation}
\tau=\tau_{\mathrm{max}}\coloneqq\frac{2\mathcal{L}^{2}}{(-W_{\mathrm{o}})},\label{eq:tau_max}
\end{equation}
the maximum power is reached

\begin{equation}
P_{\mathrm{max}}=\frac{(-W_{\mathrm{o}})^{2}}{4\mathcal{L}^{2}}.\label{eq:maximumpower_general}
\end{equation}

The efficiency of the finite-time cycle is defined as

\begin{align}
\eta & \coloneqq\frac{-W}{Q_{+}}\label{eq:55}\\
 & =-\frac{W_{\mathrm{o}}+W_{\mathrm{irr}}}{Q_{\mathrm{o},+}+Q_{\mathrm{irr},+}}.
\end{align}
In the optimal protocol on a given closed path, the irreversible work
is expressed by the thermodynamic length

\begin{equation}
W_{\mathrm{irr}}=\frac{\mathcal{L}^{2}}{\tau}.
\end{equation}
The irreversible heat flux is a negative constant $\dot{Q}_{\mathrm{irr}}=-\mathcal{L}^{2}/\tau^{2}$,
and the irreversible part of the overall heat absorbed is

\begin{equation}
Q_{\mathrm{irr},+}=-\frac{\mathcal{L}}{\tau}\mathcal{L}_{+}.
\end{equation}
Together with the quasistatic efficiency (\ref{eq:quasistaticeffciency}),
we obtain the efficiency

\begin{align}
\eta & =\frac{-W_{\mathrm{o}}-\mathcal{L}^{2}/\tau}{-W_{\mathrm{o}}/\eta_{\mathrm{o}}-\mathcal{L}_{+}\mathcal{L}/\tau}.\label{eq:efficiency_finite_time}
\end{align}
With Eq. (\ref{eq:tau_max}), the efficiency at the maximum power
is

\begin{equation}
\eta_{\mathrm{EMP}}=\frac{\eta_{\mathrm{o}}}{2-\eta_{\mathrm{o}}\mathcal{\mathcal{L}}_{+}/\mathcal{L}}.\label{eq:EMP}
\end{equation}
For the cycles with exponential connecting paths, the maximum power
and the efficiency at the maximum power are then obtained with the
above analytical results of $\mathcal{A}_{\mathrm{exp}},\eta_{\mathrm{o}},\mathcal{L}_{\mathrm{exp}}$
and $\mathcal{L}_{+}$.

\begin{figure}
\includegraphics[width=12cm]{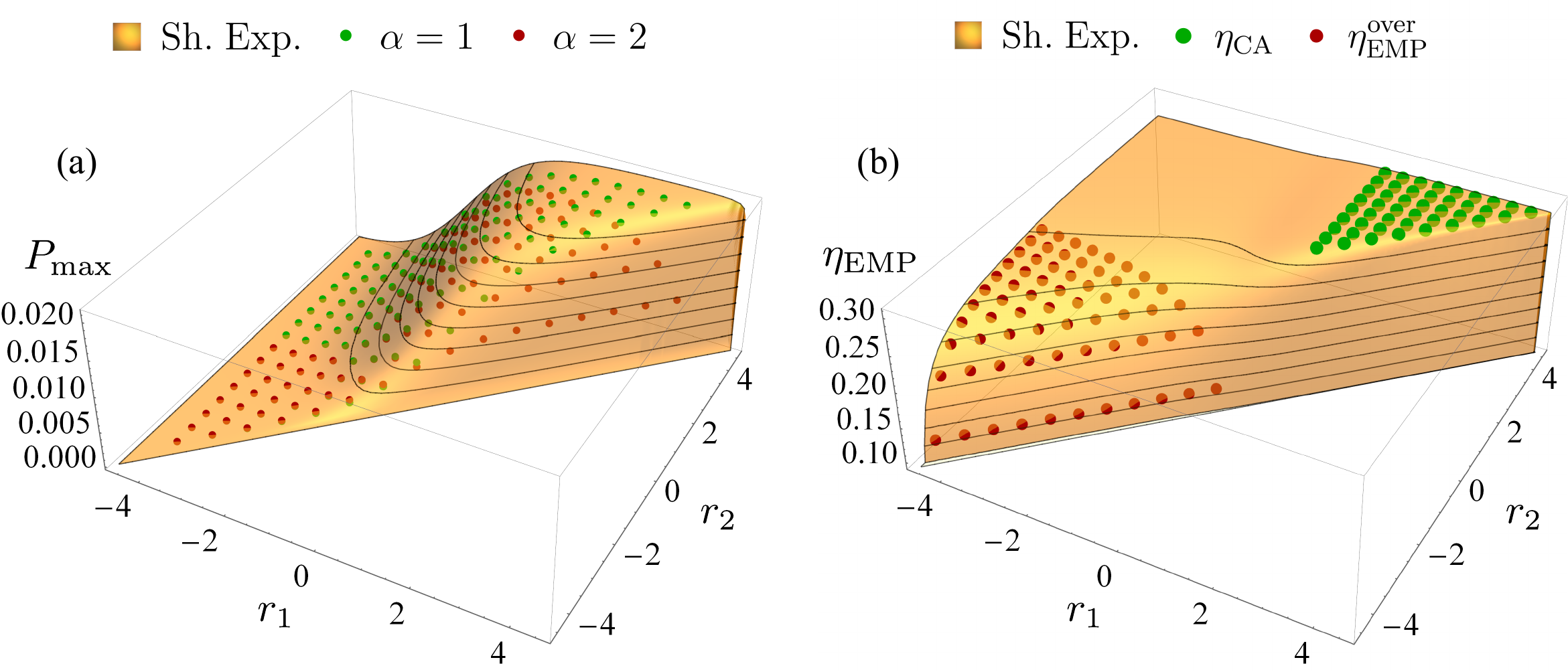}

\caption{The maximum power and the efficiency at the maximum power as functions
of $r_{1}$ and $r_{2}$. (a) The maximum power $P_{\mathrm{max}}$
of the heat-engine cycles with exponential paths, where we choose
the shortest exponential paths (orange surface) and the exponential
paths with fixed $\alpha_{1,2}=1$ (green dots) and $\alpha_{1,2}=2$
(red dots). (b) The efficiency at the maximum power $\eta_{\mathrm{EMP}}$
of the heat-engine cycle with the shortest exponential paths. It approaches
the Curzon-Ahlborn efficiency $\eta_{\mathrm{CA}}$ (green dots) in
the highly underdamped regime, and approach $\eta_{\mathrm{EMP}}^{\mathrm{over}}$
in the overdamped regime (red dots).\label{fig:The-maximum-power}}
\end{figure}

We presents the results of the maximum power $P_{\mathrm{max}}$ and
the efficiency at the maximum power $\eta_{\mathrm{EMP}}$ of different
cycles with the exponential connecting paths in Fig. \ref{fig:The-maximum-power},
where $P_{\mathrm{max}}$ and $\eta_{\mathrm{EMP}}$ are functions
of $r_{1}$ and $r_{2}$. We consider the harmonic potential $n=1$,
and the other parameters are chosen as $\lambda_{0}=1$, $m=1$, $\kappa=1$,
$T_{L}=0.5$, and $T_{H}=1$. Figure \ref{fig:The-maximum-power}(a)
shows the maximum power $P_{\mathrm{max}}$ of different heat-engine
cycles with exponential paths. The connecting paths are chosen as
the shortest exponential paths minimizing $\mathcal{L}_{\alpha_{1,2}}(r_{1,2})$
(orange surface) and the exponential paths with fixed $\alpha_{1,2}=1$
(green dots) and $\alpha_{1,2}=2$ (red dots). The maximum power of
the cycle with the shortest exponential paths is larger than that
of the cycles with fixed $\alpha_{1,2}$, and coincides with the results
by choosing the connecting paths as the shortest geodesic paths (not
shown in the figure). Figure \ref{fig:The-maximum-power}(b) shows
the efficiency at the maximum power $\eta_{\mathrm{EMP}}$ for the
cycle with the shortest exponential paths. In the highly underdamped
regime, the efficiency at the maximum power returns to the Curzon-Ahlborn
efficiency \citep{Curzon1975} $\eta_{\mathrm{CA}}=1-\sqrt{T_{L}/T_{H}}$
(green dots). In the overdamped regime, the efficiency at the maximum
power agrees with the prediction $\eta_{\mathrm{EMP}}^{\mathrm{over}}=\eta_{\mathrm{o}}^{\mathrm{over}}/(2-\eta_{\mathrm{o}}^{\mathrm{over}}/2)$
(red dots) in Ref. \citep{Schmiedl2007} where the quasistatic efficiency
$\eta_{\mathrm{o}}^{\mathrm{over}}$ {[}Eq. (\ref{eq:quasistatic_efficiency_over}){]}
includes the heat leakage induced by the kinetic energy.

\section{Maximum power at given efficiency\label{sec:Maximum-power-at}}

From Eq. (\ref{eq:efficiency_finite_time}), we can also calculate
the maximum power at given efficiency. By introducing the function

\begin{equation}
\mathcal{F}=\frac{-W_{\mathrm{o}}}{\tau}-\frac{\mathcal{L}^{2}}{\tau^{2}}+\zeta\left(\eta-\frac{-W_{\mathrm{o}}-\mathcal{L}^{2}/\tau}{-W_{\mathrm{o}}/\eta_{\mathrm{o}}-\mathcal{L}_{+}\mathcal{L}/\tau}\right),
\end{equation}
we calculate the maximum power at given efficiency through

\begin{equation}
\frac{\partial\mathcal{F}}{\partial\tau}=0,\quad\frac{\partial\mathcal{F}}{\partial\zeta}=0.
\end{equation}
The solution is 

\begin{align}
\tau & =\frac{\eta_{\mathrm{o}}(1-\eta\mathcal{L}_{+}/\mathcal{L})}{\eta_{\mathrm{o}}-\eta}\frac{\mathcal{L}^{2}}{(-W_{\mathrm{o}})},
\end{align}
and the maximum power at given efficiency is

\begin{align}
\frac{P_{\eta}}{P_{\mathrm{max}}} & =\frac{4(\eta_{\mathrm{o}}-\eta)(1-\eta_{\mathrm{o}}\mathcal{L}_{+}/\mathcal{L})\eta}{\eta_{\mathrm{o}}^{2}(1-\eta\mathcal{L}_{+}/\mathcal{L})^{2}}.
\end{align}

\begin{figure}
\includegraphics[width=7cm]{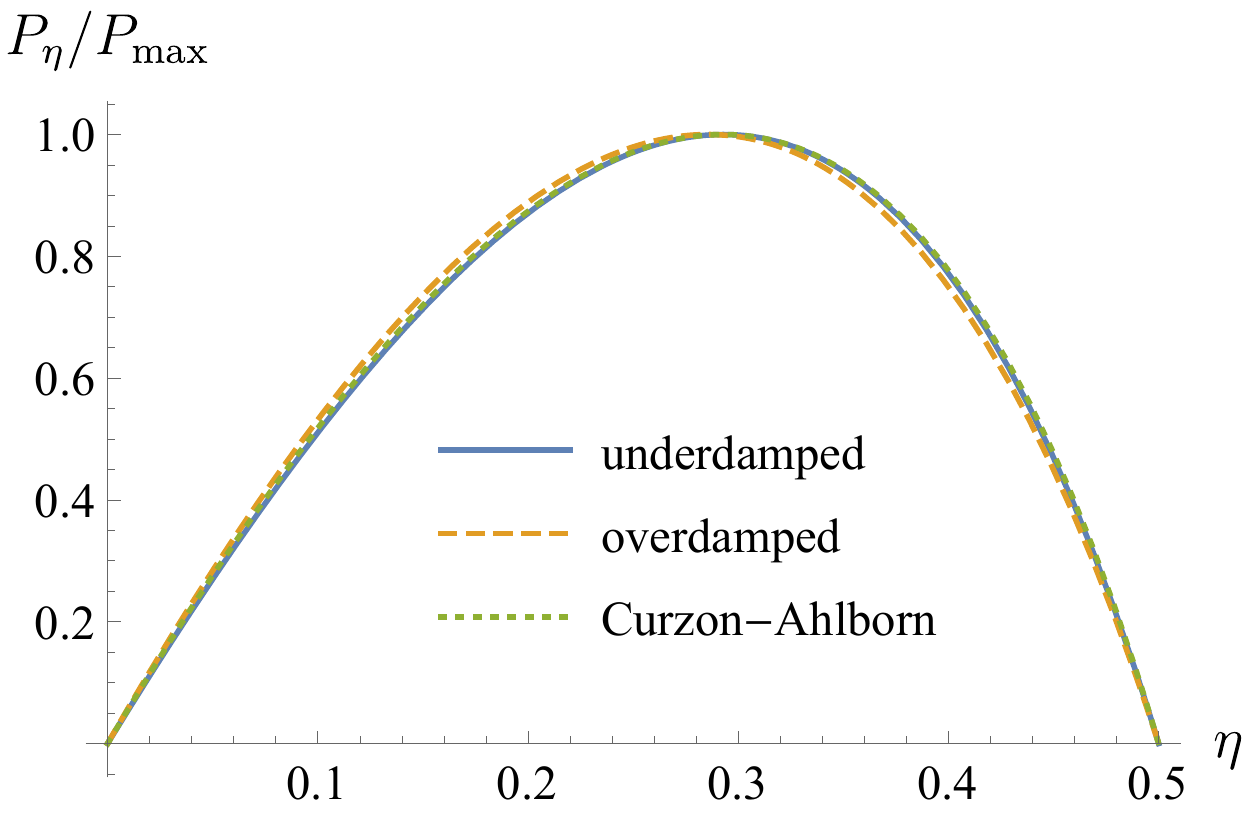}

\caption{The maximum power at given efficiency. We set $\eta_{\mathrm{C}}=1/2$
with $\eta_{\mathrm{CA}}=1-1/\sqrt{2}$ for the highly underdamped
regime and $\eta_{\mathrm{o}}^{\mathrm{over}}=1/2$ for the overdamped
regime. \label{fig:tradeoff_underdamped}}
\end{figure}

In the highly underdamped regime, the quasistatic efficiency is the
Carnot efficiency $\eta_{\mathrm{o}}^{\mathrm{under}}=\eta_{\mathrm{C}}$,
and ratio of the thermodynamic lengths is $\mathcal{L}_{+}^{\mathrm{under}}/\mathcal{L}^{\mathrm{under}}=\sqrt{T_{H}}/(\sqrt{T_{H}}+\sqrt{T_{L}})$
{[}see Eq. (\ref{eq:lengthratiounder}){]}. The maximum power at given
efficiency is

\begin{equation}
\frac{P_{\eta}^{\mathrm{under}}}{P_{\mathrm{max}}}=\frac{4\sqrt{T_{L}/T_{H}}\left(\eta_{\mathrm{C}}-\eta\right)\eta}{\left(\eta_{\mathrm{C}}-\eta\eta_{\mathrm{CA}}\right)^{2}}.\label{eq:trade-offrelation_underdamped_shortcut}
\end{equation}
Notice that the maximum power at given efficiency in the shortcut
strategy is different from the one obtained in the highly underdamped
regime \citep{Chen1989,Dechant2017,Chen2021b}
\begin{equation}
\frac{\tilde{P}_{\eta}^{\mathrm{under}}}{P_{\mathrm{max}}}=\frac{\left(\eta_{\mathrm{C}}-\eta\right)\eta}{\left(1-\eta\right)\eta_{\mathrm{CA}}^{2}},\label{eq:trade-offrelation_CA}
\end{equation}
but they are quite similar, as shown in Fig. \ref{fig:tradeoff_underdamped}.
We explain the difference in Section \ref{sec:Underdamped-limit}.

In the overdamped regime, the quasistatic efficiency $\eta_{\mathrm{o}}^{\mathrm{over}}$
is given by Eq. (\ref{eq:quasistatic_efficiency_over}), and the ratio
of the thermodynamic lengths is $\mathcal{L}_{+}^{\mathrm{over}}/\mathcal{L}^{\mathrm{over}}=1/2$.
The maximum power at given efficiency is

\begin{equation}
\frac{P_{\eta}^{\mathrm{over}}}{P_{\mathrm{max}}}=\frac{4(\eta_{\mathrm{o}}^{\mathrm{over}}-\eta)(1-\eta_{\mathrm{o}}^{\mathrm{over}}/2)\eta}{(\eta_{\mathrm{o}}^{\mathrm{over}})^{2}(1-\eta_{\mathrm{o}}^{\mathrm{over}}/2)^{2}},
\end{equation}
where $\eta_{\mathrm{o}}^{\mathrm{over}}$ is given by Eq. (\ref{eq:quasistatic_efficiency_over}).

In Fig. \ref{fig:tradeoff_underdamped}, we show the maximum power
at given efficiency in the highly underdamped and the overdamped regimes
by setting $\eta_{\mathrm{C}}=1/2$ and $\eta_{\mathrm{o}}^{\mathrm{over}}=1/2$.
We also show the maximum power at given efficiency {[}Eq. (\ref{eq:trade-offrelation_CA}){]}
for the Curzon-Ahlborn heat engine \citep{Chen1989,Dechant2017,Chen2021b},
which is quite similar to that with the shortcut strategy in the highly
underdamped regime.

\section{Highly underdamped regime $\chi\ll1$ \label{sec:Underdamped-limit}}

In the highly underdamped regime $\chi\ll1$, the auxiliary functions
are reduced to 

\begin{align}
h_{\lambda}^{\mathrm{under}} & =\frac{n+1}{2n\kappa\lambda}\left(\frac{1}{2m}p^{2}+\frac{1}{2n}m\lambda^{n+1}x^{2n}\right),\\
h_{\beta}^{\mathrm{under}} & =\frac{n+1}{2n\kappa\beta}\left(\frac{1}{2m}p^{2}+\frac{1}{2n}m\lambda^{n+1}x^{2n}\right).
\end{align}
The auxiliary Hamiltonian becomes 

\begin{equation}
H_{\mathrm{a}}^{\mathrm{under}}=\frac{n+1}{2n\kappa}\left(\frac{\dot{\lambda}}{\lambda}+\frac{\dot{\beta}}{\beta}\right)\left(\frac{1}{2m}p^{2}+\frac{1}{2n}m\lambda^{n+1}x^{2n}\right),\label{eq:H_a_underdamped}
\end{equation}
which is proportional to the original Hamiltonian. Therefore, an equilibrium
state of the original Hamiltonian is also an equilibrium state of
the total Hamiltonian with a different effective temperature $T_{\mathrm{eff}}^{\mathrm{under}}$,
which is explicitly

\begin{equation}
T_{\mathrm{eff}}^{\mathrm{under}}=\left[1+\frac{n+1}{2n\kappa}\left(\frac{\dot{\lambda}}{\lambda}+\frac{\dot{\beta}}{\beta}\right)\right]\frac{1}{\beta}.\label{eq:T_under_eff}
\end{equation}

The irreversible work and the thermodynamic length are dominated by
the first term of Eqs. (\ref{eq:Wirr}) and (\ref{eq:thermodynamic_length}),
and are explicitly

\begin{align}
W_{\mathrm{irr}}^{\mathrm{under}} & =\frac{(n+1)^{2}}{4n^{2}\kappa\beta}\int_{0}^{\tau}\left(\frac{\dot{\lambda}}{\lambda}+\frac{\dot{\beta}}{\beta}\right)^{2}dt,\label{eq:71}\\
\mathcal{L}^{\mathrm{under}} & =\frac{n+1}{2n\sqrt{\kappa\beta}}\oint\left|\frac{\lambda^{\prime}(s)}{\lambda}+\frac{\beta^{\prime}(s)}{\beta}\right|ds.\label{eq:72}
\end{align}
It is worth noting that the expressions (\ref{eq:71}) and (\ref{eq:72})
depend on $n$ through the effective degree of freedom $f_{n}=1+1/n$.
The auxiliary Hamiltonian (\ref{eq:H_a_underdamped}) is zero on the
exponential paths with $\alpha=1$, and the irreversible work and
the thermodynamic length are also zero. The optimal connecting paths
should be chosen as two of these exponential paths (IV) $T=T_{L}\exp(r-r_{1})$
and (II) $T=T_{L}\exp(r-r_{2})$ with zero thermodynamic lengths $\mathcal{L}_{\mathrm{IV}}^{\mathrm{under}}=\mathcal{L}_{\mathrm{II}}^{\mathrm{under}}=0$,
and the boundary values of the control parameters are related by

\begin{align}
r_{1}^{\prime} & =r_{1}+\ln\left(\frac{T_{H}}{T_{L}}\right),\label{eq:r1primeunder}\\
r_{2}^{\prime} & =r_{2}+\ln\left(\frac{T_{H}}{T_{L}}\right).\label{eq:r2primeunder}
\end{align}
The quasistatic processes on these paths are adiabatic since $\dot{Q}_{\mathrm{o}}=0$
{[}Eq. (\ref{eq:Qo_dot}){]}. In the highly underdamped regime, the
finite-time processes on these paths are also adiabatic, namely $Q_{\mathrm{IV}}^{\mathrm{under}}=0$
and $Q_{\mathrm{II}}^{\mathrm{under}}=0$, since $\dot{Q}=\dot{Q}_{\mathrm{o}}+\dot{Q}_{\mathrm{irr}}=0$.

The quasistatic heat of the two isothermal processes is obtained from
Eq. (\ref{eq:Qo_dot}) as

\begin{align}
Q_{\mathrm{o,I}}^{\mathrm{under}} & =-\frac{n+1}{2n}\int_{r_{1}}^{r_{2}}T_{L}dr\\
 & =-\frac{n+1}{2n}T_{L}(r_{2}-r_{1}),
\end{align}
and

\begin{align}
Q_{\mathrm{o,III}}^{\mathrm{under}} & =-\frac{n+1}{2n}\int_{r_{2}^{\prime}}^{r_{1}^{\prime}}T_{H}dr\\
 & =\frac{n+1}{2n}T_{H}(r_{2}-r_{1}),
\end{align}
Since the connecting processes are adiabatic, the overall heat absorbed
in a quasistatic cycle is $Q_{\mathrm{o,+}}^{\mathrm{under}}=Q_{\mathrm{o,III}}^{\mathrm{under}}$,
and the overall heat released is $Q_{\mathrm{o,-}}^{\mathrm{under}}=Q_{\mathrm{o,I}}^{\mathrm{under}}$.
The area of the closed path is 

\begin{equation}
\mathcal{A}^{\mathrm{under}}=(T_{H}-T_{L})(r_{2}-r_{1}),\label{eq:area_both_underdamped=000026overdamped}
\end{equation}
and the quasistatic work of the cycle is 
\begin{align}
W_{\mathrm{o}}^{\mathrm{under}} & =-\frac{n+1}{2n}(T_{H}-T_{L})(r_{2}-r_{1}).
\end{align}
The efficiency of the quasistatic cycle is the Carnot efficiency

\begin{align}
\eta_{\mathrm{o}}^{\mathrm{under}} & =-\frac{W_{\mathrm{o}}^{\mathrm{under}}}{Q_{\mathrm{o,+}}^{\mathrm{under}}}\\
 & =1-\frac{T_{L}}{T_{H}}.\label{eq:quasistatic_efficiency_under}
\end{align}

The thermodynamic lengths (\ref{eq:lengthI}) and (\ref{eq:lengthIII})
of the isothermal paths are simplified into

\begin{equation}
\mathcal{L}_{\mathrm{I}}^{\mathrm{under}}=\frac{n+1}{2n}\sqrt{\frac{T_{L}}{\kappa}}(r_{2}-r_{1}),
\end{equation}
and

\begin{equation}
\mathcal{L}_{\mathrm{III}}^{\mathrm{under}}=\frac{n+1}{2n}\sqrt{\frac{T_{H}}{\kappa}}(r_{2}-r_{1}).
\end{equation}
They satisfy the relation 
\begin{equation}
\frac{\mathcal{L}_{\mathrm{I}}^{\mathrm{under}}}{\mathcal{L}_{\mathrm{III}}^{\mathrm{under}}}=\sqrt{\frac{T_{L}}{T_{H}}}.\label{eq:lengthratiounder}
\end{equation}
The thermodynamic lengths of the heat absorbed path and the heat released
path are $\mathcal{L}_{+}^{\mathrm{under}}=\mathcal{L}_{\mathrm{III}}^{\mathrm{under}}$
and $\mathcal{L}_{-}^{\mathrm{under}}=\mathcal{L}_{\mathrm{I}}^{\mathrm{under}}$.
The thermodynamic length of the whole cycle is 
\begin{equation}
\mathcal{L}^{\mathrm{under}}=\frac{n+1}{2n}\frac{\sqrt{T_{L}}+\sqrt{T_{H}}}{\sqrt{\kappa}}(r_{2}-r_{1}).
\end{equation}
The efficiency at the maximum power {[}Eq. (21) in the main text{]}
is then obtained with the results of $\eta_{\mathrm{o}}^{\mathrm{under}}$,
$\mathcal{L}_{\mathrm{+}}^{\mathrm{under}}$ and $\mathcal{L}^{\mathrm{under}}$.

\subsection{Optimal protocol}

The optimal protocol on the closed path is given by varying the control
parameters with a constant velocity of the thermodynamic length, i.e.,

\begin{equation}
\frac{d\mathcal{L}}{ds}=\mathcal{L}^{\mathrm{under}},\label{eq:68-1}
\end{equation}
where $s\in[0,1]$ is the rescaled time. The control scheme of the
work parameter is given by

\begin{equation}
r^{\mathrm{under}}(s)=\begin{cases}
r_{1}+(r_{2}-r_{1})\left(1+\sqrt{\frac{T_{H}}{T_{L}}}\right)s & 0<s<\frac{1}{1+\sqrt{T_{H}/T_{L}}}\\
r_{2}^{\prime}+(r_{1}^{\prime}-r_{2}^{\prime})\left(1+\sqrt{\frac{T_{L}}{T_{H}}}\right)\left(s-\frac{1}{1+\sqrt{T_{H}/T_{L}}}\right) & \frac{1}{1+\sqrt{T_{H}/T_{L}}}<s<1
\end{cases},\label{eq:r_underdamped}
\end{equation}
or 

\begin{equation}
\lambda^{\mathrm{under}}(s)=\begin{cases}
\lambda_{1}\left(\frac{\lambda_{2}}{\lambda_{1}}\right)^{\left(1+\sqrt{T_{H}/T_{L}}\right)s} & 0<s<\frac{1}{1+\sqrt{T_{H}/T_{L}}}\\
\lambda_{2}^{\prime}\left(\frac{\lambda_{1}^{\prime}}{\lambda_{2}^{\prime}}\right)^{\left(1+\sqrt{T_{L}/T_{H}}\right)s-\sqrt{T_{L}/T_{H}}} & \frac{1}{1+\sqrt{T_{H}/T_{L}}}<s<1
\end{cases},\label{eq:lambdaunder-1}
\end{equation}
where $\lambda_{1}=\lambda_{0}\exp(r_{1})$, $\lambda_{2}=\lambda_{0}\exp(r_{2})$,
$\lambda_{1}^{\prime}=\lambda_{0}\exp(r_{1}^{\prime})$ and $\lambda_{2}^{\prime}=\lambda_{0}\exp(r_{2}^{\prime})$.
The control scheme of the temperature is given by

\begin{equation}
T^{\mathrm{under}}(s)=\begin{cases}
T_{L} & 0<s<\frac{1}{1+\sqrt{T_{H}/T_{L}}}\\
T_{H} & \frac{1}{1+\sqrt{T_{H}/T_{L}}}<s<1
\end{cases}.
\end{equation}
It is worth noting that the optimal protocol in the highly underdamped
regime does not depend on the shape of the potential $n$. 

The evolution of the effective temperature in a cycle according to
Eq. (\ref{eq:T_under_eff}) is 

\begin{equation}
T_{\mathrm{eff}}^{\mathrm{under}}(s)=\begin{cases}
T_{L}\left[1+\frac{n+1}{2n\kappa\tau}\left(1+\sqrt{T_{H}/T_{L}}\right)\ln\left(\lambda_{2}/\lambda_{1}\right)\right] & 0<s<\frac{1}{1+\sqrt{T_{H}/T_{L}}}\\
T_{H}\left[1+\frac{n+1}{2n\kappa\tau}\left(1+\sqrt{T_{L}/T_{H}}\right)\ln\left(\lambda_{1}^{\prime}/\lambda_{2}^{\prime}\right)\right] & \frac{1}{1+\sqrt{T_{H}/T_{L}}}<s<1
\end{cases}.\label{eq:effective_T_underdamped}
\end{equation}
The operation time consumed in each process, according to Eq. (\ref{eq:operation_time_in_each process}),
is

\begin{align}
\tau_{\mathrm{I}}^{\mathrm{under}} & =\frac{\sqrt{T_{L}}}{\sqrt{T_{L}}+\sqrt{T_{H}}}\tau\\
\tau_{\mathrm{II}}^{\mathrm{under}} & =0\\
\tau_{\mathrm{III}}^{\mathrm{under}} & =\frac{\sqrt{T_{H}}}{\sqrt{T_{L}}+\sqrt{T_{H}}}\tau\\
\tau_{\mathrm{IV}}^{\mathrm{under}} & =0.
\end{align}
Compared to the isothermal processes I and III, the connecting processes
II and IV consume negligible time.

\begin{figure}
\includegraphics[width=16cm]{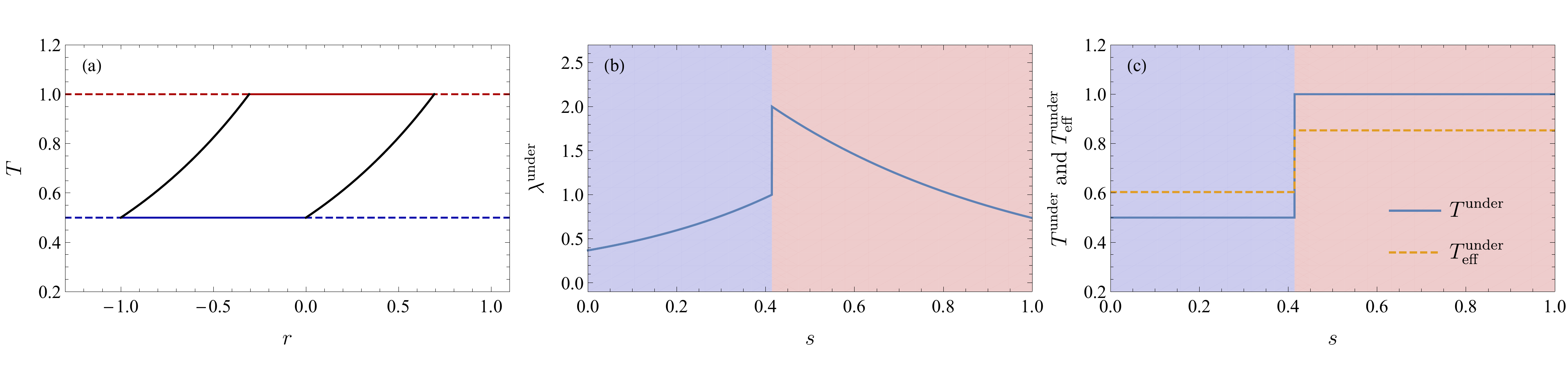}

\caption{Optimal protocol in the highly underdamped regime. (a) The $r-T$
diagram. (b) The optimal protocol to vary the frequency $\lambda^{\mathrm{under}}(s)$.
(c) The evolution of the temperatures. \label{fig:underdamped_cycle}}
\end{figure}

Figure \ref{fig:underdamped_cycle} shows the optimal protocol of
the cycle in the highly underdamped regime. We consider the harmonic
potential $n=1$ here. Figure \ref{fig:underdamped_cycle}(a) is the
$r-T$ diagram of the cycle. The boundary values of the frequency
on the low-temperature line $T=T_{L}=1/2$ are $\lambda_{1}=1/e$,
$\lambda_{2}=1$ with $e=2.718$. These parameters are the same with
those in the main text. The boundary values of the frequency on the
high-temperature line $T=T_{H}=1$ are $\lambda_{1}^{\prime}=\lambda_{1}T_{H}/T_{L}=2/e$
and $\lambda_{2}^{\prime}=\lambda_{2}T_{H}/T_{L}=2$. The friction
coefficient is set to be $\kappa=0.1$. The area of the closed path
is $\mathcal{A}^{\mathrm{under}}=1/2$, and the thermodynamic length
of the cycle is $\mathcal{L}^{\mathrm{under}}=5.40$. We set the operation
time of the cycle to be $\tau_{\mathrm{max}}^{\mathrm{under}}=116.57$
to reach the maximum output power $P_{\mathrm{max}}^{\mathrm{under}}=0.0021$.
Figure \ref{fig:underdamped_cycle}(b) shows the optimal protocol
(\ref{eq:lambdaunder-1}) to vary the frequency $\lambda^{\mathrm{under}}(s)$
in a cycle. Figure \ref{fig:underdamped_cycle}(c) shows the evolution
of the temperatures in a cycle. The effective temperature $T_{\mathrm{eff}}^{\mathrm{under}}$
defined by Eq. (\ref{eq:effective_T_underdamped}) remain unchanged
during the isothermal processes.

\subsection{Comparison to Refs. \citep{Dechant2017,Chen2021b}}

In the shortcut strategy, we need to implement the auxiliary Hamiltonian
$H_{\mathrm{a}}$ {[}Eq. (\ref{eq:H_a_underdamped}){]}. The moment
term in $H_{\mathrm{a}}$ changes the mass of the particle. Notice
that in Refs. \citep{Dechant2017,Chen2021b} only the potential can
be varied in the isothermal processes, not the mass of the particle.
As a result, the optimal protocol (\ref{eq:lambdaunder-1}) is not
exactly the same as the one obtained in Refs. \citep{Dechant2017,Chen2021b}.
The ratio of the time consumed in the two isothermal process is $\tau_{\mathrm{I}}^{\mathrm{under}}/\tau_{\mathrm{III}}^{\mathrm{under}}=\sqrt{T_{L}/T_{H}}$,
while the ratio is equal to one in the protocol of Refs. \citep{Dechant2017,Chen2021b}.
It is the same reason that the maximum power at given efficiency {[}Eq.
(\ref{eq:trade-offrelation_underdamped_shortcut}){]} is different
from the one (\ref{eq:trade-offrelation_CA}) obtained in Refs. \citep{Chen1989,Dechant2017,Chen2021b}.

We continue to compare the expressions of the output power in this
work and the one obtained in \citep{Chen2021b}. In this work, the
output power according to Eq. (\ref{eq:boundofpower}) is bounded
by 

\begin{align}
P^{\mathrm{under}} & \leq\frac{n+1}{2n}(T_{H}-T_{L})(r_{2}-r_{1})\frac{1}{\tau}-\left[\frac{n+1}{2n}\frac{\sqrt{T_{L}}+\sqrt{T_{H}}}{\sqrt{\kappa}}(r_{2}-r_{1})\right]^{2}\frac{1}{\tau^{2}}.\label{eq:power_underdamped_bound}
\end{align}
The irreversible work of the cycle is rigorously proportional to the
inverse of the operation time. In Ref. \citep{Chen2021b}, we have
also obtained the bound of the output power {[}see Eq. (13) of Ref.
\citep{Chen2021b}{]}, which in the current notation is explicitly
\begin{equation}
\tilde{P}^{\mathrm{under}}\leq\frac{\kappa}{2}\left(\frac{T_{H}}{\frac{\tau}{2}\frac{2n}{n+1}\frac{\kappa}{(r_{2}-r_{1})}+1}-\frac{T_{L}}{\frac{\tau}{2}\frac{2n}{n+1}\frac{\kappa}{(r_{2}-r_{1})}-1}\right),\label{eq:power_underdamped_bound_from_curzon_ahlborn}
\end{equation}
We have substituted the parameters $\Gamma_{H}$, $\Gamma_{C}$, $\tau_{H}$,
$\tau_{C}$, $r$, $T_{C}$ in Ref. \citep{Chen2021b} with the current
parameters 

\begin{align}
\Gamma_{H} & =\Gamma_{C}\rightarrow\frac{2n}{n+1}\kappa\\
\tau_{H} & =\tau_{C}\rightarrow\frac{\tau}{2}\\
\ln r & \rightarrow r_{2}-r_{1}\\
T_{C} & \rightarrow T_{L}.
\end{align}
One can see that the expressions (\ref{eq:power_underdamped_bound})
and (\ref{eq:power_underdamped_bound_from_curzon_ahlborn}) are different,
but share the same maximum value $P_{\mathrm{max}}^{\mathrm{under}}=\kappa(\sqrt{T_{H}}-\sqrt{T_{L}})^{2}/4$
by choosing a proper value of the operation time $\tau$.

\section{Overdamped regime $\chi\gg1$\label{sec:Overdamped-regime}}

In the overdamped regime $\chi\gg1$, the auxiliary functions are
reduced to

\begin{align}
h_{\lambda}^{\mathrm{over}} & =\frac{n+1}{4n\lambda}\kappa mx^{2},\\
h_{\beta}^{\mathrm{over}} & =\frac{1}{4n\beta}\kappa mx^{2}.
\end{align}
The auxiliary Hamiltonian becomes 

\begin{equation}
H_{\mathrm{a}}^{\mathrm{over}}=\frac{\kappa m}{4n}\left[(n+1)\frac{\dot{\lambda}}{\lambda}+\frac{\dot{\beta}}{\beta}\right]x^{2},\label{eq:H_a_over}
\end{equation}
which is just a harmonic potential of the position. 

The irreversible work and the thermodynamic length are dominated by
the second term of Eqs. (\ref{eq:Wirr}) and (\ref{eq:thermodynamic_length}),
and are explicitly 
\begin{align}
W_{\mathrm{irr}}^{\mathrm{over}} & =\frac{(n+1)^{2}}{4n^{2}}\int_{0}^{\tau}\frac{\kappa m}{\lambda}\frac{R_{n}}{(\lambda m\beta)^{1/n}}\left(\frac{\dot{\lambda}}{\lambda}+\frac{\dot{\beta}}{(n+1)\beta}\right)^{2}dt,\\
\mathcal{L}^{\mathrm{over}} & =\frac{n+1}{2n}\oint\sqrt{\frac{\kappa m}{\lambda}\frac{R_{n}}{(\lambda m\beta)^{1/n}}}\left|\frac{\lambda^{\prime}(s)}{\lambda}+\frac{\beta^{\prime}(s)}{(n+1)\beta}\right|ds.
\end{align}
On the exponential paths with $\alpha=n+1$, the auxiliary Hamiltonian
(\ref{eq:H_a_over}) is zero, and the irreversible work and the thermodynamic
length are also zero. The optimal connecting paths should be chosen
as two of these exponential paths (IV) $T=T_{L}\exp[(n+1)(r-r_{1})]$
and (II) $T=T_{L}\exp[(n+1)(r-r_{2})]$ with zero thermodynamic lengths
$\mathcal{L}_{\mathrm{IV}}^{\mathrm{over}}=\mathcal{L}_{\mathrm{II}}^{\mathrm{over}}=0$,
and the boundary values of the control parameters are related by

\begin{align}
r_{1}^{\prime} & =r_{1}+\frac{1}{n+1}\ln\left(\frac{T_{H}}{T_{L}}\right),\label{eq:r1primeover}\\
r_{2}^{\prime} & =r_{2}+\frac{1}{n+1}\ln\left(\frac{T_{H}}{T_{L}}\right).\label{eq:r2primeover}
\end{align}

We calculate the quasistatic heat of each process according to Eq.
(\ref{eq:Qo_dot}). The quasistatic heat of the two connecting processes
is

\begin{align}
Q_{\mathrm{o,II}}^{\mathrm{over}} & =\frac{n+1}{2}\int_{r_{2}}^{r_{2}^{\prime}}T_{L}\exp[(n+1)(r-r_{2})]dr\\
 & =\frac{1}{2}\left(T_{H}-T_{L}\right),
\end{align}
and

\begin{equation}
Q_{\mathrm{o,IV}}^{\mathrm{over}}=-\frac{1}{2}\left(T_{H}-T_{L}\right).
\end{equation}
The heat $Q_{\mathrm{o,II}}^{\mathrm{over}}$ and $Q_{\mathrm{o,IV}}^{\mathrm{over}}$
are equal to the change of the kinetic energy at two temperatures
$T_{L}$ and $T_{H}$. In a whole cycle, besides the heat generated
in the isothermal processes, additional heat $(T_{H}-T_{L})/2$ is
transferred from the hot bath to the cold bath. This additional heat
is the heat leakage \citep{Schmiedl2007}. The quasistatic heat of
the two isothermal processes is

\begin{align}
Q_{\mathrm{o,I}}^{\mathrm{over}} & =-\frac{n+1}{2n}T_{L}\left(r_{2}-r_{1}\right),
\end{align}
and

\begin{align}
Q_{\mathrm{o,III}}^{\mathrm{over}} & =\frac{n+1}{2n}T_{H}\left(r_{2}-r_{1}\right).
\end{align}
In a quasistatic cycle, the overall heat absorbed is

\begin{align}
Q_{\mathrm{o},+}^{\mathrm{over}} & =Q_{\mathrm{o,III}}^{\mathrm{over}}+Q_{\mathrm{o,II}}^{\mathrm{over}}\\
 & =\frac{n+1}{2n}T_{H}\left(r_{2}-r_{1}\right)+\frac{1}{2}\left(T_{H}-T_{L}\right)
\end{align}
and the overall heat released is

\begin{align}
Q_{\mathrm{o},-}^{\mathrm{over}} & =Q_{\mathrm{o,I}}^{\mathrm{over}}+Q_{\mathrm{o,IV}}^{\mathrm{over}}\\
 & =-\frac{n+1}{2n}T_{L}\left(r_{2}-r_{1}\right)-\frac{1}{2}\left(T_{H}-T_{L}\right).
\end{align}
The area of the closed path is 

\begin{equation}
\mathcal{A}^{\mathrm{over}}=\left(T_{H}-T_{L}\right)\left(r_{2}-r_{1}\right),
\end{equation}
and the quasistatic work of the cycle is 

\begin{equation}
W_{\mathrm{o}}^{\mathrm{over}}=-\frac{n+1}{2n}\left(T_{H}-T_{L}\right)\left(r_{2}-r_{1}\right).
\end{equation}
The quasistatic efficiency is

\begin{align}
\eta_{\mathrm{o}}^{\mathrm{over}} & =-\frac{W_{\mathrm{o}}^{\mathrm{over}}}{Q_{\mathrm{o},+}^{\mathrm{over}}}\\
 & =\frac{\eta_{\mathrm{C}}}{1+\frac{n}{n+1}\frac{\eta_{\mathrm{C}}}{\left(r_{2}-r_{1}\right)}}.\label{eq:quasistatic_efficiency_over}
\end{align}

The thermodynamic lengths (\ref{eq:lengthI}) and (\ref{eq:lengthIII})
of the isothermal paths are simplified into

\begin{align}
\mathcal{L}_{\mathrm{I}}^{\mathrm{over}} & =\sqrt{R_{n}\frac{\kappa m}{\lambda_{0}}\left(\frac{T_{L}}{\lambda_{0}m}\right)^{1/n}}\left(e^{-\frac{(n+1)}{2n}r_{1}}-e^{-\frac{(n+1)}{2n}r_{2}}\right),\label{eq:LIover}
\end{align}
and

\begin{align}
\mathcal{L}_{\mathrm{III}}^{\mathrm{over}} & =\sqrt{R_{n}\frac{\kappa m}{\lambda_{0}}\left(\frac{T_{H}}{\lambda_{0}m}\right)^{1/n}}\left(e^{-\frac{(n+1)}{2n}r_{1}^{\prime}}-e^{-\frac{(n+1)}{2n}r_{2}^{\prime}}\right).\label{eq:LIIIover}
\end{align}
The two thermodynamic lengths are equal $\mathcal{L}_{\mathrm{III}}^{\mathrm{over}}=\mathcal{L}_{\mathrm{I}}^{\mathrm{over}}$
by plugging Eqs. (\ref{eq:r1primeover}) and (\ref{eq:r2primeover})
into Eq. (\ref{eq:LIIIover}). Thus, in the overdamped regime, the
two isothermal processes consume the same operation time in the optimal
cycle. Since the thermodynamic lengths of the connecting paths are
zero, the thermodynamic lengths of the heat absorbed path and the
heat released path are $\mathcal{L}_{\mathrm{+}}^{\mathrm{over}}=\mathcal{L}_{\mathrm{III}}^{\mathrm{over}}$
and $\mathcal{L}_{-}^{\mathrm{over}}=\mathcal{L}_{\mathrm{I}}^{\mathrm{over}}$.
The thermodynamic length of the whole cycle is

\begin{equation}
\mathcal{L}^{\mathrm{over}}=2\sqrt{R_{n}\frac{\kappa m}{\lambda_{0}}\left(\frac{T_{L}}{\lambda_{0}m}\right)^{1/n}}\left(e^{-\frac{(n+1)}{2n}r_{1}}-e^{-\frac{(n+1)}{2n}r_{2}}\right).
\end{equation}
The efficiency at the maximum power {[}Eq. (26) in the main text{]}
is then obtained with the results of $\eta_{\mathrm{o}}^{\mathrm{over}}$,
$\mathcal{L}_{\mathrm{+}}^{\mathrm{over}}$ and $\mathcal{L}^{\mathrm{over}}$. 

\subsection{Optimal protocol}

The optimal protocol is solved from

\begin{equation}
\frac{d\mathcal{L}}{ds}=\mathcal{L}^{\mathrm{over}}.\label{eq:68}
\end{equation}
We plug the expressions (\ref{eq:LIover}) and (\ref{eq:LIIIover})
into Eq. (\ref{eq:68}), and obtain the control scheme of the work
parameter

\begin{equation}
r^{\mathrm{over}}(s)=\begin{cases}
-\frac{2n}{n+1}\ln[(1-2s)e^{-\frac{(n+1)}{2n}r_{1}}+2se^{-\frac{(n+1)}{2n}r_{2}}] & 0<s<\frac{1}{2}\\
-\frac{2n}{n+1}\ln[(2-2s)e^{-\frac{(n+1)}{2n}r_{2}^{\prime}}+(2s-1)e^{-\frac{(n+1)}{2n}r_{1}^{\prime}}] & \frac{1}{2}<s<1
\end{cases},\label{eq:67}
\end{equation}
or
\begin{equation}
\lambda^{\mathrm{over}}(s)=\begin{cases}
\left[(1-2s)\lambda_{1}^{-\frac{(n+1)}{2n}}+2s\lambda_{2}^{-\frac{(n+1)}{2n}}\right]^{-\frac{2n}{n+1}} & 0<s<\frac{1}{2}\\
\left[(2-2s)\lambda_{2}^{\prime-\frac{(n+1)}{2n}}+(2s-1)\lambda_{1}^{\prime-\frac{(n+1)}{2n}}\right]^{-\frac{2n}{n+1}} & \frac{1}{2}<s<1
\end{cases}.\label{eq:lambda_over-1}
\end{equation}
The control scheme of the temperature is given by

\begin{equation}
T^{\mathrm{over}}(s)=\begin{cases}
T_{L} & 0<s<\frac{1}{2}\\
T_{H} & \frac{1}{2}<s<1
\end{cases}.
\end{equation}
The variance of position in a cycle according to Eq. (\ref{eq:x^2_average})
is 

\begin{equation}
\left\langle x^{2}\right\rangle ^{\mathrm{over}}(s)=\begin{cases}
R_{n}\left(\frac{T_{L}}{m}\right)^{1/n}\left[(1-2s)\lambda_{1}^{-\frac{(n+1)}{2n}}+2s\lambda_{2}^{-\frac{(n+1)}{2n}}\right]^{2} & 0<s<\frac{1}{2}\\
R_{n}\left(\frac{T_{H}}{m}\right)^{1/n}\left[(2-2s)\lambda_{2}^{\prime-\frac{(n+1)}{2n}}+(2s-1)\lambda_{1}^{\prime-\frac{(n+1)}{2n}}\right]^{2} & \frac{1}{2}<s<1
\end{cases}.
\end{equation}
It can be verified that $\left\langle x^{2}\right\rangle ^{\mathrm{over}}(s)$
is continuous at the piecewise points $s=0$ and $s=1/2$. In the
two connecting processes, the control parameters are switched suddenly.
The distribution of the momentum changes to the equilibrium state
at the new temperature, while the distribution of the coordinate remains
unchanged. 

\begin{figure}
\includegraphics[width=16cm]{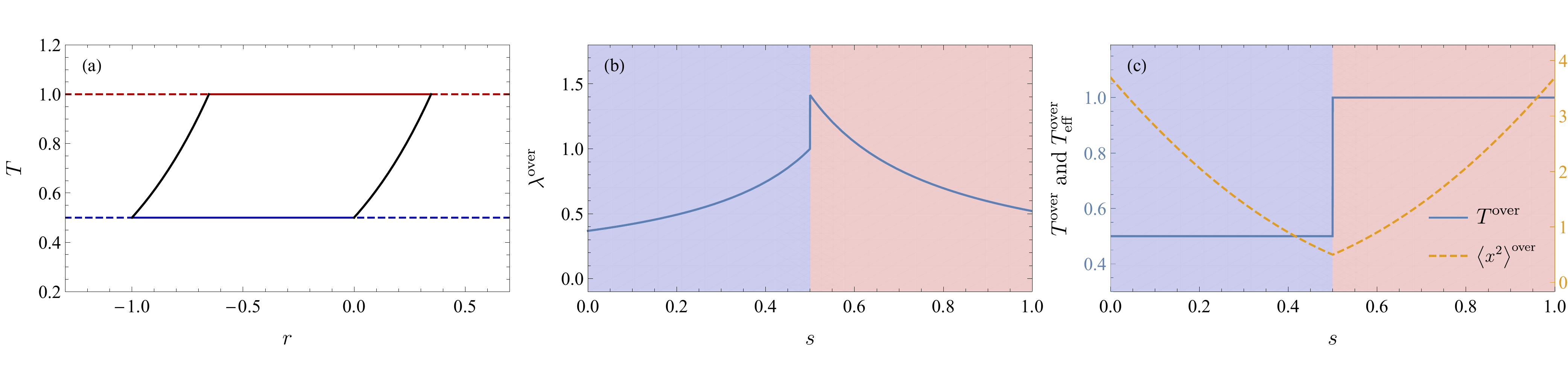}

\caption{Optimal protocol in the overdamped regime. (a) The $r-T$ diagram.
(b) The optimal protocol to vary the frequency $\lambda^{\mathrm{over}}(s)$.
(c) The evolution of the temperatures. \label{fig:overdamped_cycle}}
\end{figure}

Figure \ref{fig:overdamped_cycle} shows the optimal protocol of the
cycle in the overdamped regime. We still consider the harmonic potential
$n=1$. Figure \ref{fig:overdamped_cycle}(a) is the $r-T$ diagram
of the cycle. The boundary values of the frequency on the low-temperature
line $T=T_{L}=1/2$ are $\lambda_{1}=1/e$, $\lambda_{2}=1$ with
$e=2.718$. The boundary values of the frequency on the high-temperature
line $T=T_{H}=1$ are $\lambda_{1}^{\prime}=\lambda_{1}\sqrt{T_{H}/T_{L}}=\sqrt{2}/e$
and $\lambda_{2}^{\prime}=\lambda_{2}\sqrt{T_{H}/T_{L}}=\sqrt{2}$.
The friction coefficient is set to be $\kappa=10$ for the overdamped
regime. The area of the closed path is $\mathcal{A}^{\mathrm{over}}=1/2$,
and the thermodynamic length of the cycle is $\mathcal{L}^{\mathrm{over}}=7.68.$
We set the operation time of the cycle $\tau_{\mathrm{max}}^{\mathrm{over}}=236.20$
to reach the maximum output power $P_{\mathrm{max}}^{\mathrm{over}}=0.0011$.
Figure \ref{fig:overdamped_cycle}(b) shows the optimal protocol (\ref{eq:lambda_over_harmonic})
to vary the frequency $\lambda^{\mathrm{over}}(s)$ in a cycle. Figure
\ref{fig:overdamped_cycle}(c) shows the evolution of the temperature
$T^{\mathrm{over}}(s)$ and the variance of the position $\left\langle x^{2}\right\rangle ^{\mathrm{over}}(s)$
in a cycle. 

\subsection{Comparison to Ref. \citep{Schmiedl2007}}

For the harmonic potential $n=1$, the auxiliary Hamiltonian (\ref{eq:H_a_over})
is explicitly

\begin{equation}
H_{\mathrm{a}}^{\mathrm{over}}=\frac{1}{2}m\kappa\left(\frac{\dot{\lambda}}{\lambda}+\frac{\dot{\beta}}{2\beta}\right)x^{2}.
\end{equation}
The implementation of the auxiliary Hamiltonian is equivalent to changing
the frequency of the harmonic potential into

\begin{equation}
\Lambda^{\mathrm{over}}=\sqrt{\lambda^{2}+\kappa\left(\frac{\dot{\lambda}}{\lambda}+\frac{\dot{\beta}}{2\beta}\right)}.
\end{equation}
Therefore, the optimal protocol with the shortcut strategy in the
overdamped regime converges to the control scheme obtained in Ref.
\citep{Schmiedl2007}. The control scheme (\ref{eq:lambda_over-1})
to vary the work parameter is explicitly

\begin{equation}
\lambda^{\mathrm{over}}(s)=\begin{cases}
\frac{\lambda_{1}}{1-2s+2s\lambda_{1}/\lambda_{2}} & 0<s<\frac{1}{2}\\
\frac{\lambda_{2}^{\prime}}{2-2s+(2s-1)\lambda_{2}^{\prime}/\lambda_{1}^{\prime}} & \frac{1}{2}<s<1
\end{cases},\label{eq:lambda_over_harmonic}
\end{equation}
 The variance of the particle position $\left\langle x^{2}\right\rangle $
is explicitly

\begin{equation}
\left\langle x^{2}\right\rangle ^{\mathrm{over}}(s)=\begin{cases}
\frac{T_{L}}{\lambda_{1}^{2}m}(1-2s+2s\lambda_{1}/\lambda_{2})^{2} & 0<s<\frac{1}{2}\\
\frac{T_{H}}{\lambda_{2}^{\prime2}m}[2-2s+(2s-1)\lambda_{2}^{\prime}/\lambda_{1}^{\prime}]^{2} & \frac{1}{2}<s<1
\end{cases},
\end{equation}
which is the same as the result obtained in Ref. \citep{Schmiedl2007}. 

An effective temperature of the position degree of freedom can be
defined in respective to the altered frequency
\begin{equation}
T_{\mathrm{eff}}^{\mathrm{over}}\coloneqq m(\Lambda^{\mathrm{over}})^{2}\left\langle x^{2}\right\rangle ^{\mathrm{over}}.
\end{equation}
Its evolution in a cycle is

\begin{equation}
T_{\mathrm{eff}}^{\mathrm{over}}(s)=\begin{cases}
\left[1-\frac{2\kappa}{\lambda_{1}^{2}\lambda_{2}^{2}\tau}\left(\lambda_{1}-\lambda_{2}\right)\left[\lambda_{2}+2\left(\lambda_{1}-\lambda_{2}\right)s\right]\right]T_{L} & 0<s<\frac{1}{2}\\
\left[1+\frac{2\kappa}{\lambda_{1}^{\prime2}\lambda_{2}^{\prime2}\tau}\left(\lambda_{1}^{\prime}-\lambda_{2}^{\prime}\right)\left[2\lambda_{1}^{\prime}-\lambda_{2}^{\prime}+2\left(\lambda_{2}^{\prime}-\lambda_{1}^{\prime}\right)s\right]\right]T_{H} & \frac{1}{2}<s<1
\end{cases},\label{eq:Teffover}
\end{equation}
which is not a constant during the isothermal processes.

\bibliographystyle{apsrev4-1_addtitle}
\bibliography{ref}